\documentclass[12pt]{article}
\usepackage{graphicx}
\usepackage{subcaption}
\usepackage{amsmath,amssymb,physics}
\usepackage{url}
\usepackage{caption}
\usepackage{comment}
\usepackage{hyperref}
\usepackage{cite}
\usepackage{tabularray}
\usepackage[export]{adjustbox}
\usepackage[dvipsnames]{xcolor}

\allowdisplaybreaks[4]

\usepackage{float}
\usepackage{blkarray}

\usepackage{rotating}
\usepackage{pdflscape}

\definecolor{darkblue}{rgb}{0, 0, 0.7} 

\begin{document}

\begin{center}

{\large
A minimal implementation of Yang--Mills theory}\\
\vspace{2mm}
{\large
on a digital quantum computer}
\end{center}

\vspace{0.2cm}
\begin{center}
Georg Bergner$\, ^{\rm a}$, Masanori Hanada$\, ^{\rm b}$, Emanuele Mendicelli$\, ^{\rm c}$
\end{center}

\begin{center}

$^{\rm a}$\, Institute for Theoretical Physics, University of Jena\\
Max-Wien-Platz 1, 07743 Jena, Germany\\
$^{\rm a}$\, Leibniz Institute of Photonic Technology, Albert-Einstein-Str. 9, 07745, Jena, Germany\\
\vspace{1mm}
$^{\rm b}$\, School of Mathematical Sciences, Queen Mary University of London\\
Mile End Road, London, E1 4NS, United Kingdom\\
\vspace{1mm}
$^{\rm c}$\, Department of Mathematical Sciences, University of Liverpool\\
Liverpool L69 7ZL, United Kingdom\\
\end{center}

\vspace{0.5cm}

\begin{center}
  {\bf Abstract}
\end{center}
We present a minimal implementation of SU($N$) pure Yang-Mills theory in $3+1$ dimensions for digital quantum simulation, designed to enable quantum advantage. Building on the orbifold lattice simulation protocol with logarithmic scaling in the local Hilbert-space truncation, we introduce further simplified Hamiltonians. Furthermore, we test simple methods that improve the convergence to the infinite mass limit, thereby removing the requirement of a large scalar mass to obtain the Kogut-Susskind Hamiltonian. For the SU(2) theory, we can cut the resource requirement further by utilizing the embedding of $\mathrm{SU}(2)\cong\mathrm{S}^3$ into $\mathbb{R}^4$. Monte Carlo simulations of the Euclidean path integral were used to benchmark the accuracy of these new analytical improvements to the theory. These results provide further support for the noncompact-variable-based approach as a practical framework for quantum simulation of non-Abelian gauge theories.
\vspace{2 cm}

\newpage
\tableofcontents
\section{Introduction}
Yang-Mills theory~\cite{Yang:1954ek} sits at the center of high-energy physics and string theory, providing the framework for the Standard Model and its description of fundamental interactions~\cite{Han:1965pf,Fritzsch:1973pi,Gross:1973id,Politzer:1973fx}\cite{Glashow:1961tr,Weinberg:1967tq,Salam:1968rm}\cite{tHooft:1971qjg,tHooft:1972tcz}. It also plays a central role in approaches to quantum gravity via dualities between supersymmetric Yang-Mills theories and string/M-theory, most notably the AdS/CFT correspondence~\cite{Maldacena:1997re}, where black hole solutions in the gravitational description are mapped to thermal states in the dual gauge theory.

Research efforts toward the quantum simulation of Yang-Mills theories have significant transformative potential, with far-reaching implications for both the development of quantum algorithms and the advancement of quantum hardware. For a long time, it has been anticipated that quantum simulations could address classically challenging problems, including real-time dynamics, systems at non-zero chemical potential, and more generally scenarios affected by the sign problem. In practice, however, progress has been limited. Until very recently, no established and scalable formalism existed for deploying quantum resources to simulate (3+1)-dimensional Yang-Mills theories. Recent developments based on the orbifold lattice formulation~\cite{Kaplan:2002wv,Buser:2020cvn,Bergner:2024qjl,Halimeh:2024bth,Bergner:2025zkj,Halimeh:2025ivn} represent an important step toward addressing this longstanding gap.

Indeed, over the past two decades, efforts toward the quantum simulation of Yang-Mills theory have largely been shaped by the work of Byrnes and Yamamoto~\cite{Byrnes:2005qx}, which demonstrated that, in principle, Yang-Mills theory can be quantum simulated using the Kogut-Susskind Hamiltonian~\cite{Kogut:1974ag}. The approach relies on the use of compact unitary link variables, which are natural for the analytical formulation of the theory. A second advantage is the correspondence with the standard Wilson action in lattice gauge theory. On the other hand, the realization of compact unitary link variables on a quantum computer is highly nontrivial, which became the main obstacle in subsequent efforts to develop quantum simulation protocols~\cite{Martinez:2016yna, Klco:2018kyo, Gorg:2018xyc, Lewis:2019wfx, Schweizer_2019, Mil:2019pbt, Yang:2020yer, Atas:2021ext, Zhou:2021kdl, Wang:2021xra, Su:2022glk, ARahman:2022tkr, Wang:2022dpp, ARahman:2021ktn, Ciavarella:2021nmj, Ciavarella:2021lel, Farrell:2022vyh, Farrell:2022wyt, Atas:2022dqm, Mendicelli:2022ntz, Zhang:2023hzr, Zhu:2024dvz, Kavaki:2024ijd, Ciavarella:2024fzw, Gustafson:2024kym, Than:2024zaj, Li:2025sgo, Froland:2025bqf, Yao:2025cxs, Santra:2025dsm, Jiang:2025ufg, Yang:2025edn} based on this framework. In special cases such as Abelian theories or $(1+1)$-dimensional theories, practically useful solutions have been found. Unfortunately, for the general non-Abelian gauge theory in $3+1$ dimensions, the identification of a suitable representation of the Hamiltonian and translating it into a quantum circuit remains highly challenging, typically requiring impractically large classical and quantum computational resources~\cite{Hanada:2025yzx}. As a consequence, there is no clear roadmap to reach the relevant scaling regime toward the continuum limit of the theory. The goal of practical quantum simulation of lattice gauge theories and quantum advantage in this research field depends not only on the availability of fault-tolerant digital quantum computers, but also on conceptual improvements of the lattice formulation of gauge theories. For further details and recent reviews of quantum simulations of lattice gauge theories (LGT) see~\cite{Banuls:2019bmf,Zohar:2021nyc, Klco:2021lap, Bauer:2022hpo, Halimeh:2025vvp}.

A fundamental challenge in the implementation of gauge theories relates to the nature of the \emph{compact unitary link variables} that represent the gauge fields on the lattice. In analytic approaches based on strong coupling methods, these provide technically viable solutions. In numerical algorithms and weak coupling calculations, however, they give rise to considerable complications. In the development of quantum simulation protocols, the group variables have been identified as a limiting factor even for the relatively simple case of SU(2). This can be illustrated by the natural attempt to represent elements of the group manifold $\mathrm{SU}(2) \cong \mathrm{S}^3$ by polar coordinates or spherical harmonics. Such approaches are not straightforward on a quantum computer, as they require encoding continuous variables into qubits and handling nontrivial operators such as the Laplacian with native gates on a quantum device. This technical issue becomes severer for SU($N$) with $N\ge 3$. What may appear as a minor issue has proven to be highly nontrivial, as the detailed, complicated computations in Ref.~\cite{Byrnes:2005qx} showed. 

The obvious solution would be to abandon the constraints of the group manifold. However, this has far-reaching consequences if done in an arbitrary, uncontrolled manner. The main rationale is to embed the group manifold into a larger space and allow controllable fluctuations around its constrained subspace. These fluctuations represent an additional field such that gauge invariance is never actually broken, once the gauge transformation of this field is taken into account. On the lattice, the links are represented by modified matrices (decomposed into two matrices): the standard compact gauge links are not abandoned, but extended by scalar degrees of freedom.  This is the essential idea behind the orbifold lattice construction. It is summarized in later sections and further details can be found in \cite{Kaplan:2002wv,Buser:2020cvn,Bergner:2024qjl,Halimeh:2024bth,Bergner:2025zkj,Halimeh:2025ivn}. The orbifold lattice formulation might seem unconventional at first glance, unless one has experience in formal QFT. The current work aims to present it as part of a more general set of approaches that embed the group manifold into a larger space and present appropriate Lagrangian or Hamiltonian formulations of the corresponding link variables. It is shown that the lattice formulation appears in this context as a natural and simple solution. 

In this work, the presentation starts from the orbifold lattice Hamiltonian and successively simplifies the approach. However, it can also be regarded in the opposite direction: starting with the Kogut-Susskind Hamiltonian and exploring minimal embeddings of the group into $\mathbb{R}^{n}$. This leads to a theory with additional fields analogous to scalar (which we simply call ``scalar" from now on), but, if applied for a simple Kogut-Susskind formulation (as shown later in \eqref{eq:H'_2}), the Hamiltonian will in general be unbounded from below. It can nevertheless be remedied by a large mass parameter for these fields. By adding further terms to the action, a stable theory can be obtained even at lower masses. The group embedding of the orbifold approach, while not being the minimal one, is the simplest with respect to possible extensions to larger gauge groups. A simple observation of this work is that other approaches with simpler Hamiltonians work as well provided the scalar masses are sufficiently large, with no sign of non-smooth  behaviors when the mass is increased. This indicates that the modifications of the theory by the additional scalar fields are frozen out, and even the details of the scalar sector are not relevant. This observation is important since it opens up the possibility of further simplifications of the theory by removing terms that are not relevant in the large mass limit. The resulting simplified theories can be more easily implemented on a quantum computer (in the sense that the number of terms in the Hamiltonian is smaller), while still capturing the relevant physics of Yang-Mills theory in the continuum limit. 

The orbifold lattice approach~\cite{Kaplan:2002wv,Buser:2020cvn,Bergner:2024qjl,Halimeh:2024bth,Bergner:2025zkj,Halimeh:2025ivn} embeds the SU($N$) group manifold into $\mathbb{C}^{N^2} \cong \mathbb{R}^{2N^2}$. Unitary link variables $U$ become complex link variables $Z$ whose components can be treated as real numbers (real and imaginary parts of matrix entries). In the original orbifold-lattice Hamiltonian, the extra scalars have a well-defined continuum limit and contribute to the final field theory. These extra scalar degrees of freedom, corresponding to the radial coordinate in $Z$, decouple from the low-energy physics. Simpler Hamiltonians investigated in this work have the same low-energy behaviors if the mass parameters are sufficiently large. Although the same low-energy physics is described, there is a significant technical difference when it comes to quantum simulations. In simple terms, it can be phrased as \emph{now we can use Cartesian coordinates instead of polar coordinates}. From this perspective, it might seem like a trivial trick that eliminates the need for dealing with a complicated group manifold. It allows us to write the truncated Hamiltonian and construct quantum circuits analytically~\cite{Bergner:2024qjl,Halimeh:2024bth,Halimeh:2025ivn}. Significant speedups and scaling laws can be derived explicitly since full circuit constructions are available \textit{analytically}, in terms of quantum Fourier transform, CNOT gates, and single-qubit rotations for the case of the Hamiltonian time evolution~\cite{Hanada:2025yzx, Halimeh:2024bth}.

To illustrate this simplicity of the construction, we discuss the explicit universal form of the Hamiltonian and how it is related to its representation on a quantum device. The general form of the Hamiltonian is
\begin{align}
\hat{H}
=
\frac{1}{2}\sum_a\hat{p}_a^2
+
V(\hat{x})\, , 
\label{eq:universal_Hamiltonian}
\end{align}
where the position operator $\hat{x}_a$ and momentum operator $\hat{p}_a$ satisfy the canonical commutation relations
\begin{align}
[\hat{x}_a,\hat{p}_b]
=
\mathrm{i}\delta_{ab}\, , 
\qquad
[\hat{x}_a,\hat{x}_b]
=
[\hat{p}_a,\hat{p}_b]
=
0\, . 
\end{align}
$V(\hat{x})$ is an at most quartic order polynomial in the cases of SU(2) theories studied in this paper. 
When $Q$ qubits are assigned to each boson, we can write the truncated version of the position operator in the position basis as 
\begin{align}
\hat{x}_a
=
-
\delta_x\cdot
\left(
\frac{\sigma_{z;a,1}}{2}
+
2\cdot\frac{\sigma_{z;a,2}}{2}
+
\cdots
+
2^{Q-1}\cdot\frac{\sigma_{z;a,Q}}{2}
\right)\, , 
\label{eq:x_with_Paulis}
\end{align}
where $\sigma_{z;a,j}$ is the Pauli-Z operator acting on the $j$-th qubit of the $a$-th boson, so that the value of $x_a$ can take $\pm\frac{1}{2}\delta_x$, $\pm\frac{3}{2}\delta_x$, ..., $\pm\frac{2^Q-1}{2}\delta_x$. By imposing the periodic boundary condition with period $2^Q\delta_x$, we can access the momentum basis via the quantum Fourier transform. The truncated version of the momentum operator $\hat{p}_a$ is written in the momentum basis as 
\begin{align}
\hat{p}_a
=
-
\delta_p\cdot
\left(
\frac{\sigma_{z;a,1}}{2}
+
2\cdot\frac{\sigma_{z;a,2}}{2}
+
\cdots
+
2^{Q-1}\cdot\frac{\sigma_{z;a,Q}}{2}
\right)\, , 
\label{eq:p_with_Paulis}
\end{align}
where $\delta_p=2\pi/(2^Q\delta_x)$. Therefore, $\frac{1}{2}\sum_a\hat{p}_a^2$ and $V(\hat{x})$ can be written as a sum of tensor products of Pauli-Z operators in the momentum and position bases, respectively. Such a simple form enables the explicit construction of efficient quantum circuits using only analytical methods. Therefore, we can \textit{prove} a quantum advantage when a fault-tolerant digital quantum computer with a sufficient number of qubits becomes available. Quantum advantage in this context means that problems like the sign problem or real time dynamics in four dimensional SU(3) can be addressed with a procedure that allows continuum extrapolation and is not targeting a theory that merely shares some properties with the problems under consideration.

This article is a continuation of our explorations along the idea of controlled group space embeddings leading to non-compact link variables. The main goal is to explain and investigate in what respects the orbifold Hamiltonian/Lagrangian is unique and whether simpler representations of the approach can be found, in particular regarding even simpler simulation protocols for Yang-Mills theory in (3+1)-dimensions. The key observation is that many terms in the orbifold-lattice Hamiltonian reduce to constant terms or a rescaling of couplings when the complex matrix $Z$ is restricted to be a unitary matrix $U$ and scalar fluctuations are integrated out. They are hence higher order corrections in an expansion around the infinite mass limit. Since such contributions do not affect the remaining effective theory, there is a certain freedom in the construction of the Hamiltonian. This can be exploited to find alternative but equivalent Hamiltonian formulations, while substantially simplifying the resulting quantum circuits.

The orbifold-lattice construction uses an embedding of the gauge group into flat space. Such an embedding need not be unique. Therefore, we consider simplifications of the embeddings for the case of SU(2): instead of $\mathbb{R}^{2N^2}=\mathbb{R}^{8}$, it can also be embedded into $\mathbb{R}^{4}$, reflecting the fact that $\mathrm{SU}(2) \cong \mathrm{S}^3\subset\mathbb{R}^4$. The validity of this embedding, and the consequent use of Hamiltonians with simpler quantum circuits involving half the number of scalar degrees of freedom, can be assessed through classical Monte Carlo simulations, which test for the absence of obstructions in the extrapolation to the Kogut-Susskind limit.

This paper is organized as follows. In Section~\ref{sec:orbifold_review}, we review the orbifold-lattice approach, presenting the Hamiltonian and Lagrangian, and explain the procedure for obtaining the Kogut-Susskind Hamiltonian as the infinite-mass limit. While this limit is not necessary when we consider the original orbifold lattice, it is necessary when we consider simplified Hamiltonians in later sections.
In Section~\ref{sec:simplifications}, we explore alternative lattice formulations: simplified Hamiltonians obtained by discarding terms that become negligible in the Kogut-Susskind limit (Section~\ref{sec:removing_terms}), a more efficient embedding of SU(2) into $\mathbb{R}^{4}$ that halves the number of scalar degrees of freedom per link (Section~\ref{sec:embedding_R4}), and two strategies to improve convergence to the Kogut-Susskind limit without requiring a large scalar mass --- a linear counter-term (Section~\ref{sec:ZZbar_counterterm}) and a tuning of the effective lattice spacing (Section~\ref{sec:tune_bare_spacing}). Section~\ref{sec:minimal_Hamiltonians} presents the explicit minimal Hamiltonian constructions used for quantum circuit implementation. The numerical Monte Carlo results that validate these analytical developments are presented in Section~\ref{sec:numerical_results}. While we focus on SU(2), similar simulations at $N\ge 3$ are straightforward and there is no reason to expect a significant difference there. We conclude in Section~\ref{sec:conclusions_f_directions} by summarizing our key findings and highlighting directions for future research.

\section{Orbifold lattice Hamiltonian and Lagrangian}\label{sec:orbifold_review}
The orbifold-lattice Hamiltonian and Lagrangian, which was originally derived for supersymmetric theories~\cite{Kaplan:2002wv},
has already been presented in previous works. Although this is a well-established framework, it may seem exotic to some researchers from the lattice QCD community, and some of the resulting misconceptions are clarified in~\cite{Hanada:2026zab}. We briefly review the orbifold lattice approach in this section for comparison with the simplifications introduced in later sections.
\subsection{Hamiltonian}
The orbifold lattice Hamiltonian for pure Yang-Mills theory in $D=d+1$ spacetime dimensions is~\cite{Kaplan:2002wv,Buser:2020cvn,Bergner:2024qjl}
\begin{align}
\hat{H}
&=
\sum_{\vec{n}}
{\rm Tr}\Biggl(
\sum_{j=1}^{D-1} \hat{P}_{j,\vec{n}} \,  \hat{\bar{P}}_{j,\vec{n}}
+
\frac{g_D^2}{2a^{D-1}}\left|\sum_{j=1}^{D-1}
\left(
\hat{Z}_{j,\vec{n}} \, \hat{\bar{Z}}_{j,\vec{n}} -\hat{\bar{Z}}_{j,\vec{n}-\hat{j}}\hat{Z}_{j,\vec{n}-\hat{j}}
\right)
\right|^2 
\nonumber\\
&\qquad\qquad\qquad
+
\frac{2g_D^2}{a^{D-1}}\sum_{j<k}
\left|
\hat{Z}_{j,\vec{n}} \, \hat{Z}_{k,\vec{n}+\hat{j}}
-
\hat{Z}_{k,\vec{n}} \, \hat{Z}_{j,\vec{n}+\hat{k}}
\right|^2
 \Biggl)
 \, +\,  
 \Delta\hat{H}\, , 
\label{eq:Hamiltonian_orbifold}
\end{align}
where $a$ is the lattice spacing, $g_D$ is the bare coupling constant, and the bar denotes the Hermitian conjugate of an $N\times N$ matrix, i.e.,  $(\bar{Z}_{j,\vec{n}})_{ab}=[(Z_{j,\vec{n}})_{ba}]^\ast$ and $(\bar{P}_{j,\vec{n}})_{ab}=[(P_{j,\vec{n}})_{ba}]^\ast$. The operators respect the standard canonical commutation relation
\begin{align}
    \left[
\hat{Z}_{j,\vec{n};ab},\hat{\bar{P}}_{k,\vec{n}';cd}
    \right]
    =
    \left[
    \hat{Z}_{j,\vec{n};ab},\left(\hat{P}_{k,\vec{n}';dc}\right)^\dagger
    \right]
    =\mathrm{i}\delta_{jk}\delta_{\vec{n}\vec{n}'}\delta_{ad}\delta_{bc}\, .
    \label{Z-P-commutator}
\end{align}
By introducing the real and imaginary parts of $\hat{Z}$, denoted by $\hat{X}$ and $\hat{Y}$, and those of $\hat{P}$, denoted by $\hat{P}_X$ and $\hat{P}_Y$, as
\begin{align}
\hat{Z}_{j,\vec{n};ab}
=
\frac{1}{\sqrt{2}}\left(
\hat{X}_{j,\vec{n};ab}
+
\mathrm{i}\hat{Y}_{j,\vec{n};ab}
\right)\, , 
\qquad
\hat{P}_{j,\vec{n};ab}
=
\frac{1}{\sqrt{2}}\left(
\hat{P}_{X;j,\vec{n};ab}
+
\mathrm{i}\hat{P}_{Y;j,\vec{n};ab}
\right)\, , 
\end{align}
we can see that 
\begin{align}
    \left[
\hat{X}_{j,\vec{n};ab},\hat{P}_{X;k,\vec{n}';cd}
    \right]
    =
        \left[
\hat{Y}_{j,\vec{n};ab},\hat{P}_{Y;k,\vec{n}';cd}
    \right]
    =
\mathrm{i}\delta_{jk}\delta_{\vec{n}\vec{n}'}\delta_{ad}\delta_{bc}\, , 
\end{align}
and all other commutators vanish. Therefore, the momentum conjugate of $\hat{X}_{j,\vec{n};ab}$ (resp., $\hat{Y}_{j,\vec{n};ab}$) is $\hat{P}_{X;j,\vec{n};ba}$ (resp., $\hat{P}_{Y;j,\vec{n};ba}$). These variables can be identified with $\hat{x}$ and $\hat{p}$ appeared in the introduction, and hence, the orbifold lattice Hamiltonian \eqref{eq:Hamiltonian_orbifold} takes the universal form \eqref{eq:universal_Hamiltonian}. 

$\Delta\hat{H}$ is chosen appropriately depending on the details of the strategy. Ref.~\cite{Bergner:2025zkj} used 
\begin{align}
\Delta\hat{H}
&\equiv
\frac{m^2g_D^2}{2a^{D-3}}\sum_{\vec{n}}\sum_{j=1}^{D-1}
{\rm Tr}
\left|\hat{Z}_{j,\vec{n}}\hat{\bar{Z}}_{j,\vec{n}} -\frac{a^{D-3}}{2g_D^2}\right|^2
\nonumber\\
&\quad
+
\frac{m^2_{\rm U(1)}a^{D-3}}{2g_D^2}
\sum_{\vec{n}}\sum_{j=1}^{D-1}
\left|
    \left(\frac{a^{D-3}}{2g_D^2}\right)^{-N/2}\det(\hat{Z}_{j,\vec{n}})-1
    \right|^2\, . 
    \label{eq:H_add}
\end{align}
With this choice, in the large-mass limit $m^2,m^2_{\rm U(1)}\to\infty$, the orbifold lattice Hamiltonian reduces to the Kogut-Susskind Hamiltonian~\cite{Kogut:1974ag}. Hence, we call this limit the \textit{Kogut-Susskind limit}.

The orbifold lattice Hamiltonian is designed so that Yang-Mills theory coupled to scalars is obtained when the complex link variable $Z$ is expanded about an appropriate background~\cite{Kaplan:2002wv}. 
The complex link variable $Z$ is a product of a positive-definite Hermitian site variable $W$ and a unitary link variable $U$,
\begin{align}
Z_{j,\vec{n}}=\sqrt{\frac{a^{D-3}}{2g_D^2}}W_{j,\vec{n}}U_{j,\vec{n}}\, .  
\label{eq:Z-W-U}
\end{align}
They describe an adjoint scalar field $\phi_j$ and gauge field $A_j$, respectively, as 
\begin{align}
W_{j,\vec{n}}=\exp\left(ag_D\phi_{j,\vec{n}}\right)\, , 
\qquad
U_{j,\vec{n}}=\exp(\mathrm{i}ag_DA_{j,\vec{n}})\, . 
\end{align}
It is quite obvious that the first term on the right-hand side of \eqref{eq:H_add} gives scalar mass term proportional to $m^2\mathrm{Tr}(W^2-1)^2$. In the limit $m^2\to\infty$, this forces $W$ to the identity and the scalars $\phi$ completely decouple at any energy scale. (Note that, strictly speaking, there is no need for restricting $W$ to identity; see Section~\ref{sec:no_large_mass}.) Note that,  a priori, the determinant of $U$  is not equal to 1. 
There are two options: either U($N$) gauge theory is considered keeping the U(1) part, or this part is removed by the second term of $\Delta H$ in \eqref{eq:H_add} that enforces $\det U=1$, thereby describing the standard SU($N$) gauge theory. For pure Yang-Mills theory, the U(1) part is free and decoupled from the SU($N$) part, and hence both approaches lead to the same final theory. For the numerical simulations in this paper, we have chosen to remove the U(1) part. 
\subsection{Gauge symmetry}
The Hamiltonian and commutation relation are invariant under the local U($N$) transformation, 
\begin{align}
\hat{Z}_{j,\vec{n}}\, ,\  \hat{P}_{j,\vec{n}}
\quad\rightarrow\quad
    \Omega^{-1}_{\vec{n}}
     \hat{Z}_{j,\vec{n}}
     \Omega_{\vec{n}+\hat{j}}\, ,\  
     \Omega^{-1}_{\vec{n}}
     \hat{P}_{j,\vec{n}}
     \Omega_{\vec{n}+\hat{j}}\, , 
\end{align}
where $\Omega_{\vec{n}}$ is an $N\times N$ unitary matrix. For an infinitesimal transformation $\Omega_{\vec{n}}=e^{\mathrm{i}\epsilon_{\vec{n}}}$ at point $\vec{n}$, we have:
\begin{align}
\Delta\hat{Z}_{j,\vec{n}}=-\mathrm{i}\epsilon_{\vec{n}}\hat{Z}_{j,\vec{n}}\, , 
\quad
\Delta\hat{P}_{j,\vec{n}}=-\mathrm{i}\epsilon_{\vec{n}}\hat{P}_{j,\vec{n}}\, , 
\quad
\Delta\hat{Z}_{j,\vec{n}-\hat{j}}=    \mathrm{i}
     \hat{Z}_{j,\vec{n}-\hat{j}}\epsilon_{\vec{n}}\,
\quad     
\Delta\hat{P}_{j,\vec{n}-\hat{j}}=\mathrm{i}
\hat{P}_{j,\vec{n}-\hat{j}}\epsilon_{\vec{n}}\, . 
\end{align}
The generator that reproduces this transformation is 
\begin{align}
\hat{G}_{\vec{n},pq}
\equiv
\mathrm{i}\sum_{j=1}^{D-1}
\left(
-
\hat{Z}_{j,\vec{n}}\hat{\bar{P}}_{j,\vec{n}}
+
\hat{P}_{j,\vec{n}}\hat{\bar{Z}}_{j,\vec{n}}
-
\hat{\bar{Z}}_{j,\vec{n}-\hat{j}}\hat{P}_{j,\vec{n}-\hat{j}}
+
\hat{\bar{P}}_{j,\vec{n}-\hat{j}}\hat{Z}_{j,\vec{n}-\hat{j}}
\right)_{pq}. 
\label{eq:gauge-generators}
\end{align}
The Hamiltonian is invariant under local U($N$) transformations:
\begin{align}
[\hat{G}_{\vec{n}},\hat{H}]=0\, . 
\end{align}
Among them, we gauge only SU($N$).  To obtain SU($N$) generators, we remove the U(1) part, $\frac{1}{N}\mathrm{Tr}\left(\hat{G}_{\vec{n}}\right)$. If one prefers to remove non-singlet modes from the low-energy spectrum (though, usually there is no reason to do so), a penalty term of the following form can be added:
\begin{align}
\hat{H}^{\rm (penalty)}
 \equiv
c_{\rm penalty}\,\cdot 
\mathrm{Tr}
\left(
\hat{G}_{\vec{n},pq}
-
\frac{1}{N}
\mathrm{Tr}\, \hat{G}_{\vec{n}}
\right)^2
=
c_{\rm penalty}\,\cdot 
\left(
\mathrm{Tr}
\left(
\hat{G}_{\vec{n}}^2
\right)
-
\frac{1}{N}
\left(
\mathrm{Tr}
\hat{G}_{\vec{n}}
\right)^2\right)\, . 
\label{gauge-penalty}
\end{align}
Here, $c_{\rm penalty}$ is a large positive value. 
\subsection{Lagrangian}\label{sec:Lagrangian}
The Lagrangian of the orbifold lattice Yang-Mills theory follows from the Hamiltonian. Since we will perform Monte Carlo simulations, we consider the Euclidean formulation
\begin{align}
	L
	&=
	\sum_{\vec{n}}
	{\rm Tr}\Biggl(
	\sum_{j=1}^{D-1} |D_tZ_{j,\vec{n}}|^2
	+
	\frac{g_D^2}{2a^{D-1}}\left|\sum_{j=1}^{D-1}
	\left(
	Z_{j,\vec{n}} \, \bar{Z}_{j,\vec{n}} -\bar{Z}_{j,\vec{n}-\hat{j}}Z_{j,\vec{n}-\hat{j}}
	\right)
	\right|^2 
	\nonumber\\
	&\qquad\qquad\qquad
	+
	\frac{2g_D^2}{a^{D-1}}\sum_{j<k}
	\left|
	Z_{j,\vec{n}} \, Z_{k,\vec{n}+\hat{j}}
	-
	Z_{k,\vec{n}} \, Z_{j,\vec{n}+\hat{k}}
	\right|^2
	\Biggl)
	+ 
	\Delta L\, , 
	\label{eq:Lagrangian_orbifold}
\end{align}
where 
\begin{align}
	\Delta L
	&\equiv
	\frac{m^2g_D^2}{2a^{D-3}}\sum_{\vec{n}}\sum_{j=1}^{D-1}
	{\rm Tr}
	\left|Z_{j,\vec{n}}\bar{Z}_{j,\vec{n}} -\frac{a^{D-3}}{2g_D^2}\right|^2
	\nonumber\\
	&\quad
	+
	\frac{m^2_{\rm U(1)}a^{D-3}}{2g_D^2}
	\sum_{\vec{n}}\sum_{j=1}^{D-1}
	\left|
	\left(\frac{a^{D-3}}{2g_D^2}\right)^{-N/2}\det(Z_{j,\vec{n}})-1
	\right|^2\, . 
\end{align} 
So far, time is continuous. For Monte Carlo simulations, we discretize time and introduce a spacetime lattice. To represent the gauge covariant time derivative $D_t$, unitary links $U_{0,n}=\exp(ia_t g_d A_0)$ in time directions are introduced. Note that these are just common SU($N$) link variables with the gauge potential $A_0$ in time directions and the anisotropy leads to a different lattice spacing $a_t$; specifically, we do not use complex links in the time direction. The lattice version of the covariant time derivative term is hence
\begin{align}
\sum_{n}
	{\rm Tr}
	\sum_{j=1}^{D-1} |D_tZ_{j,n}|^2=	\sum_{n}
	\frac{1}{a_t^2}{\rm Tr}
	\sum_{j=1}^{D-1} |U_{0,n}Z_{j,n+\hat{0}}-Z_{j,n}U_{0,n+\hat{0}}|^2
\end{align}
Like for the Hamiltonian, the Kogut-Susskind limit (or Wilson limit) is $m^2,m^2_{\rm U(1)}\to\infty$. In this limit the ordinary Wilson gauge action is reproduced.

\section{Alternative lattice formulations}\label{sec:simplifications}
The orbifold lattice approach has originally been developed in the context of supersymmetric gauge
theories on the lattice~\cite{Kaplan:2002wv,Cohen:2003xe,Cohen:2003qw,Kaplan:2005ta}. In the current context it has been derived to provide a simpler implementation of the
theory on the lattice due to the additional scalar field, which allows to lift the constraints of the group manifold.
In the infinite mass limit the constraints are recovered and the theory behaves as pure Yang-Mills theory.
In this section we explore generalizations of the approach. Since no supersymmetry is required in our applications, we can remove terms from the action and change the embedding. More generally, anything that provides only higher order corrections in the infinite mass limit can be added. These modifications can be optimized for convergence to the infinite mass limit.

We explore three different routes: first, simplifications of the lattice action by eliminating interaction terms that vanish in the Kogut-Susskind limit (Section~\ref{sec:removing_terms}); second, more efficient embeddings of the group manifold, in particular $\mathrm{SU}(2)\subset\mathbb{R}^4$ instead of $\mathrm{SU}(2)\subset\mathbb{R}^8$ (Section~\ref{sec:embedding_R4}); and third, tuning the effective lattice spacing to improve convergence without requiring large scalar masses (Section~\ref{sec:no_large_mass}).

Because these models are not exactly the orbifold lattice --- they are not obtained by orbifold projection from a matrix model --- we sometimes call them \textit{orbifold-ish} models.
\subsection{Eliminating some interaction terms}
\label{sec:removing_terms}
When we see the orbifold lattice Hamiltonian as a theory describing scalars and gluons, we cannot change the Hamiltonian arbitrarily. However, in the Kogut-Susskind limit, scalars decouple completely from the Hamiltonian already at the lattice-regularized level. Specifically, in the limit of $m^2\to\infty$, the link variables behave as $Z_{j,\vec{n}} \hat{\bar{Z}}_{j,\vec{n}} = \frac{a^{D-3}}{2g_D^2}W_{j,\vec{n}}^2 \to \frac{a^{D-3}}{2g_D^2} \textbf{1}_N$, and hence, $W_{j,\vec{n}}\to\textbf{1}_N$. 
As a result, we can omit certain terms in the Hamiltonian without affecting the low-energy physics.

For example, we can drop the entire second term in \eqref{eq:Hamiltonian_orbifold}, because
\begin{align}
\hat{Z}_{j,\vec{n}} \, \hat{\bar{Z}}_{j,\vec{n}} -\hat{\bar{Z}}_{j,\vec{n}-\hat{j}}\hat{Z}_{j,\vec{n}-\hat{j}}
\, \rightarrow\, 
\textbf{0}_N\, . 
\end{align}
Then, we obtain a simpler Hamiltonian, 
\begin{align}
\hat{H}_1
&=
\sum_{\vec{n}}
{\rm Tr}\Biggl(
\sum_{j=1}^{D-1} \hat{P}_{j,\vec{n}} \,  \hat{\bar{P}}_{j,\vec{n}}
+
\frac{2g_{\rm D}^2}{a^{D-1}}\sum_{j<k}
\left|
\hat{Z}_{j,\vec{n}} \, \hat{Z}_{k,\vec{n}+\hat{j}}
-
\hat{Z}_{k,\vec{n}} \, \hat{Z}_{j,\vec{n}+\hat{k}}
\right|^2
 \Biggl)
 \, +\,  
 \Delta\hat{H}\, . 
 \label{eq:H'_1}
\end{align}

Furthermore, using the same observation, the above Hamiltonian can be further simplified by expanding the modulus and eliminating the terms proportional to the identity, resulting in an even simpler Hamiltonian:

\begin{align}
\hat{H}_2
&=
\sum_{\vec{n}}
{\rm Tr}\Biggl(
\sum_{j=1}^{D-1} \hat{P}_{j,\vec{n}} \,  \hat{\bar{P}}_{j,\vec{n}}
-
\frac{2g_{\rm D}^2}{a^{D-1}}\sum_{j<k}
\left(
\hat{Z}_{j,\vec{n}} \, \hat{Z}_{k,\vec{n}+\hat{j}}
\hat{\bar{Z}}_{j,\vec{n}+\hat{k}} \, \hat{\bar{Z}}_{k,\vec{n}}
+
{\rm h.c.}
\right)
 \Biggl)
 \, +\,  
 \Delta\hat{H}\, . 
 \label{eq:H'_2}
\end{align}
These expressions were previously noted in Ref.~\cite{Bergner:2024qjl}, although it was not the focus of a detailed study.
In this paper, we conduct a detailed numerical investigation of these simplified versions for SU(2). 

There is also room for improvement of $\Delta\hat{H}$, as we will discuss in Section~\ref{sec:no_large_mass}.

\subsection{Embedding SU(2) into $\mathbb{R}^4$}\label{sec:embedding_R4}
In the orbifold lattice formulation, SU($N$) or U($N$) is embedded into a flat space $\mathbb{C}^{N^2} \cong \mathbb{R}^{2N^2}$ or torus $\mathrm{T}^{2N^2}$. By embedding SU($N$) into a lower-dimensional space, more efficient quantum simulations may be achievable. Let us investigate this possibility.
\subsubsection{More efficient embedding of group manifold into flat space}

\subsubsection*{U(1) as $\mathrm{S}^1$}
U(1) is the same as $\mathrm{S}^1$, which can be seen as a one-dimensional torus $\mathrm{T}^1$. Therefore, the universal framework can be applied without introducing a scalar. 
We can go back and forth between the coordinate (magnetic) basis and the momentum (electric) basis by the quantum Fourier transform. 
\subsubsection*{Embedding SU(2) into $\mathbb{R}^{4}$}
An embedding of SU(2) into $\mathbb{R}^4$ follows from the observation that a generic element of SU(2) admits a simple parametrization
\begin{align}
U
=
\begin{pmatrix}
    \alpha & -\beta^\ast\\
    \beta & \alpha^\ast
\end{pmatrix}\, , 
\qquad
|\alpha|^2+|\beta|^2=1. 
\end{align}
This parametrization makes $\mathrm{SU}(2) \cong \mathrm{S}^3$ manifest. 
The Haar measure on the SU(2) group manifold is the pull-back of the flat measure on $\mathbb{R}^4$. We can see it because the constraint $|\alpha|^2+|\beta|^2=1$ is invariant under $U\to gU$.

This embedding corresponds to setting $W$ in \eqref{eq:Z-W-U} to be proportional to the identity matrix $\mathbf{1}$. 
This embedding constitutes one of the main objectives of the present work.

When SU(2) is embedded into $\mathbb{R}^8$, $W=e^{a\phi}$ is two-by-two positive-definite Hermitian and $\phi$ is two-by-two Hermitian. As previously discussed, the infinite mass limit decouples the scalar and leads to a pure Yang-Mills limit. 

When SU(2) is embedded into $\mathbb{R}^4$, the SU(2) part of $\phi$ is gone, and we only have the U(1) part. The unitary part is SU(2), without the U(1) phase. 
\subsubsection*{Embedding SU($N$) into $\mathbb{R}^{N^2+N-2}$}
Using a well-known relation $\textrm{SU}(N)/\textrm{SU}(N-1)\cong\textrm{S}^{2N-1}\subset\mathbb{R}^{2N}$, SU (2) can be parameterized using $\mathbb{R}^4$, SU (3) can be parameterized using $\mathbb{R}^{6+4}=\mathbb{R}^{10}$, and in general, SU($N$) can be parametrized by using $\mathbb{R}^{N^2+N-2}$. 
However, for $N\ge 2$, the embedding is technically involved and the circuit may not be simple; at least, we do not know a simple way yet. 
\subsubsection{Detailed construction of SU(2) embedded into $\mathbb{R}^4$}
From here on, we focus on $N=2$. 
Let us write $Z_{j,\vec{n}}$ using two complex vectors $\vec{v}_{j,\vec{n}},\vec{w}_{j,\vec{n}}\in\mathbb{C}^2$, as $Z_{j,\vec{n}}=(\vec{v}_{j,\vec{n}},\vec{w}_{j,\vec{n}})$.  
$\Delta L$ enforces the following constraints:
\begin{align}
|\vec{v}_{j,\vec{n}}|^2=|\vec{w}_{j,\vec{n}}|^2=\frac{a^{D-3}}{2g_D^2}\, \qquad
\vec{v}_{j,\vec{n}}^{\, \ast}\cdot\vec{w}_{j,\vec{n}}=0\, , 
    \qquad
    \det Z_{j,\vec{n}}= \frac{a^{D-3}}{2g_D^2}\, . 
\end{align}
We can write them into five independent constraints. Firstly, there are three conditions that makes $Z_{j,\vec{n}}$ unitary up to a scalar factor:
\begin{align}
|\vec{v}_{j,\vec{n}}|^2=|\vec{w}_{j,\vec{n}}|^2\, ,  \qquad
\mathrm{Re}\left(\vec{v}_{j,\vec{n}}^{\, \ast}\cdot\vec{w}_{j,\vec{n}}\right)
=
0\, , 
\qquad
\mathrm{Im}\left(\vec{v}_{j,\vec{n}}^{\, \ast}\cdot\vec{w}_{j,\vec{n}}\right)
=
0\, . 
\end{align}
After imposing these three conditions, we can write $Z_{j,\vec{n}}$ as $re^{\mathrm{i}\theta}U$, where $U\in\mathrm{SU}(2)$, $r>0$, and $\theta\in\mathbb{R}$. Then, we have two more constraints, 
\begin{align}
r=\sqrt{\frac{a^{D-3}}{2g_D^2}}, \qquad
\theta=0\, . 
\end{align}
To have embedding of SU(2) into $\mathbb{R}^4$, we solve four of them first, leaving $r=\sqrt{a^{D-3}/2g_D^2}$. 
Then, we can write $Z_{j,\vec{n}}$ as 
\begin{align}
Z_{j,\vec{n}}
=
\begin{pmatrix}
    \alpha_{j,\vec{n}} & -\beta_{j,\vec{n}}^\ast\\
    \beta_{j,\vec{n}} & \alpha_{j,\vec{n}}^\ast
\end{pmatrix}\, . 
\end{align}
The kinetic part becomes
\begin{align}
2
\sum_{\vec{n}}
{\rm Tr}\left(
\sum_{j=1}^{D-1}
\dot{\alpha}_{j,\vec{n}}
\dot{\alpha}^\ast_{j,\vec{n}}
+
\dot{\beta}_{j,\vec{n}}
\dot{\beta}^\ast_{j,\vec{n}}
\right)
\end{align}
To have the standard normalizations for real variables, we introduce $x_{j,\vec{n};a}$, $a=1,2,3,4$, as
\begin{align}
    \alpha_{j,\vec{n}}
    =
    \frac{x_{j,\vec{n};4}+\mathrm{i}x_{j,\vec{n};3}}{2}\, 
    \qquad
    \beta_{j,\vec{n}}
    =
\frac{-x_{j,\vec{n};2}+\mathrm{i}x_{j,\vec{n};1}}{2}\, , 
\end{align}
so that
\begin{align}
Z_{j,\vec{n}}
=
\frac{1}{2}
\left(
x_{j,\vec{n};4}\cdot\textbf{1}_2
+
\mathrm{i}\sum_{\ell=1}^3x_{j,\vec{n};\ell}\sigma_\ell
\right)
\equiv
\frac{\mathrm{i}}{2}\sum_{\ell=1}^4x_{j,\vec{n};\ell}\sigma_\ell\, , 
\label{eq:Z-and-x}
\end{align}
where $\sigma_{1,2,3}$ are Pauli matrices and $\sigma_4=-\mathrm{i}\cdot\textbf{1}_2$. 
Then, we have the standard normalization of the kinetic term,
\begin{align}
\frac{1}{2}
\sum_{\vec{n}}
\sum_{j=1}^{D-1}
\sum_{a=1}^4
\left(
\dot{x}_{j,\vec{n};a}
\right)^2\, . 
\end{align}
we can use the same form of the Lagrangian \eqref{eq:Lagrangian_orbifold}, except that the constraint term $\Delta L$ is replaced with 
\begin{align}
\Delta L
&\equiv
-
\frac{m^2g_D^2}{16a^{D-3}}
\sum_{\vec{n}}\sum_{j=1}^{D-1}
\left(
    \sum_{a=1}^4(x_{j,\vec{n};a})^2
    -
    \frac{2a^{D-3}}{g_D^2}
    \right)^2\, . 
\end{align}
\subsubsection*{Hamiltonian}
The new Hamiltonian in terms of SU(2) embedded into $\mathbb{R}^4$ is 
\begin{align}
\hat{H}
&=
\sum_{\vec{n}}
{\rm Tr}\Biggl(
\frac{1}{2}\sum_{j=1}^{D-1}
\sum_{a=1}^4
\left(
\hat{p}_{j,\vec{n};a}
\right)^2
+
\frac{g_D^2}{2a^{D-1}}\left|\sum_{j=1}^{D-1}
\left(
\hat{Z}_{j,\vec{n}} \, \hat{\bar{Z}}_{j,\vec{n}} -\hat{\bar{Z}}_{j,\vec{n}-\hat{j}}\hat{Z}_{j,\vec{n}-\hat{j}}
\right)
\right|^2
\nonumber\\
&\qquad\qquad\qquad
+
\frac{2g_D^2}{a^{D-1}}\sum_{j<k}
\left|
\hat{Z}_{j,\vec{n}} \, \hat{Z}_{k,\vec{n}+\hat{j}}
-
\hat{Z}_{k,\vec{n}} \, \hat{Z}_{j,\vec{n}+\hat{k}}
\right|^2
 \Biggl)
 + 
 \Delta\hat{H}\, , 
\end{align}
\begin{align}
\Delta\hat{H}
&\equiv
\frac{m^2g_D^2}{16a^{D-3}}
\sum_{\vec{n}}\sum_{j=1}^{D-1}
\left(
    \sum_{a=1}^4(\hat{x}_{j,\vec{n};a})^2
    -
    \frac{2a^{D-3}}{g_D^2}
    \right)^2\, , 
\end{align}
where $\hat{Z}$ is defined by 
\begin{align}
\hat{Z}_{j,\vec{n}}
=
\frac{1}{2}
\begin{pmatrix}
\hat{x}_{j,\vec{n};4}+\mathrm{i}\hat{x}_{j,\vec{n};3} & \hat{x}_{j,\vec{n};2}+\mathrm{i}\hat{x}_{j,\vec{n};1}\\ 
-\hat{x}_{j,\vec{n};2}+\mathrm{i}\hat{x}_{j,\vec{n};1} & \hat{x}_{j,\vec{n};4}-\mathrm{i}\hat{x}_{j,\vec{n};3}
\end{pmatrix}\, , 
\label{Z-in-terms-of-x}
\end{align}
with the canonical commutation relation 
\begin{align}
[\hat{x}_{j,\vec{n};a},\hat{p}_{j',\vec{n}';a'}]=\mathrm{i}\delta_{jj'}\delta_{\vec{n}\vec{n}'}\delta_{aa'}\, . 
\end{align}

If we use 
\begin{align}
\hat{P}_{j,\vec{n}}
=
\begin{pmatrix}
\hat{p}_{j,\vec{n};4}+\mathrm{i}\hat{p}_{j,\vec{n};3} & \hat{p}_{j,\vec{n};2}+\mathrm{i}\hat{p}_{j,\vec{n};1}\\ 
-\hat{p}_{j,\vec{n};2}+\mathrm{i}\hat{p}_{j,\vec{n};1} & \hat{p}_{j,\vec{n};4}-\mathrm{i}\hat{p}_{j,\vec{n};3}
\end{pmatrix}\, , 
\label{P-in-terms-of-p}
\end{align}
we can keep the canonical commutation relation \eqref{Z-P-commutator}. We can use the same expressions \eqref{eq:gauge-generators}, \eqref{gauge-penalty} for gauge generators, while the kinetic term in the Hamiltonian becomes $\frac{1}{2}\mathrm{Tr}(\hat{P}\hat{\bar{P}})$.  

Note that, in the Kogut-Susskind limit ($m^2\to\infty$), there is ambiguity of the choice of the Hamiltonian, such as \eqref{eq:H'_1} and \eqref{eq:H'_2}. 
\subsection{Controlling the effective lattice spacing}\label{sec:no_large_mass}
In Hamiltonians and Lagrangians defined above, the large scalar mass term forces $W=e^{a\phi}$ to be close to $\textbf{1}$ and, equivalently, $\phi$ to be stay close to zero. This defines the Kogut-Susskind limit of the theory. There are corrections to this limit in an expansion around the infinite mass limit. In order to improve the convergence to this limit, one can optimize the coefficients of these terms.

In the Hamiltonians we consider in this paper, the ``lattice spacing" $a$ appears explicitly. We call this $a$ ``lattice spacing" because, when $W$ fluctuates around the identity matrix $\textbf{1}_N$, a lattice with the lattice spacing $a$ is obtained~\cite{Kaplan:2002wv}. Then, how about $W$ fluctuates around $\alpha\textbf{1}_N$, with $\alpha\neq 1$? Then, we can simply consider a fluctuation around this point and obtain a lattice with different \textit{effective} lattice spacing $a_{\rm eff}$.\footnote{
This may be a confusing point for beginners; see ref.~\cite{Hanada:2026zab}.}

As will be discussed below and confirmed numerically in Section~\ref{sec:numerics_eliminate_large_mass}, properly accounting for an effective lattice spacing $a_{\rm eff}$ significantly accelerates convergence to the Kogut-Susskind limit. In the following we discuss these two approaches: adding a counter-term to cancel linear corrections (Section~\ref{sec:ZZbar_counterterm}) and using effective lattice spacing for the identification with Wilson or Kogut-Susskind (Section~\ref{sec:tune_bare_spacing}).
\subsubsection{Linear counter terms}\label{sec:ZZbar_counterterm}
There is no symmetry between $\phi\leftrightarrow-\phi$ because $\phi$ is the radial coordinate and not the phase; hence the odd powers of $\phi$ is not forbidden by a symmetry. Therefore the lowest order correction in $\phi$ that we expect, is a linear term or, equivalently, a shift in $\phi$. We expect the effective potential of the form  $c\, \mathrm{Tr}\phi+\frac{c'+m_{\rm bare}^2}{2}\mathrm{Tr}\phi^2+\cdots$, where $c$ is of order $a^{-3}$ and $c'$ is of order $a^{-2}$ for $D=4$. The linear term $c\, \mathrm{Tr}\phi$ leads to a nonzero vev of the scalar that effectively shifts the lattice spacing. A detailed analysis of the origin of this term in the change-of-variables Jacobian introduced by the fields $\phi$ will appear in a forthcoming publication~\cite{Bergner:2026jac}. A large mass is needed to make the shift smaller. When we consider the Kogut-Susskind limit, the comparison was made with the Kogut-Susskind Hamiltonian with the same ``lattice spacing". However, the shift of the lattice spacing in the orbifold(-ish) Hamiltonian leads to a mismatch. 

A simple way to eliminate the nonzero vev of the scalar and accelerate the convergence to the Kogut-Susskind limit is to introduce a term proportional to $\mathrm{Tr}(Z\bar{Z})\sim\mathrm{Tr}\phi+\mathrm{const}$ to cancel $c\, \mathrm{Tr}\phi$~\cite{Halimeh:2024bth}. To find the appropriate value of the parameter, we tune it so that $\langle\mathrm{Tr}(W-\mathrm{1})\rangle$ becomes zero. 

For $\mathbb{R}^4$-embedding of SU(2), we want to fix the radius at
$\sum_{j} x_j^2 =2a^{D-3}/g_D^2$. Therefore, the following term is added to the Hamiltonian:
\begin{align}
    -\gamma\sum_{j,\vec{n}}\mathrm{Tr}(\hat{Z}_{j,\vec{n}}\hat{\bar{Z}}_{j,\vec{n}})\, .
\end{align}
 While to the Euclidean action, we add: \begin{align}
    -\gamma\sum_{j,\vec{n}}\mathrm{Tr}(Z_{j,\vec{n}}\bar{Z}_{j,\vec{n}})
\end{align}

Concretely, this means that Monte Carlo simulations with actions corresponding to the Hamiltonians $\hat{H}$ and $\hat{H}_1$ can be performed by tuning the parameter $\gamma$ to positive values. This tuning leads to an agreement between the orbifold actions and the Wilson action, with the $m^2$ values being several orders of magnitude smaller than those used in simulations of the orbifold action without the correction. Similarly, for the action derived from $\hat{H}_2$, a comparable reduction in the value of $m^2$ is observed, but this is achieved by tuning $\gamma$ to negative values.
\subsubsection{Using effective lattice spacing}\label{sec:tune_bare_spacing}
An even simpler approach to improve convergence to the infinite mass limit is the tuning of the bare lattice spacing. In the orbifold theory this is in fact equivalent to a constant diagonal shift in $\phi$ and hence to additional linear counterterms.

In the KS limit of SU($N$) gauge theory, the scalar field turns to zero, i.e., $\frac{1}{N}\langle\mathrm{Tr}W\rangle=1$. A constant shift $\phi(x)\rightarrow \phi_0+\phi(x)$ leads to  $c\equiv\frac{1}{N}\langle\mathrm{Tr}W\rangle$. Such a shift can be absorbed in a rescaling $\widetilde{W}\equiv c^{-1}W$ such that $\frac{1}{N}\langle\mathrm{Tr}\widetilde{W}\rangle=1$. Substituting the rescaled variables into the action, the change of variables from $W$ to $\widetilde{W}$ is compensated by a rescaling of the lattice spacing.

To see the correspondence precisely, we recall how the Wilson action is obtained on Euclidean spacetime (Section~\ref{sec:Lagrangian}) and set $W = c\, \textbf{1}_N$, expressing the Lagrangian in terms of the unitary link variable $U$. For generic $c$, which is not necessarily $1$, the coefficients in front of the spatial and temporal plaquettes become
\begin{align}
c^4\cdot\frac{a_ta^{D-5}}{g_D^2}\, , 
\qquad
c^2\cdot\frac{a_t^{-1}a^{D-3}}{g_D^2}\, . 
\end{align}
Let us consider $D=3$, which will be studied numerically later in this paper. We set $g_D^2$, which has a mass dimension and hence sets the unit of the energy scale of interest, to be $1$. Then, those coefficients become
\begin{align}
\frac{c^4a_t}{a^2}\, , 
\qquad
\frac{c^2}{a_t}\, . 
\end{align}
We can identify this theory with the Wilson action with spatial and temporal lattice spacings $a_{\rm eff}$ and $a_{{\rm eff},t}$, requiring 
\begin{align}
\frac{c^4a_t}{a^2}
=
\frac{a_{{\rm eff},t}}{a_{\rm eff}^2}\, , 
\qquad
\frac{c^2}{a_t}
=
\frac{1}{a_{{\rm eff},t}}\, ,
\end{align}
or equivalently, 
\begin{align}
a_{\rm eff}=c^{-3}a\, , 
\qquad
a_{{\rm eff},t}=c^{-2}a_t\, . 
\end{align}
Therefore, we can simply interpret the effective lattice spacings as the ``physical" value used for the comparison to the Kogut-Susskind limit. There is no reason why we have to introduce a large mass to make the bare and effective values precisely the same. 
\section{Minimal Hamiltonians}\label{sec:minimal_Hamiltonians}
In this section, we present the explicit forms of the orbifold Hamiltonians. We provide constructions for $\hat{H}$, $\hat{H}_1$, and $\hat{H}_2$, both with and without the counter-term.
\subsection{Minimal Hamiltonians for SU(2) pure Yang-Mills}
Let us start with the case of $\mathrm{SU}(2)$, using the embedding into $\mathbb{R}^4$. 
Regardless of the choice of the Hamiltonian, the link operators $\hat{Z}_{j,\vec{n}}$ can be written as \eqref{eq:Z-and-x}. Let us repeat \eqref{eq:Z-and-x} here:
$\hat{Z}_{j,\vec{n}}=\frac{\mathrm{i}}{2}\sum_{\ell=1}^4 \hat{x}_{j,\vec{n};\ell} \, \sigma_\ell$
where $\sigma_{1,2,3}$ are Pauli matrices and $\sigma_4=-\mathrm{i}\cdot\textbf{1}_2$. The Hamiltonian consists of two kinds of terms. Namely, quadratic terms 
\begin{align}
\mathrm{Tr}(\hat{Z}\hat{\bar{Z}})
=
\frac{1}{4}
\sum_{a,b}
\hat{x}_a
\hat{x}_b
\mathrm{Tr}(\sigma_a\sigma_b^\dagger)
=
\frac{1}{2}
\sum_{a}
\hat{x}_a^2\, , 
\end{align}
and quartic terms that can be written schematically as 
\begin{align}
\mathrm{Tr}(\hat{Z}^{(1)}\hat{Z}^{(2)}\hat{\bar{Z}}^{(3)}\hat{\bar{Z}}^{(4)})
=
\frac{1}{16}
\sum_{a,b,c,d}
\hat{x}^{(1)}_a
\hat{x}^{(2)}_b
\hat{x}^{(3)}_c
\hat{x}^{(4)}_d
\mathrm{Tr}(\sigma_a\sigma_b\sigma_c^\dagger\sigma_d^\dagger) 
\equiv
\frac{1}{16}
\sum_{a,b,c,d}
\hat{x}^{(1)}_a
\hat{x}^{(2)}_b
\hat{x}^{(3)}_c
\hat{x}^{(4)}_d
I_{abcd}\, . 
\end{align}

Here, $I_{abcd}\equiv\mathrm{Tr}(\sigma_a\sigma_b\sigma_c^\dagger\sigma_d^\dagger)$, where $a, b$ run over $1,2,3,4$ and $i,j$ run over $1,2,3$, can be computed explicitly as follows, obtaining 64 non-zero entries: 

\begin{itemize}
    \item 
    When all indices are the same ($a=b=c=d$), $I_{aaaa}=2$.
(4 non-zero entries.)
    \item 
    If exactly three of the indices are the same, $I_{aaab}=I_{aaba}=I_{abaa}=I_{baaa}=0$. This is because $\sigma_a^3\propto\sigma_a$, and $\mathrm{Tr}(\sigma_a\sigma_b)=0$ for $a\neq b$.

    \item
    If there are two different sets of indices: for $i,j=1,2,3$, 
    $I_{iijj}=-I_{ijij}=I_{ijji}=2$ ($i\neq j$) and $I_{44ii}=I_{ii44}=-2$, $I_{4i4i}=I_{i44i}=I_{4ii4}=I_{i4i4}=2$.  
(36 non-zero entries.)

     \item
    If there are three different indices: then for $a,b,c$ with $a\neq b$, $b\neq c$, $c\neq a$

    $I_{aabc},I_{abac},I_{abca}\propto\mathrm{Tr}(\sigma_b\sigma_c)=0$

    \item
    If all four indices are different, direct calculation gives:

$I_{1234}=-2$, 
$I_{1243}=-2$, 
$I_{1324}=+2$, 
$I_{1342}=+2$,
$I_{1423}=+2$,
$I_{1432}=-2$,

$I_{2134}=+2$, 
$I_{2143}=+2$, 
$I_{2314}=-2$, 
$I_{2341}=-2$, 
$I_{2413}=-2$, 
$I_{2431}=+2$, 

$I_{3124}=-2$, 
$I_{3142}=-2$, 
$I_{3214}=+2$, 
$I_{3241}=+2$, 
$I_{3412}=+2$, 
$I_{3421}=-2$,

$I_{4123}=+2$, 
$I_{4132}=-2$, 
$I_{4213}=-2$,
$I_{4231}=+2$,
$I_{4312}=+2$,
$I_{4321}=-2$\, . 

(24 non-zero entries.)
\end{itemize}

Using these, the Hamiltonian can be written in terms of $\hat{x}$ and $\hat{p}$ straightforwardly. For example, the counter-term:
\begin{align}
    -\gamma\cdot
\sum_{j,\vec{n}}\mathrm{Tr}(\hat{Z}_{j,\vec{n}}\hat{\bar{Z}}_{j,\vec{n}}) = -\frac{\gamma}{2} \cdot
 \sum_{j,\vec{n}} 
\sum_{\ell = 1}^4
\hat{x}_{j,\vec{n};\ell}^2\, .  
\end{align}

Note that $\mathrm{Tr}(\hat{Z}^{(1)}\hat{Z}^{(2)}\hat{\bar{Z}}^{(3)}\hat{\bar{Z}}^{(4)})$ is self-adjoint for any combinations of link variables. This comes from the pseud-reality of the link variable, which is inherited from the SU(2) unitary link. Specifically, because 
\begin{align}
(\mathrm{i}\sigma_\ell)^\ast = \mathrm{i}\sigma_2\sigma_\ell\sigma_2\, , 
\end{align}
we have
\begin{align}
\left(\hat{Z}_{ab}\right)^\dagger
=
\left(\sigma_2\hat{Z}\sigma_2\right)_{ab}\, ,
\end{align}
and hence, 
\begin{align}
\left(
    \mathrm{Tr}(\hat{Z}^{(1)}\hat{Z}^{(2)}\hat{\bar{Z}}^{(3)}\hat{\bar{Z}}^{(4)})
\right)^\dagger
&=
\mathrm{Tr}(\sigma_2\hat{Z}^{(1)}\sigma_2\cdot\sigma_2\hat{Z}^{(2)}\sigma_2\cdot\sigma_2\hat{\bar{Z}}^{(3)}\sigma_2\cdot\sigma_2\hat{\bar{Z}}^{(4)}\sigma_2)
\nonumber\\
&=
\mathrm{Tr}(\hat{Z}^{(1)}\hat{Z}^{(2)}\hat{\bar{Z}}^{(3)}\hat{\bar{Z}}^{(4)})\, .
\end{align}

Therefore the plaquette term is 
\begin{align}
\left(
\hat{Z}_{j,\vec{n}} \, \hat{Z}_{k,\vec{n}+\hat{j}}
\hat{\bar{Z}}_{j,\vec{n}+\hat{k}} \, \hat{\bar{Z}}_{k,\vec{n}}
+
{\rm h.c.}
\right) 
=
\frac{1}{8}\sum_{\ell_1,\ell_2,\ell_3,\ell_4}
\hat{x}_{j,\vec{n};\ell_1} \, \hat{x}_{k,\vec{n}+\hat{j};\ell_2}
\hat{x}_{j,\vec{n}+\hat{k};\ell_3} \, \hat{x}_{k,\vec{n};\ell_4}
I_{\ell_1\ell_2\ell_3\ell_4}\, . 
\end{align}
Using \eqref{eq:x_with_Paulis}, we can write each quartic interaction term as 
\begin{align}
\hat{x}_{a_1}\hat{x}_{a_2}\hat{x}_{a_3}\hat{x}_{a_4}    
=
\sum_{j_1,j_2,j_3,j_4=1}^Q 
\delta_x^4\cdot 2^{j_1+j_2+j_3+j_4-4}\cdot\sigma_{z;a_1,j_1}\sigma_{z;a_2,j_2}\sigma_{z;a_3,j_3}\sigma_{z;a_4,j_4}
\end{align}
in the position basis. 

Having demonstrated how to explicitly express the fundamental building blocks for coding the Hamiltonians — $ \mathrm{Tr}(\hat{Z}\hat{\bar{Z}})$ and $\mathrm{Tr}(\hat{Z}\hat{Z}\hat{\bar{Z}}\hat{\bar{Z}})$ --- the original orbifold Hamiltonian $H$, and the two simplified Hamiltonians $H_1$ and $H_2$,
can be directly written in terms of $\hat{x}$ and $\hat{p}$, making the universal framework~\cite{Halimeh:2024bth} applicable.
Let us repeat the Hamiltonian, assuming the SU(2) embedded into $\mathbb{R}^4$:
\begin{align}
\hat{H}
&=
\sum_{\vec{n}}
\Biggl\{
\frac{1}{2}\sum_{j=1}^{D-1}
\sum_{a=1}^4
\left(
\hat{p}_{j,\vec{n};a}
\right)^2
-
\frac{2g_D^2}{a^{D-1}}\sum_{j\neq k}
{\rm Tr}\left(
\hat{Z}_{j,\vec{n}} \, \hat{Z}_{k,\vec{n}+\hat{j}}
\hat{\bar{Z}}_{j,\vec{n}+\hat{k}} \, \hat{\bar{Z}}_{k,\vec{n}}
\right)
\nonumber\\
&\qquad
\mathcolor{red}{
+
\frac{2g_D^2}{a^{D-1}}\sum_{j\neq k}
{\rm Tr}\left(
\hat{\bar{Z}}_{j,\vec{n}}\, \hat{Z}_{j,\vec{n}}\, \hat{Z}_{k,\vec{n}+\hat{j}}\, \hat{\bar{Z}}_{k,\vec{n}+\hat{j}}
\right)
}
\nonumber\\
&\qquad
\mathcolor{blue}{
+
\frac{g_D^2}{2a^{D-1}}
{\rm Tr}\left(
\left|\sum_{j=1}^{D-1}
\left(
\hat{Z}_{j,\vec{n}} \, \hat{\bar{Z}}_{j,\vec{n}} -\hat{\bar{Z}}_{j,\vec{n}-\hat{j}}\hat{Z}_{j,\vec{n}-\hat{j}}
\right)
\right|^2 
\right)
}
\Biggl\}
+
 \Delta\hat{H}\, , 
 \label{eq:minimal-H-SU(2)}
\end{align}
where $\Delta\hat{H}$ is 
\begin{align}
\Delta\hat{H}
&\equiv
\frac{m^2g_D^2}{2a^{D-3}}\sum_{\vec{n}}\sum_{j=1}^{D-1}
{\rm Tr}
\left|\hat{Z}_{j,\vec{n}}\hat{\bar{Z}}_{j,\vec{n}} -\frac{a^{D-3}}{2g_D^2}\right|^2
\mathcolor{ForestGreen}{
-
\gamma\cdot
\mathrm{Tr}(\hat{Z}_{j,\vec{n}}\hat{\bar{Z}}_{j,\vec{n}})
}\, .
 \label{eq:minimal-dH-SU(2)}
\end{align}
Dropping the term in blue, we obtain $\hat{H}_1$. Further dropping the term in red, we obtain $\hat{H}_2$. In $\Delta\hat{H}$, the counter-term in green is optional. 
\subsection{Minimal Hamiltonians for SU($N$) pure Yang-Mills}
For generic $N$, we can use the embedding of SU($N$) into $\mathbb{R}^{2N^2}$. Let us introduce real and imaginary part of $\hat{Z}_{j,\vec{n};ab}$ and $\hat{P}_{j,\vec{n};ab}$ as
\begin{align}
\hat{Z}_{j,\vec{n};ab}
=
\frac{1}{\sqrt{2}}\left(
\hat{X}_{j,\vec{n};ab}
+
\mathrm{i}\hat{Y}_{j,\vec{n};ab}
\right)\, , 
\qquad
\hat{P}_{j,\vec{n};ab}
=
\frac{1}{\sqrt{2}}\left(
\hat{P}_{X,j,\vec{n};ab}
+
\mathrm{i}\hat{P}_{Y,j,\vec{n};ab}
\right)\, , 
\end{align}
in such a way that $\hat{X}_{j,\vec{n};ab}$, $\hat{Y}_{j,\vec{n};ab}$, $\hat{P}_{X,j,\vec{n};ab}$, and $\hat{P}_{Y,j,\vec{n};ab}$ are self-adjoint. Then, the commutation relation \eqref{Z-P-commutator} becomes
\begin{align}
    \left[
\hat{X}_{j,\vec{n};ab},\hat{P}_{X,j',\vec{n}';a'b'}
    \right]
    =
    \left[
\hat{Y}_{j,\vec{n};ab},\hat{P}_{Y,j',\vec{n}';a'b'}
    \right]
    =
    i\delta_{jj'}\delta_{\vec{n}\vec{n}'}\delta_{aa'}\delta_{bb'}
\end{align}
and
\begin{align}
    \left[
\hat{X}_{j,\vec{n};ab},\hat{P}_{Y,j',\vec{n}';a'b'}
    \right]
    =
    \left[
\hat{Y}_{j,\vec{n};ab},\hat{P}_{X,j',\vec{n}';a'b'}
    \right]
    =
    0\, .
\end{align}
Therefore, we can regard $\hat{X}_{j,\vec{n};ab}$ and $\hat{Y}_{j,\vec{n};ab}$ as bosons $\hat{x}$ in the universal Hamiltonian \eqref{eq:universal_Hamiltonian}. 
The Hamiltonian \eqref{eq:minimal-H-SU(2)} can be used, with a trivial replacement of the kinetic term with 
\begin{align}
\frac{1}{2}
\sum_{\vec{n}}
\sum_{j=1}^{D-1}
\sum_{a,b}
\left(
\hat{P}_{X,j,\vec{n};ab}^2
+
\hat{P}_{Y,j,\vec{n};ab}^2
\right)\, . 
\end{align}
The mass term \eqref{eq:minimal-dH-SU(2)} can be used without change. Note that we do not need to remove the U(1) part as long as we consider pure-Yang Mills because they decouple from the SU($N$) part in pure Yang-Mills theory. 
\section{Numerical Monte Carlo validation}\label{sec:numerical_results}
We use lattice Monte Carlo simulations to verify that the three Hamiltonians $H$, $H_1$, and $H_2$ admit a smooth approach to the Kogut-Susskind (KS) limit in the infinite-mass extrapolation. Specifically, we check that no singularities or phase transitions obstruct the limit, and that the extrapolated observables agree with the Wilson action. The absence of such obstructions was previously demonstrated for the full orbifold Hamiltonian~\eqref{eq:Hamiltonian_orbifold} in~\cite{Bergner:2025zkj}; here we extend the analysis to the simplified Hamiltonians $H_1$ and $H_2$ defined in Eqs.~\eqref{eq:H'_1} and~\eqref{eq:H'_2}.

Simulations are performed in $(1{+}2)$ dimensions on lattices of size $8^3$ and $16^3$, with lattice spacings $a = a_t \in \{0.1,\, 0.3\}$. For each value of $m^2$, we generate 6500 Monte Carlo configurations, discard the first 1000 for thermalization, and compute observables on the remaining 5500. Errors are estimated using the binned jackknife method with bin sizes ranging from 5 to 200 in increments of 5. The KS limit is extracted by fitting observables as quadratic functions of $1/m^2$ and extrapolating to $1/m^2\to 0$. The main text shows results for the $8^3$ lattice; results for the $16^3$ lattice are collected in Appendices~\ref{app:R8A_larger_lattice} and~\ref{app:R4_larger_lattice}.

We measure the following observables, each averaged over all links (or plaquettes), spacetime sites, and Monte Carlo configurations:
\begin{itemize}
\item The \textbf{complexified plaquette} $\langle\mathrm{Tr}(ZZ\bar{Z}\bar{Z})\rangle$, constructed from the spatial plaquette of complex links $Z_{j,\vec{n}}$.
\item The \textbf{spatial unitary plaquette} $\langle\mathrm{Tr}(UU\bar{U}\bar{U})\rangle_{\rm spatial}$ and the \textbf{temporal unitary plaquette} $\langle\mathrm{Tr}(UU\bar{U}\bar{U})\rangle_{\rm temporal}$, constructed from the unitary parts $U_{j,\vec{n}}$ obtained via polar decomposition.
\item The \textbf{scalar deviation} $\langle\mathrm{Tr}(W-\textbf{1}_N)^2\rangle$, which measures how far $W = (Z\bar{Z})^{1/2}$ is from the identity and is expected to vanish as $m^2\to\infty$.
\item The \textbf{determinant} $\langle\det U\rangle$ as a consistency check (identically 1 for SU(2)).
\end{itemize}

In all figures throughout this section, we use a common plotting convention: blue circles, red squares, and green hexagons denote measurements for $H$, $H_1$, and $H_2$, respectively. Solid lines of the same colors show quadratic fits, and triangles mark the extrapolated infinite-mass values. The horizontal orange dashed line (with shaded error band) indicates the corresponding Wilson action value.\\

\noindent
\textbf{Disclaimer:} To avoid clumsy wording, we say something like ``Plot of ... versus ... for $H$, $H_1$, and $H_2$" although, in fact, we use the Lagrangians corresponding to the Hamiltonians $H$, $H_1$, and $H_2$ and Euclidean path integral for Monte Carlo simulations. 
\subsection{Standard embedding into $\mathbb{R}^8$} \label{sec:Embedding_R8}
\begin{figure}[H]
    \centering
    \begin{subfigure}{0.48\textwidth}
        \centering
        \includegraphics[width=\linewidth]{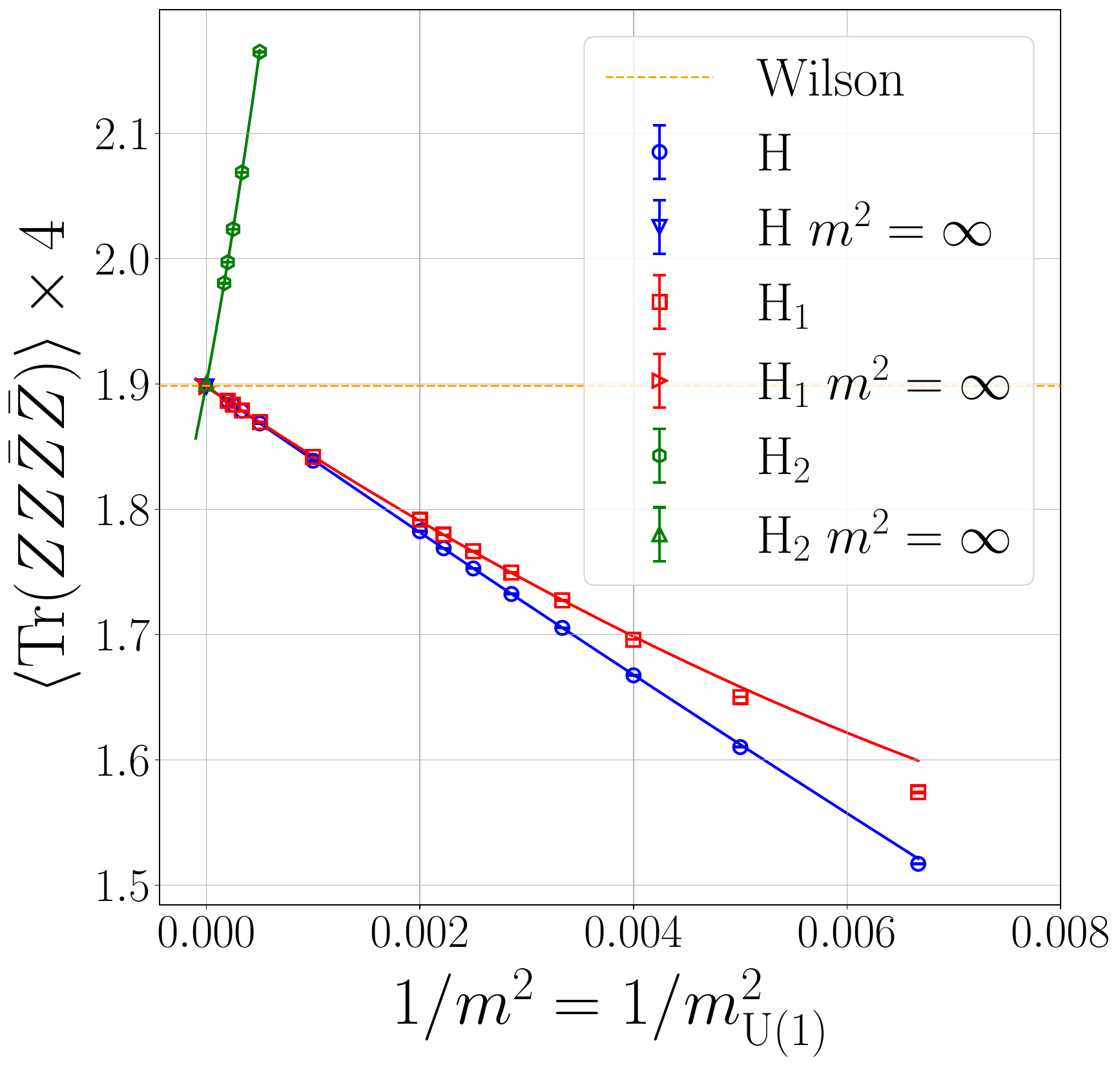}
        {$a = a_t = 0.1$}
    \end{subfigure}
    \hfill
    \begin{subfigure}{0.49\textwidth}
        \centering
    \includegraphics[width=\linewidth]{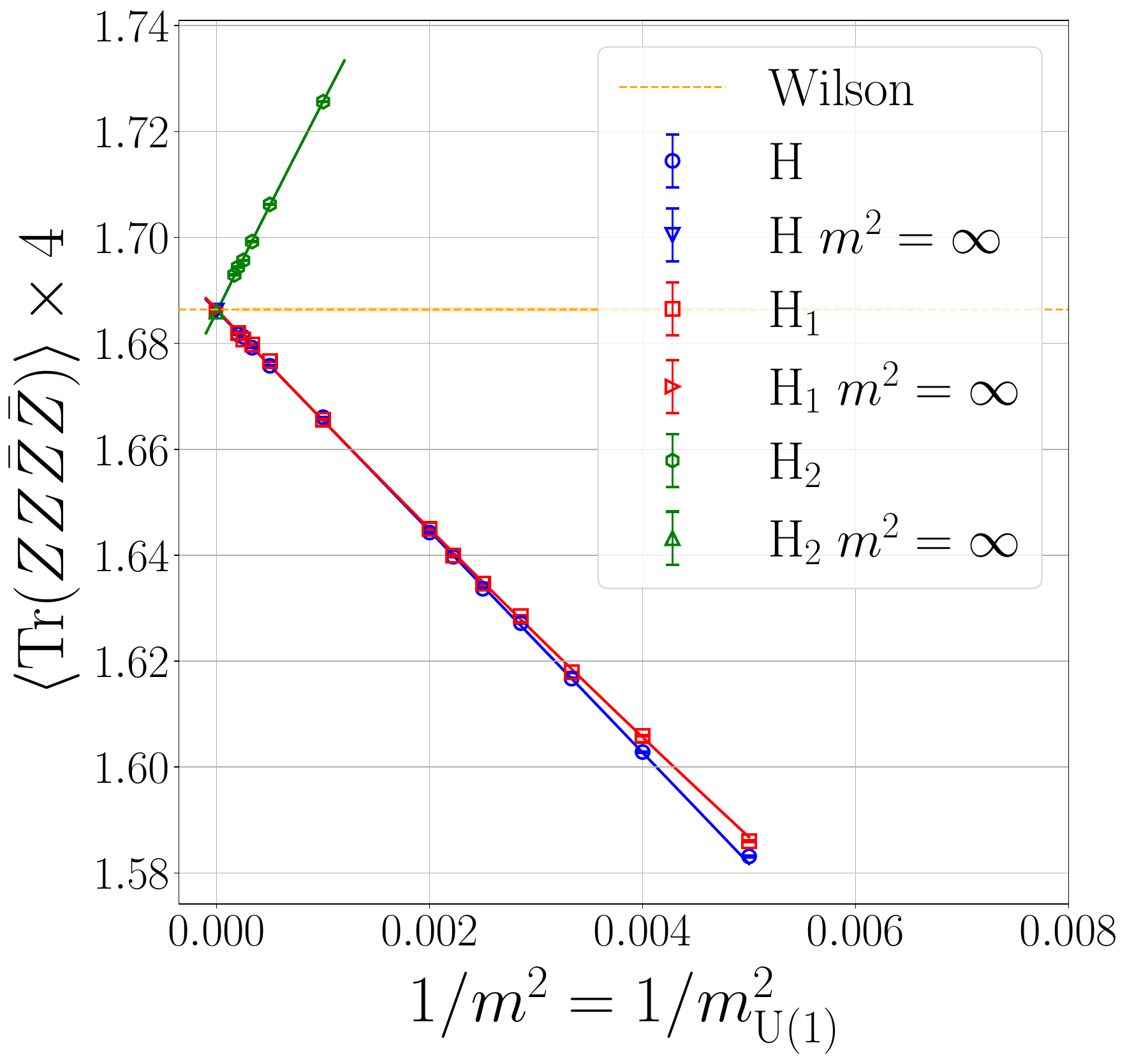}
    {$a=a_t=0.3$}
    \end{subfigure}
    \caption{Plot of $\langle\mathrm{Tr}(ZZ\bar{Z}\bar{Z})\rangle$ versus $1/m^2$ for $H$, $H_1$, and $H_2$ embedded in $\mathbb{R}^8$. Monte Carlo simulation measurements are shown for a lattice size of $8^3$ with two different lattice spacings: $a_t = a = 0.1$ [\textbf{left}] and $0.3$ [\textbf{right}].
    The multiplicative factor of 4 accounts for the difference in coupling conventions between the orbifold and Wilson formulations. The blue, red, and green solid lines show quadratic fits to these measurements, which are used to extract the infinite-mass values, indicated by blue-down, red-right, and green-up triangles for $H$, $H_1$, and $H_2$, respectively. The horizontal orange dashed line represents the corresponding observable value obtained from Wilson action simulations, with the shaded band indicating the associated jackknife error.} \label{fig:Pla_Z_8q_atas_0.1_0.3_R8}
\end{figure}

\begin{figure}[H]
    \centering
    \begin{subfigure}{0.48\textwidth}
        \centering
        \includegraphics[width=\linewidth]{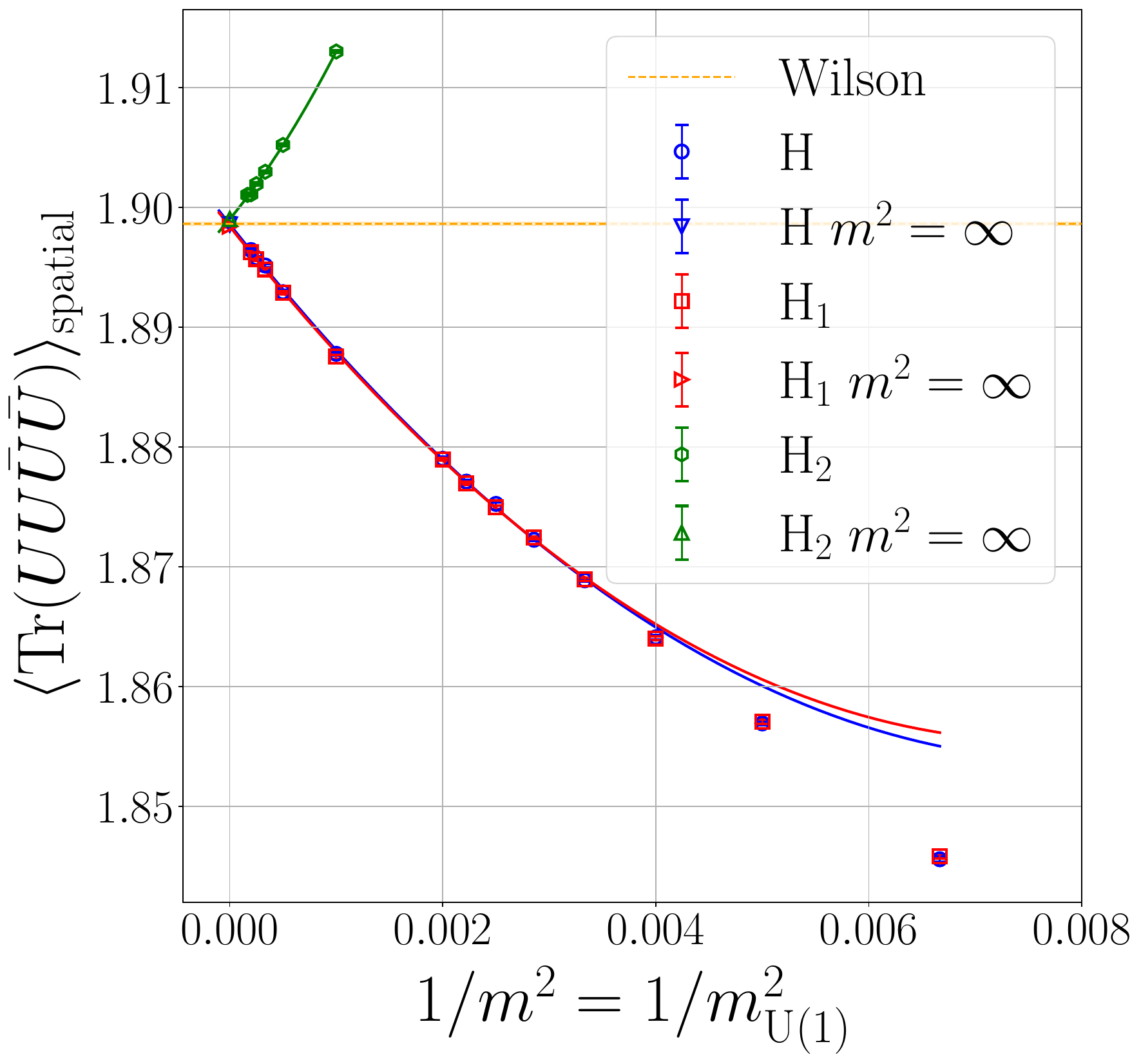}
        {\centering $a_t = a = 0.1$}
    \end{subfigure}
    \hfill
    \begin{subfigure}{0.49\textwidth}
        \centering
    \includegraphics[width=\linewidth]{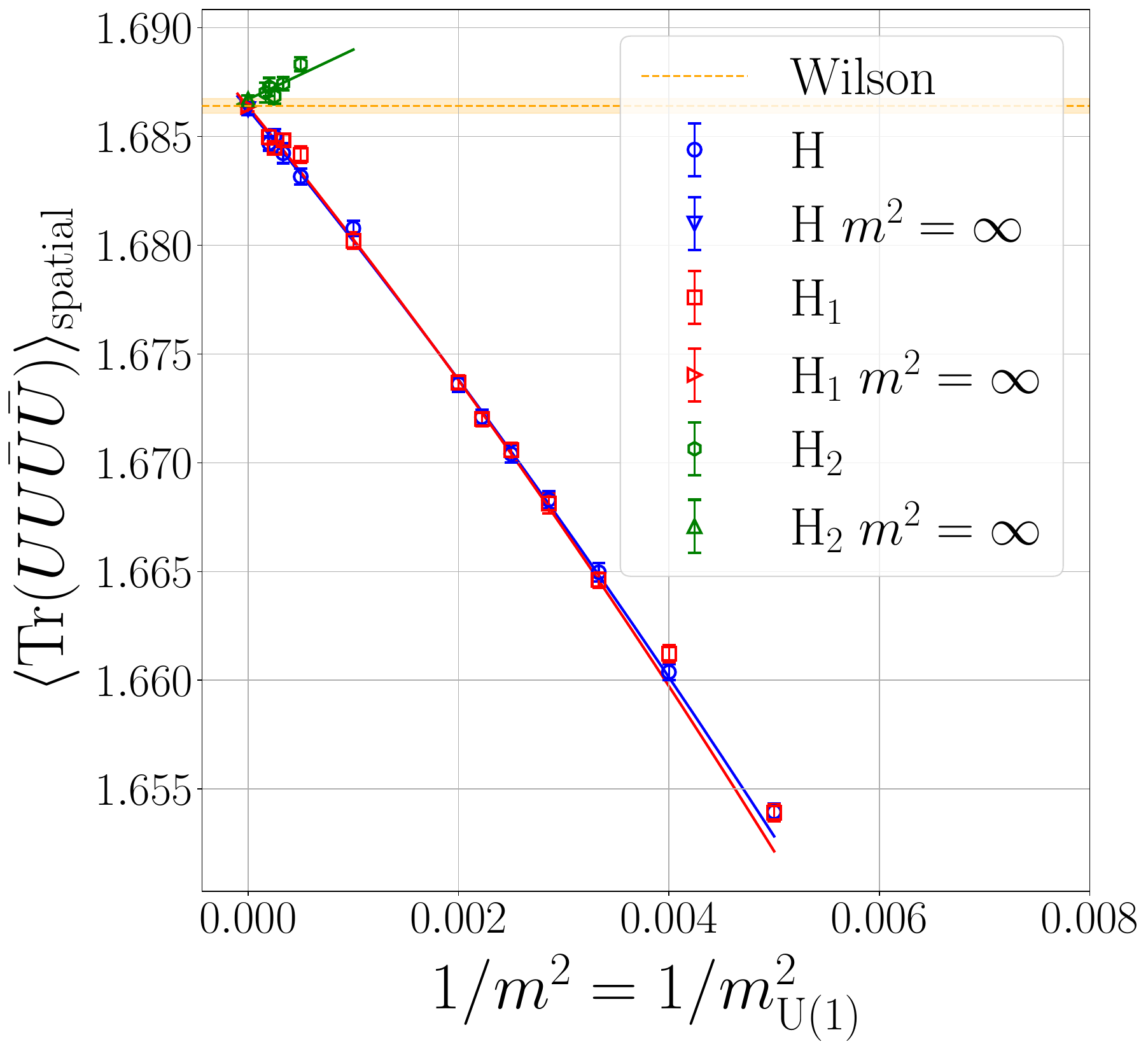}
    {\centering $a_t = a = 0.3$}
    \end{subfigure}
    \caption{Plot of $\langle\mathrm{Tr}(UU\bar{U} \bar{U})\rangle_{\rm spatial}$ versus $1/m^2$ for $H$, $H_1$, and $H_2$ embedded in $\mathbb{R}^8$. Monte Carlo simulation measurements are shown for a lattice size of $8^3$ with two different lattice spacings: $a_t = a = 0.1$ [\textbf{left}] and $0.3$ [\textbf{right}].} \label{fig:spatial_Pla_U_8q_atas_0.1_0.3_R8}
\end{figure}

\begin{figure}[H]
    \centering
    \begin{subfigure}{0.48\textwidth}
        \centering
        \includegraphics[width=\linewidth]{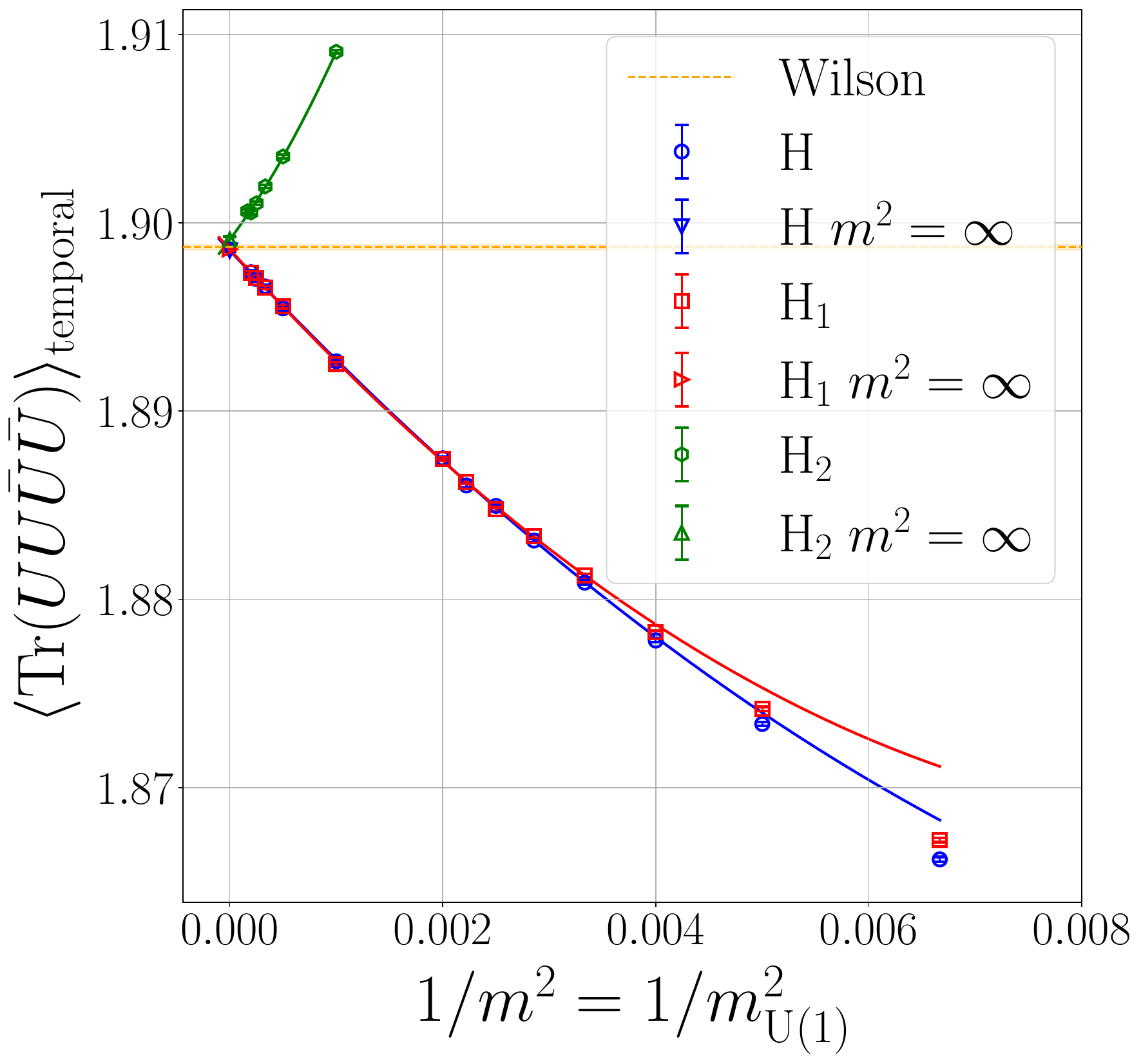}
        {\centering $a_t = a = 0.1$}
    \end{subfigure}
    \hfill
    \begin{subfigure}{0.49\textwidth}
        \centering
    \includegraphics[width=\linewidth]{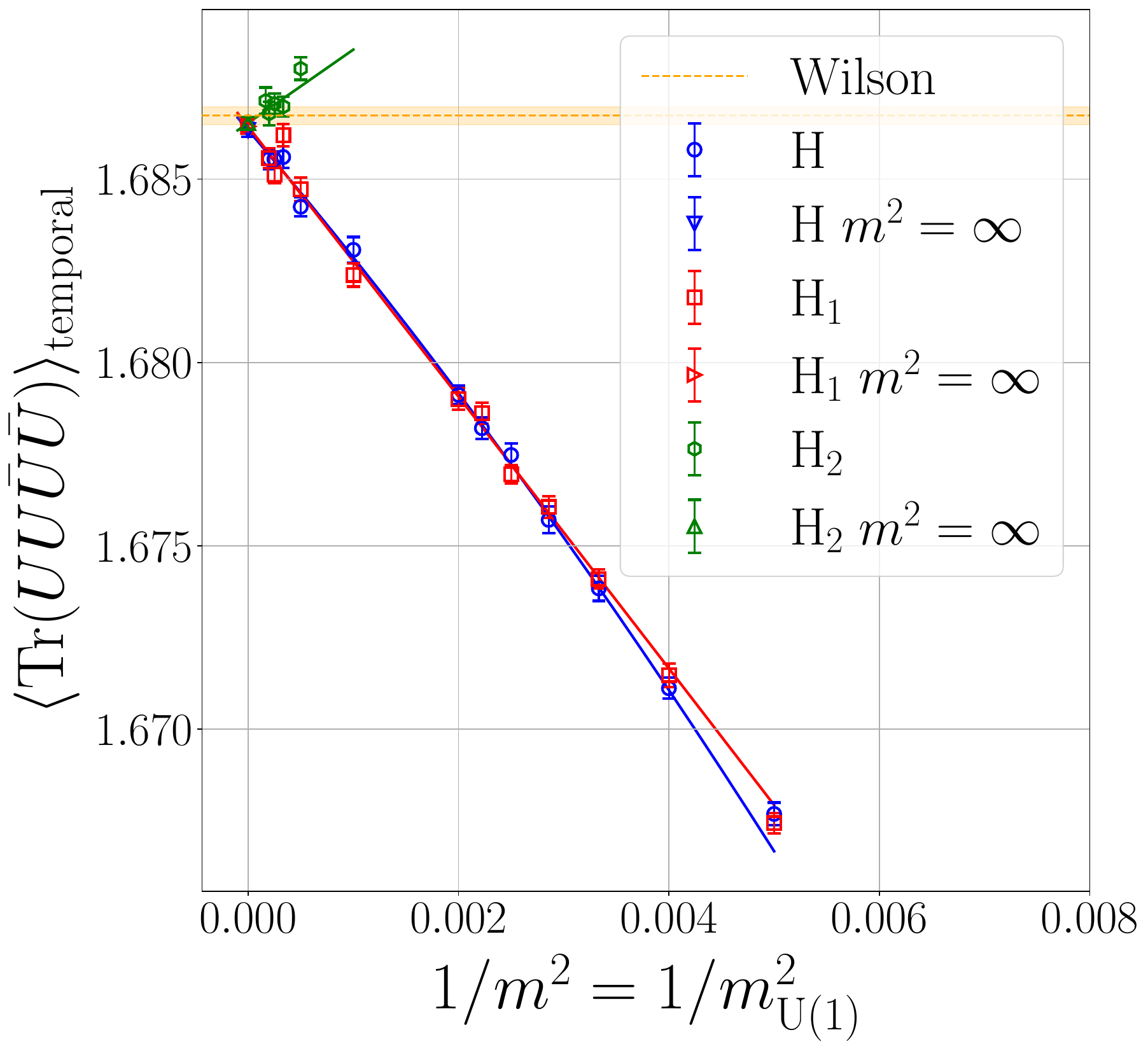}
    {\centering $a_t = a = 0.3$}
    \end{subfigure}
    \caption{ Plot of $\langle\mathrm{Tr}(UU\bar{U} \bar{U})\rangle_{\rm temporal}$ versus $1/m^2$ for $H$, $H_1$, and $H_2$ embedded in $\mathbb{R}^8$. Monte Carlo simulation measurements are shown for a lattice size of $8^3$ with two different lattice spacings: $a_t = a = 0.1$ [\textbf{left}] and $0.3$ [\textbf{right}].} \label{fig:temporal_Pla_U_8q_atas_0.1_0.3_R8}
\end{figure}

\begin{figure}[H]
    \centering
    \begin{subfigure}{0.48\textwidth}
        \centering
        \includegraphics[width=\linewidth]{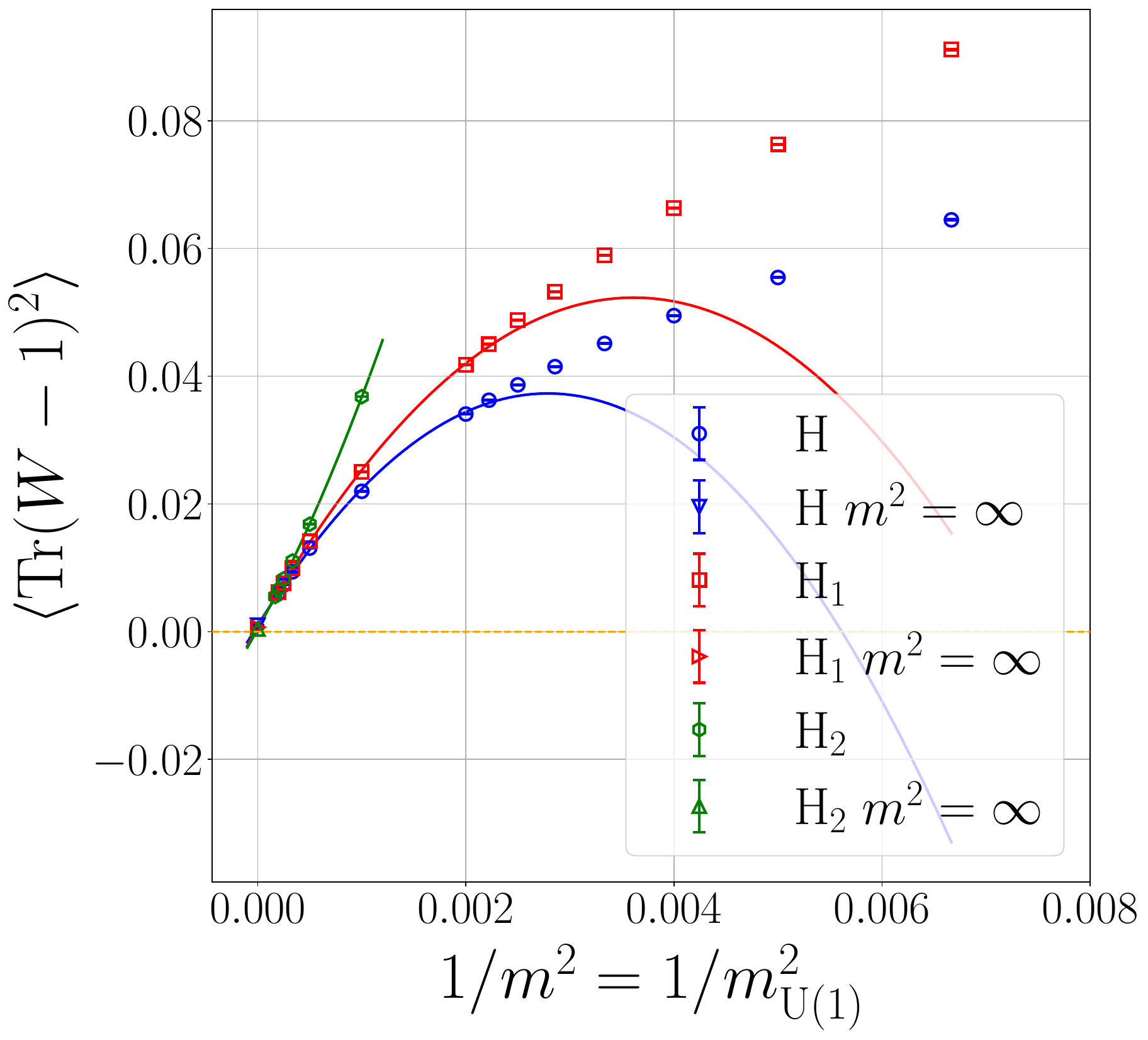}
        {$a=a_t = 0.1$}
    \end{subfigure}
    \hfill
    \begin{subfigure}{0.47\textwidth}
        \centering
    \includegraphics[width=\linewidth]{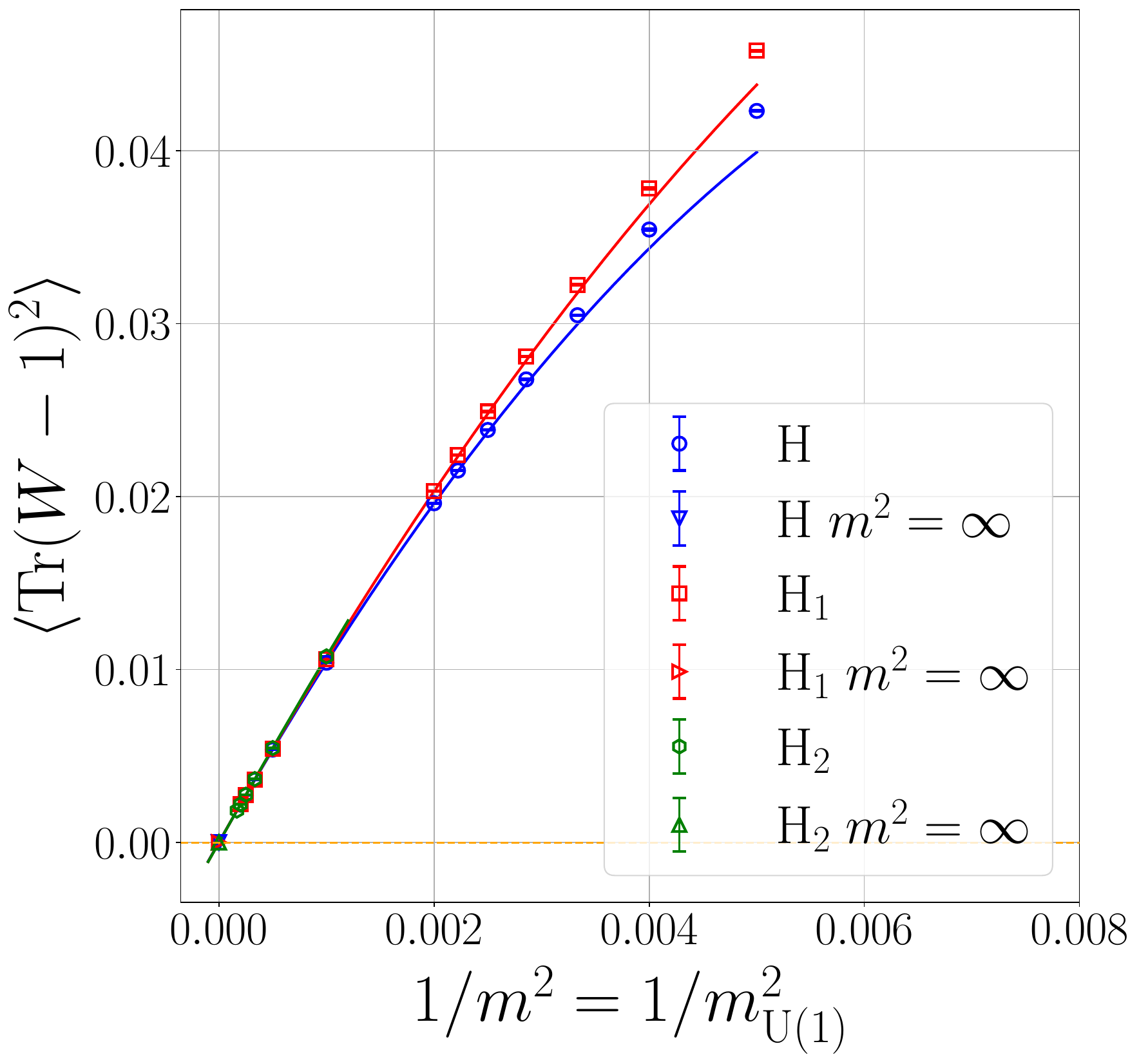}
    {\centering $a_t = a = 0.3$}
    \end{subfigure}
    \caption{Plot of $\langle\mathrm{Tr}(W-\textbf{1}_N)^2\rangle$ versus $1/m^2$ for $H$, $H_1$, and $H_2$ embedded in $\mathbb{R}^8$. Monte Carlo simulation measurements are shown for a lattice size of $8^3$ with two different lattice spacings: $a_t = a = 0.1$ [\textbf{left}] and $0.3$ [\textbf{right}]. } \label{fig:TrWm1sq_8q_atas_0.1_0.3_R8}
\end{figure}

Figures~\ref{fig:Pla_Z_8q_atas_0.1_0.3_R8}--\ref{fig:TrWm1sq_8q_atas_0.1_0.3_R8} show the four plaquette-type observables and the scalar deviation $\langle\mathrm{Tr}(W-\textbf{1}_N)^2\rangle$ as functions of $1/m^2$, for all three Hamiltonians embedded in $\mathbb{R}^8$, at lattice spacings $a = a_t = 0.1$ and $0.3$.

All three plaquette observables converge smoothly to the Wilson action values as $m^2$ increases, with no sign of phase transitions or singularities. The extrapolated infinite-mass values are consistent with the Wilson action for all three Hamiltonians at both lattice spacings.

The scalar deviation $\langle\mathrm{Tr}(W-\textbf{1}_N)^2\rangle$ (Fig.~\ref{fig:TrWm1sq_8q_atas_0.1_0.3_R8}) decreases monotonically with $m^2$ and extrapolates to zero, confirming that $W\to\textbf{1}_N$ and the scalars decouple, leaving pure Yang-Mills theory.

\subsection{Minimal embedding into $\mathbb{R}^4$}
\label{sec:Embedding_R4}
\begin{figure}[H]
    \centering
    \begin{subfigure}{0.48\textwidth}
        \centering
        \includegraphics[width=\linewidth]{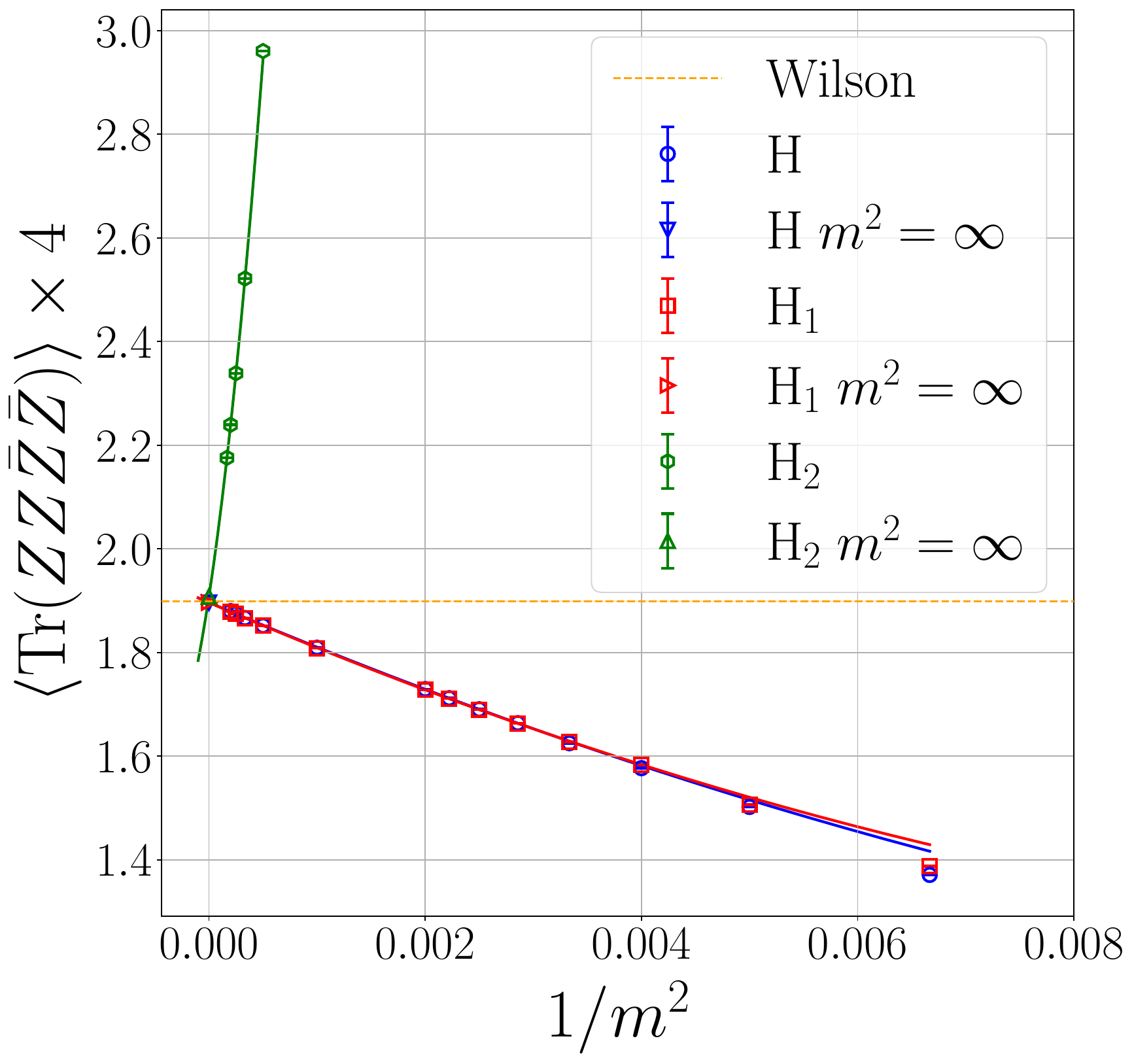}
        {$a=a_t=0.1$}
    \end{subfigure}
    \hfill
    \begin{subfigure}{0.48\textwidth}
        \centering
    \includegraphics[width=\linewidth]{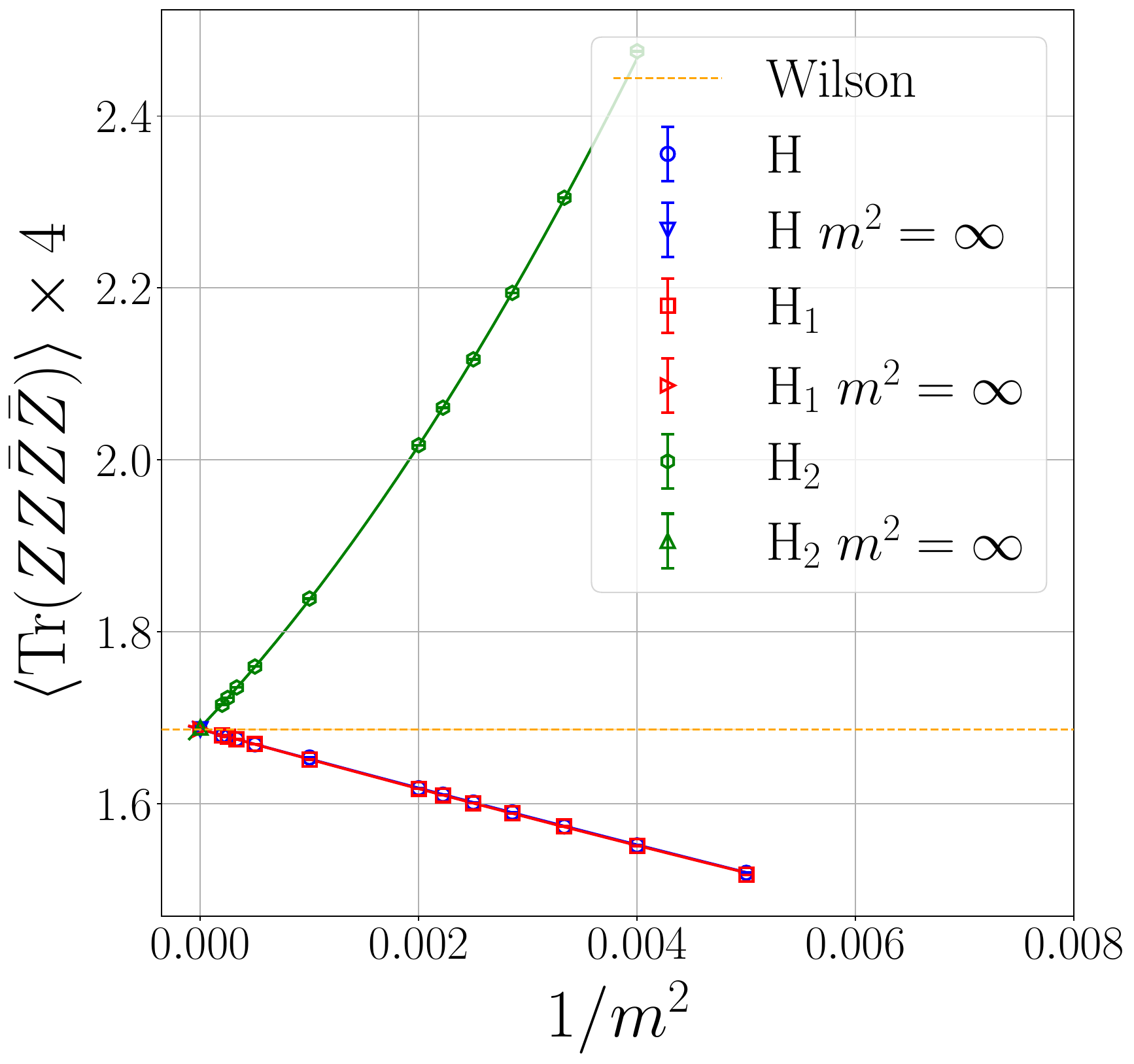}
    {$a=a_t=0.3$}
    \end{subfigure}
    \caption{Plot of $\langle\mathrm{Tr}(ZZ\bar{Z}\bar{Z})\rangle$ versus $1/m^2$ for $H$, $H_1$, and $H_2$ embedded in $\mathbb{R}^4$. Monte Carlo simulation measurements are shown for a lattice size of $8^3$ with two different lattice spacings: $a_t = a = 0.1$ [\textbf{left}] and $0.3$ [\textbf{right}].
    The multiplicative factor of 4 accounts for the difference in coupling conventions between the orbifold and Wilson formulations.} \label{fig:Pla_Z_8q_atas_0.1_0.3}
\end{figure}

\begin{figure}[H]
    \centering
    \begin{subfigure}{0.48\textwidth}
        \centering
        \includegraphics[width=\linewidth]{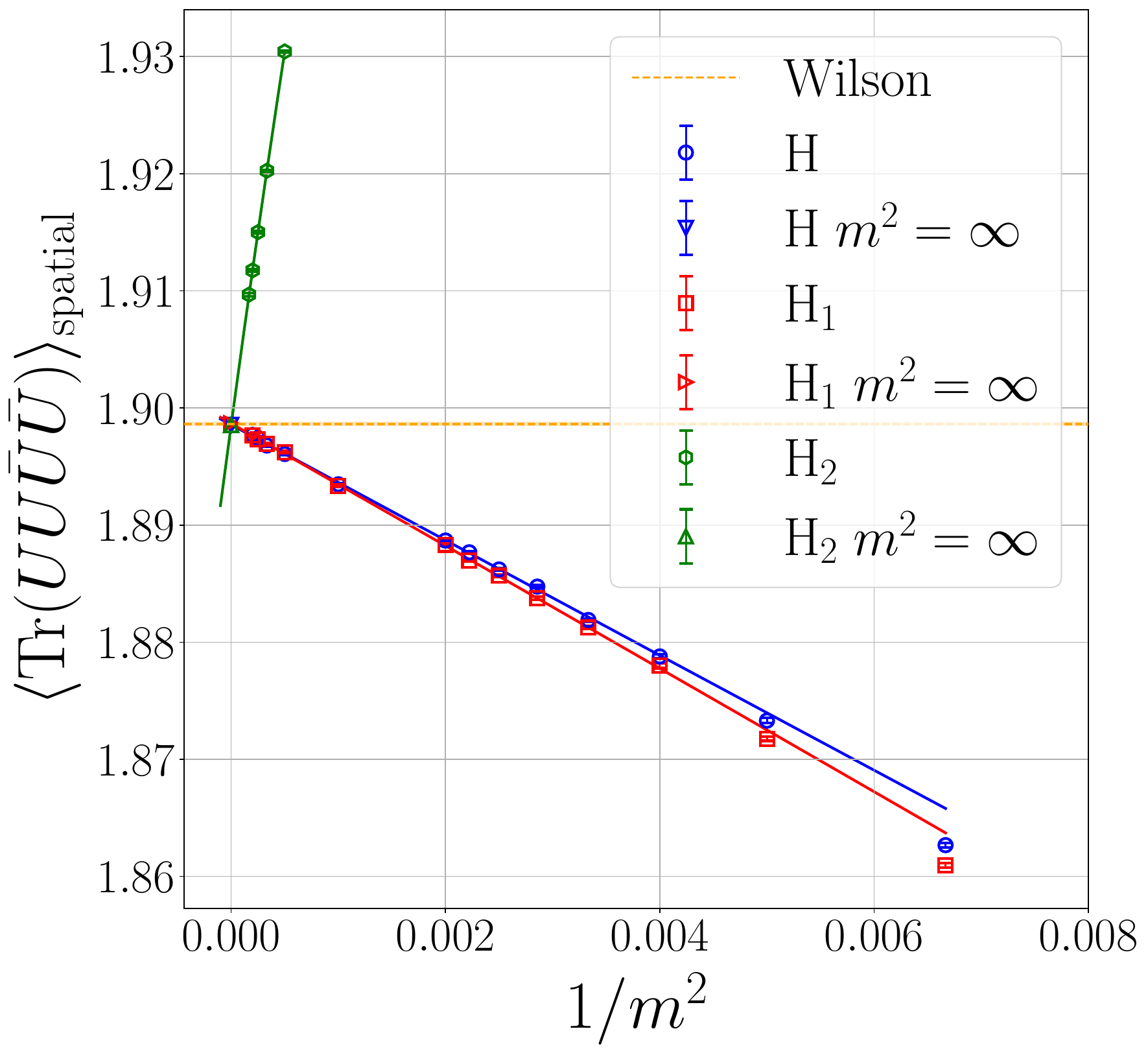}
        { $a_t=a=0.1$}
    \end{subfigure}
    \hfill
    \begin{subfigure}{0.48\textwidth}
        \centering
    \includegraphics[width=\linewidth]{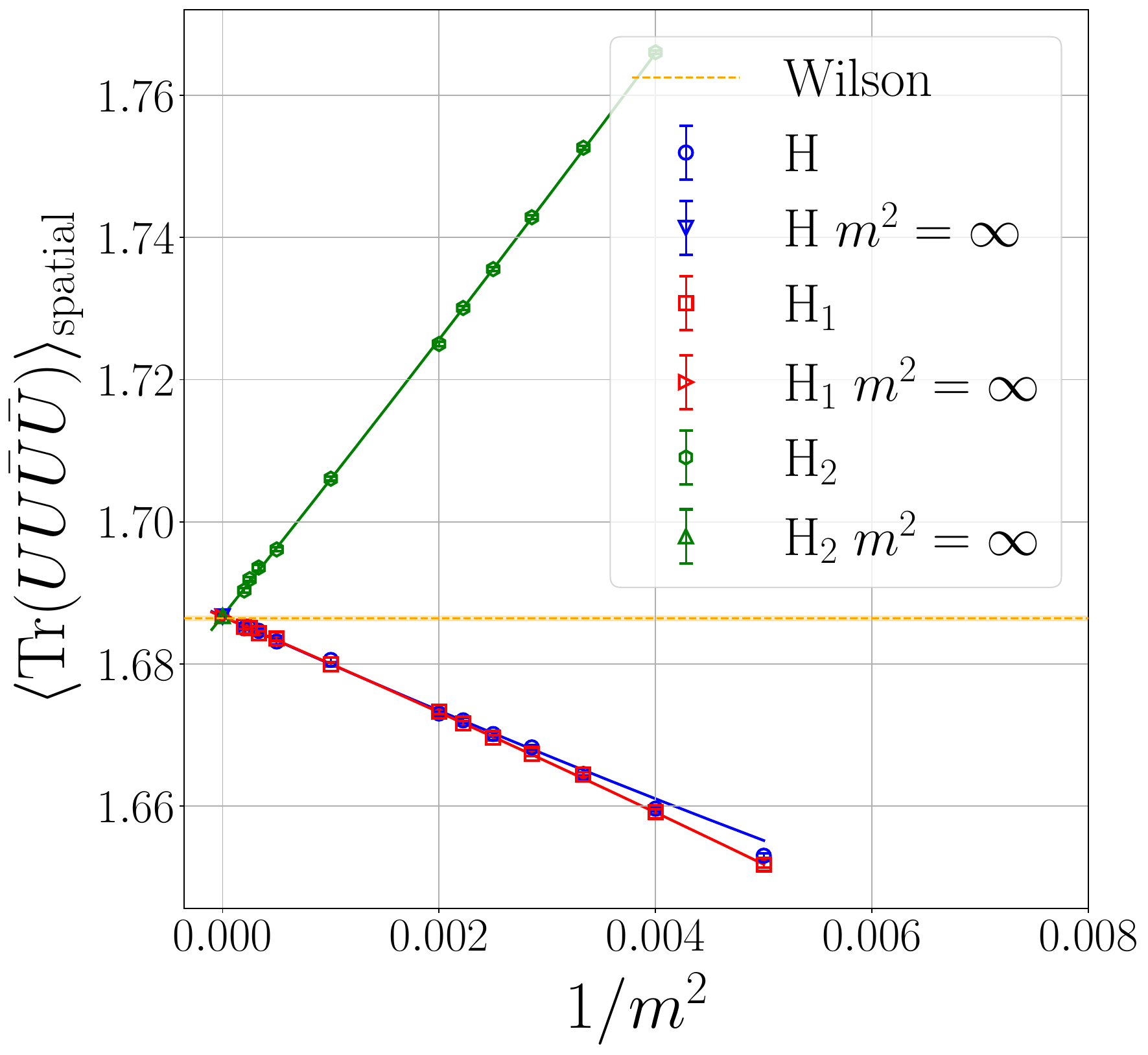}
    {$a_t=a=0.3$}
    \end{subfigure}
    \caption{ Plot of $\langle\mathrm{Tr}(UU\bar{U} \bar{U})\rangle_{\rm spatial}$ versus $1/m^2$ for $H$, $H_1$, and $H_2$ embedded in $\mathbb{R}^4$. Monte Carlo simulation measurements are shown for a lattice size of $8^3$ with two different lattice spacings: $a_t = a = 0.1$ [\textbf{left}] and $0.3$ [\textbf{right}].} \label{fig:spatial_Pla_U_8q_atas_0.1_0.3}
\end{figure}

\begin{figure}[H]
    \centering
    \begin{subfigure}{0.48\textwidth}
        \centering
        \includegraphics[width=\linewidth]{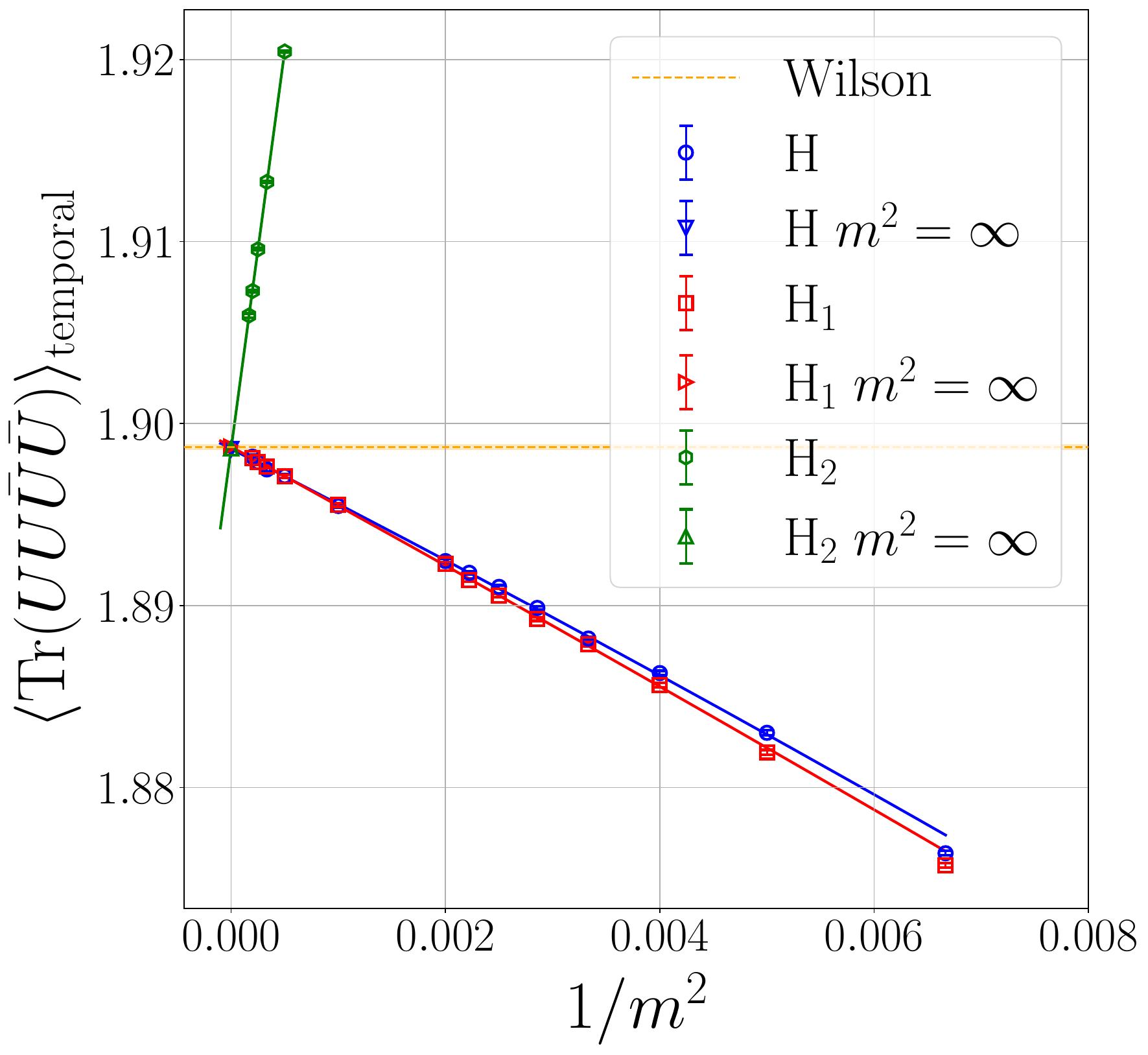}
        {$a = a_t = 0.1$}
    \end{subfigure}
    \hfill
    \begin{subfigure}{0.48\textwidth}
        \centering
    \includegraphics[width=\linewidth]{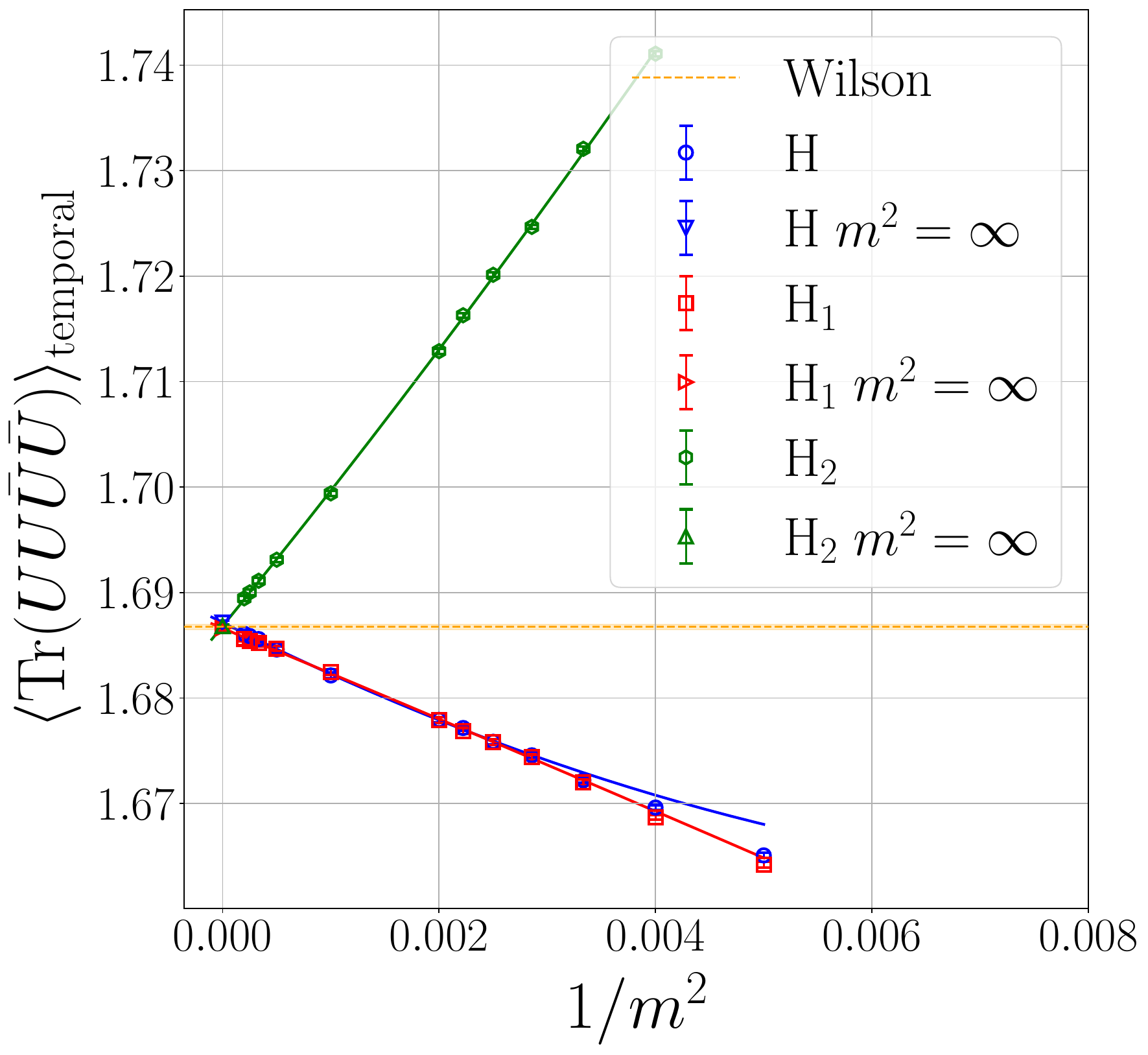}
    { $a = a_t = 0.3$}
    \end{subfigure}
    \caption{ Plot of $\langle\mathrm{Tr}(UU\bar{U} \bar{U})\rangle_{\rm temporal}$ versus $1/m^2$ for $H$, $H_1$, and $H_2$ embedded in $\mathbb{R}^4$. Monte Carlo simulation measurements are shown for a lattice size of $8^3$ with two different lattice spacings: $a_t = a = 0.1$ [\textbf{left}] and $0.3$ [\textbf{right}].} \label{fig:temporal_Pla_U_8q_atas_0.1_0.3}
\end{figure}

\begin{figure}[H]
    \centering
    \begin{subfigure}{0.48\textwidth}
        \centering
        \includegraphics[width=\linewidth]{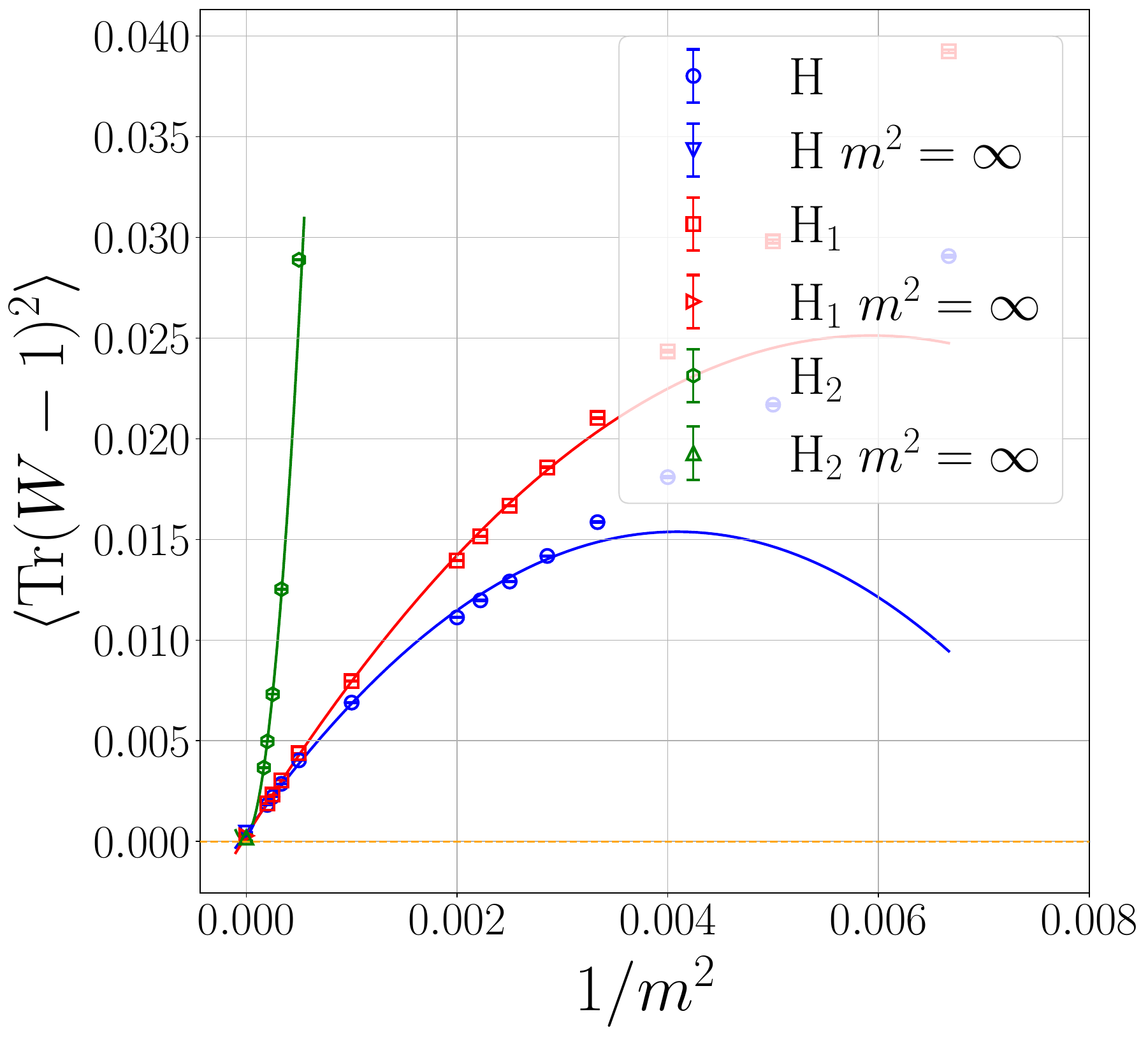}
        {$a_t=a=0.1$}
    \end{subfigure}
    \hfill
    \begin{subfigure}{0.48\textwidth}
        \centering
    \includegraphics[width=\linewidth]{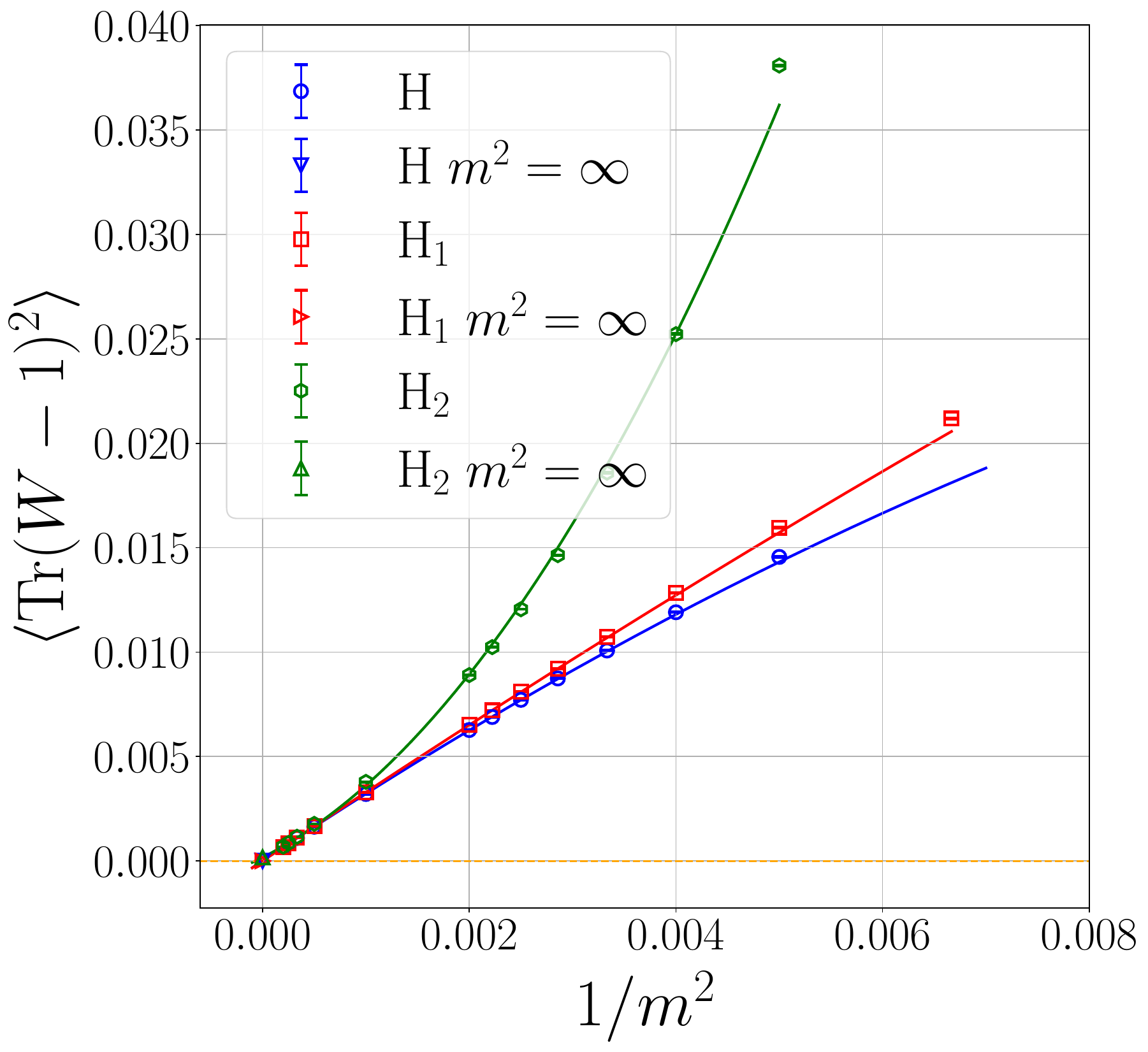}
    {$a_t=a=0.3$}
    \end{subfigure}
    \caption{Plot of $\langle\mathrm{Tr}(W-\textbf{1}_N)^2\rangle$ versus $1/m^2$ for $H$, $H_1$, and $H_2$ embedded in $\mathbb{R}^4$. Monte Carlo simulation measurements are shown for a lattice size of $8^3$ with two different lattice spacings: $a_t = a = 0.1$ [\textbf{left}] and $0.3$ [\textbf{right}]. } \label{fig:TrWm1sq_8q_atas_0.1_0.3}
\end{figure}

We repeat the analysis of the previous section with SU(2) embedded into $\mathbb{R}^4$ instead of $\mathbb{R}^8$, using the same simulation parameters. The results (Figs.~\ref{fig:Pla_Z_8q_atas_0.1_0.3}--\ref{fig:TrWm1sq_8q_atas_0.1_0.3}) show the same qualitative behavior: all plaquette observables converge smoothly to the Wilson action, and $\langle\mathrm{Tr}(W-\textbf{1}_N)^2\rangle$ extrapolates to zero. The more efficient $\mathbb{R}^4$ encoding thus introduces no new obstructions in the KS limit. Results for the $16^3$ lattice are provided in Appendix~\ref{app:R4_larger_lattice}.

\subsection{Reducing the required scalar mass}\label{sec:numerics_eliminate_large_mass}
We now test the two strategies proposed in Secs.~\ref{sec:ZZbar_counterterm} and~\ref{sec:tune_bare_spacing} to reduce the scalar mass $m^2$ needed to reach the KS limit: the linear counter-term and the tuning of the bare lattice spacing. The simulation parameters are the same as above.
\subsubsection{Adding counter-term: $\mathbb{R}^8$ embedding }\label{sec:results_ZZbar_counterterm_R8}
\begin{figure}[H]
    \centering
    \begin{subfigure}{0.49\textwidth}
        \centering
        \includegraphics[width=\linewidth]{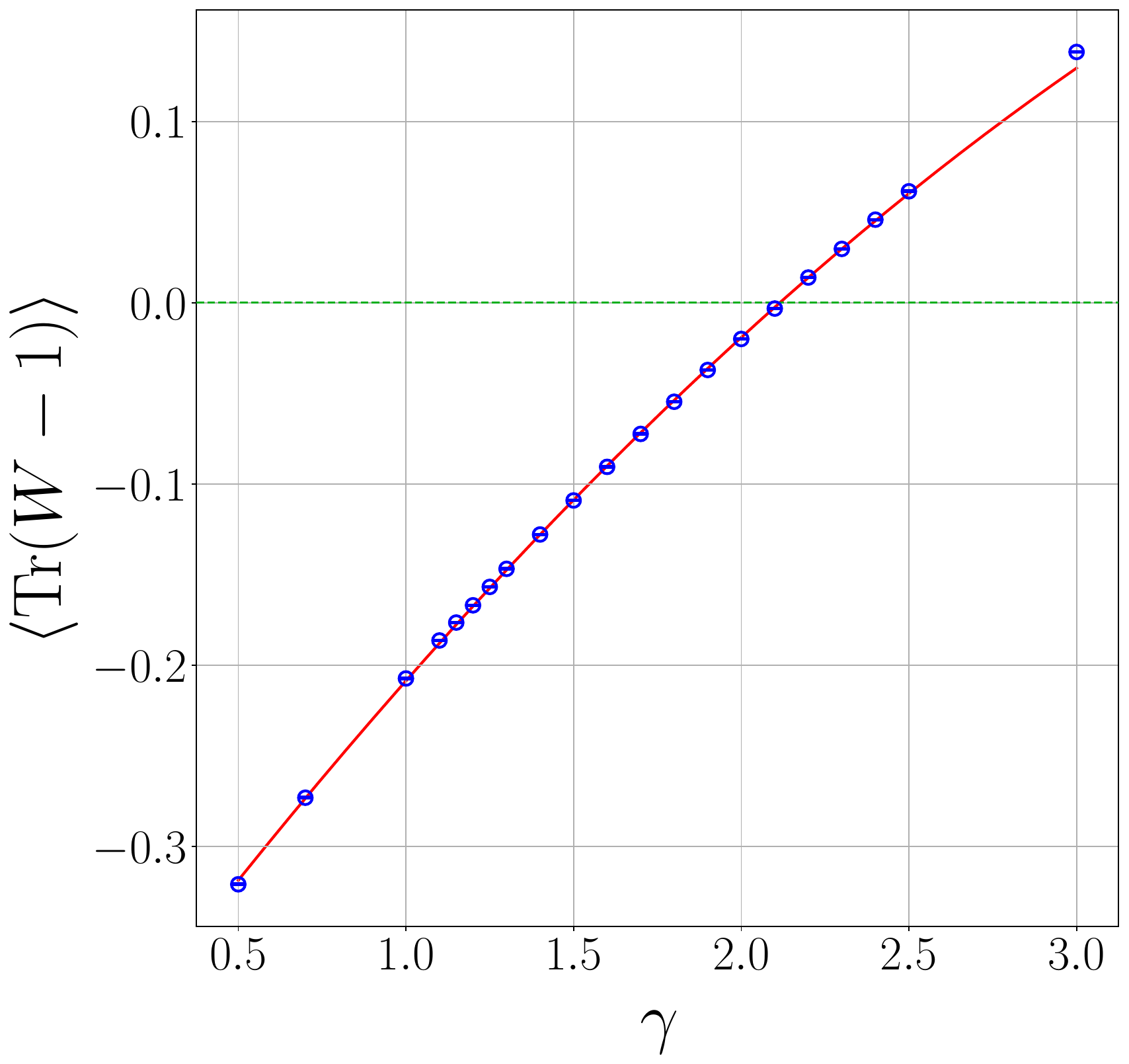}
    \end{subfigure}
    \hfill
    \begin{subfigure}{0.48\textwidth}
        \centering
    \includegraphics[width=\linewidth]{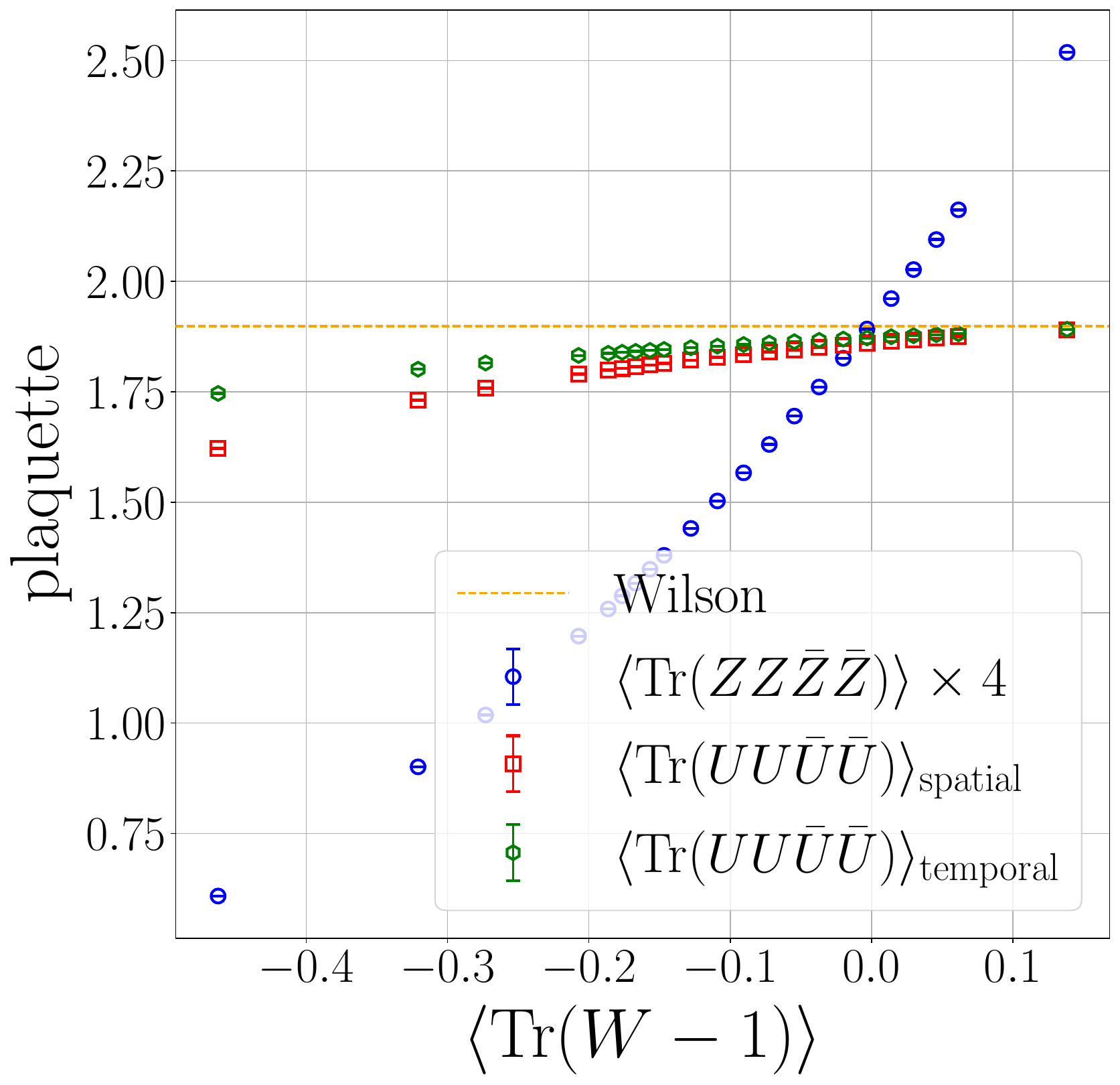}
    \end{subfigure}
    \caption{The $\hat{H}$ Hamiltonian embedded in $\mathbb{R}^8$, on $8^3$ and $a=a_t =0.1$, with $m^2=50$. [\textbf{Left}] $\langle \mathrm{Tr}(W-\mathbf{1}_N)\rangle$ versus $\gamma$. The red line is a quadratic fit. The green dashed line indicates the target value of zero.  
[\textbf{Right}] Plaquette expectation value versus $\langle \mathrm{Tr}(W-\mathbf{1}_N)\rangle$. Symbols denote the three plaquette observables (see text for color conventions).} \label{fig:TrW_1_gamma_R8_H}
\end{figure}

\begin{figure}[H]
    \centering
    \begin{subfigure}{0.50\textwidth}
        \centering
        \includegraphics[width=\linewidth]{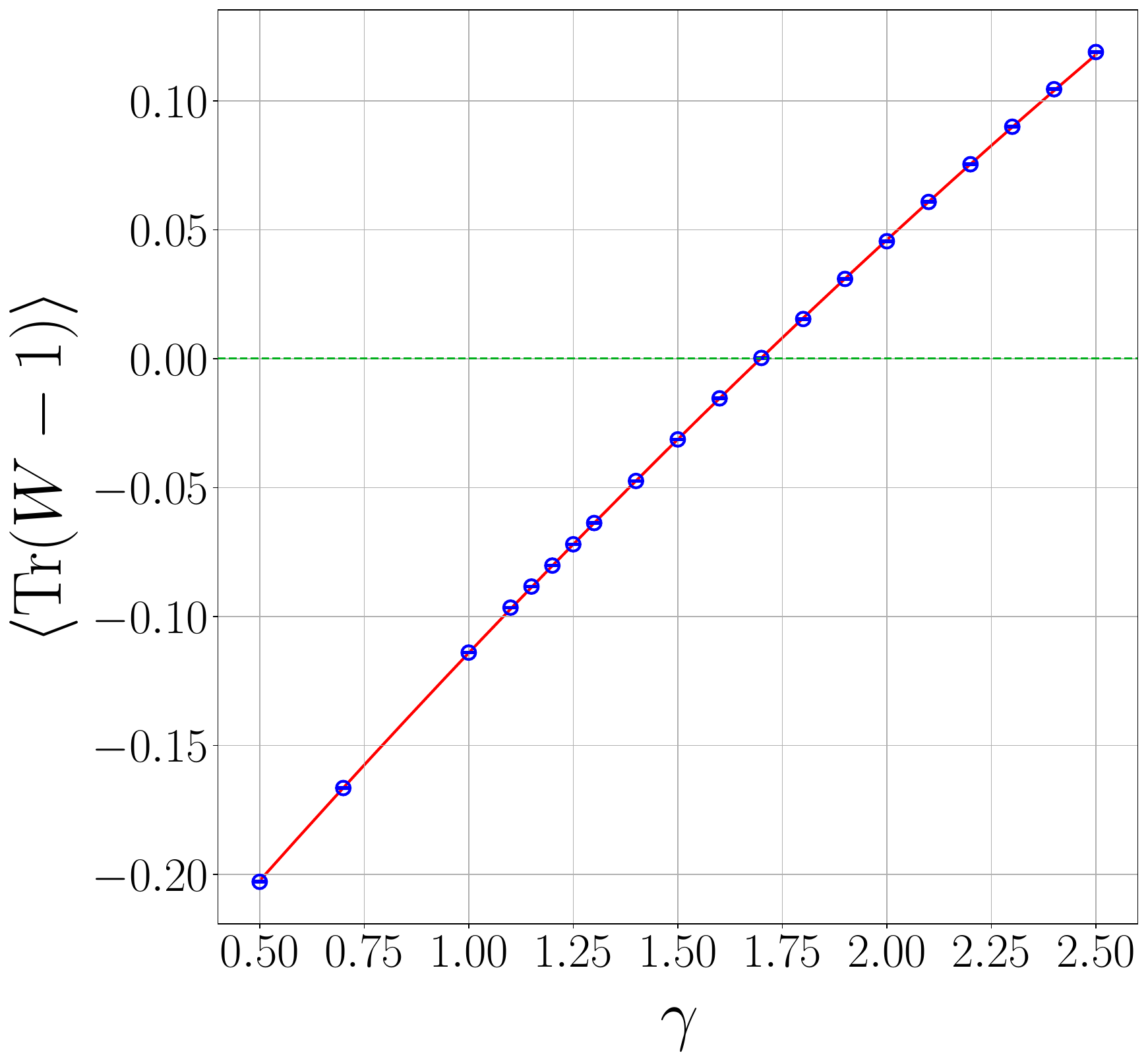}
    \end{subfigure}
    \hfill
    \begin{subfigure}{0.47\textwidth}
        \centering
    \includegraphics[width=\linewidth]{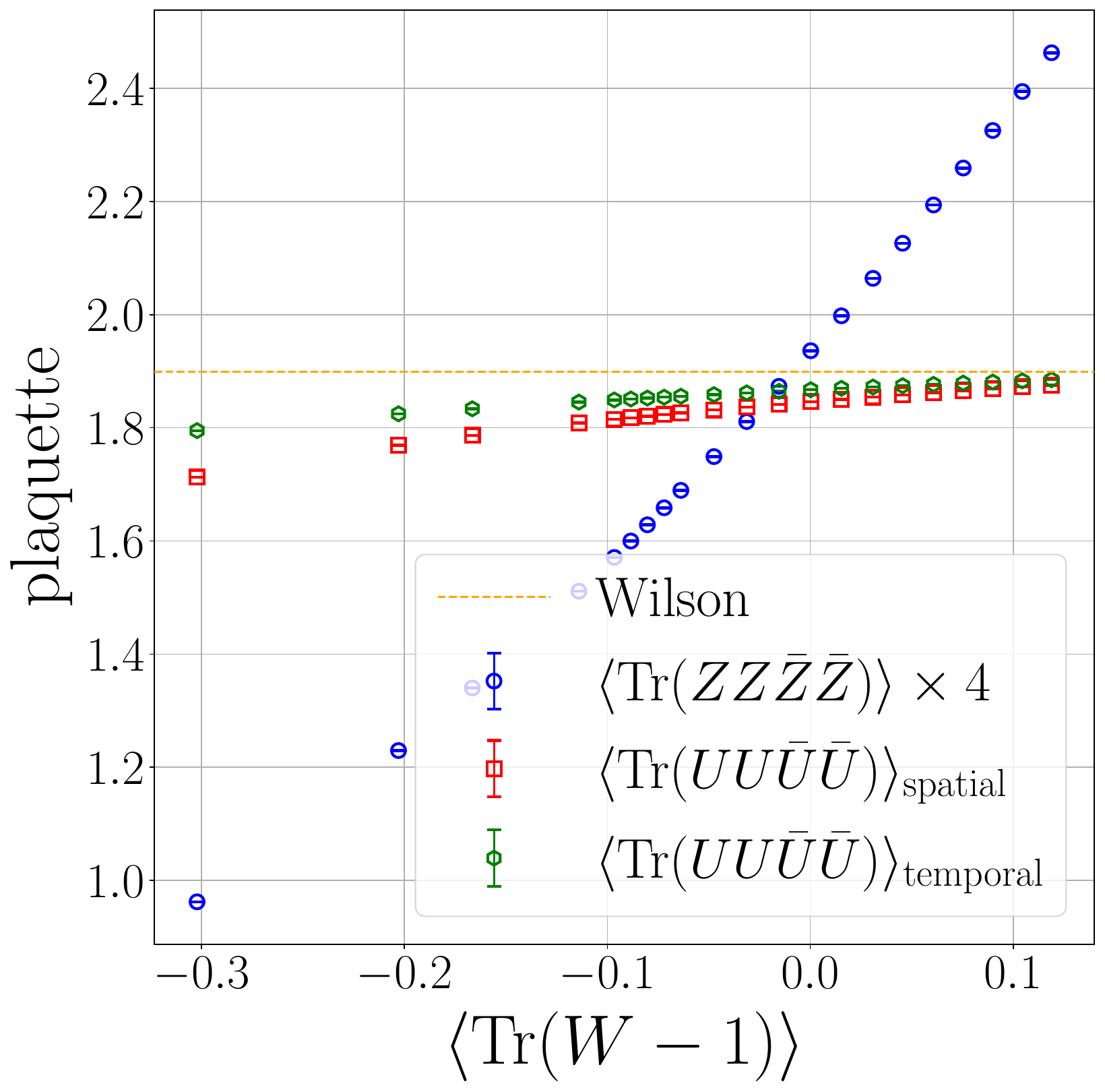}
    \end{subfigure}
    \caption{The $\hat{H}_1$ Hamiltonian embedded in $\mathbb{R}^8$, on $8^3$ and $a=a_t =0.1$, with $m^2=50$. [\textbf{Left}] $\langle \mathrm{Tr}(W-\mathbf{1}_N)\rangle$ versus $\gamma$. The red line is a quadratic fit. The green dashed line indicates the target value of zero.  
[\textbf{Right}] Plaquette expectation value versus $\langle \mathrm{Tr}(W-\mathbf{1}_N)\rangle$. Symbols denote the three plaquette observables (see text for color conventions). } \label{fig:TrW_1_gamma_R8_H1}
\end{figure}

\begin{figure}[H]
    \centering
    \begin{subfigure}{0.51\textwidth}
        \centering
        \includegraphics[width=\linewidth]{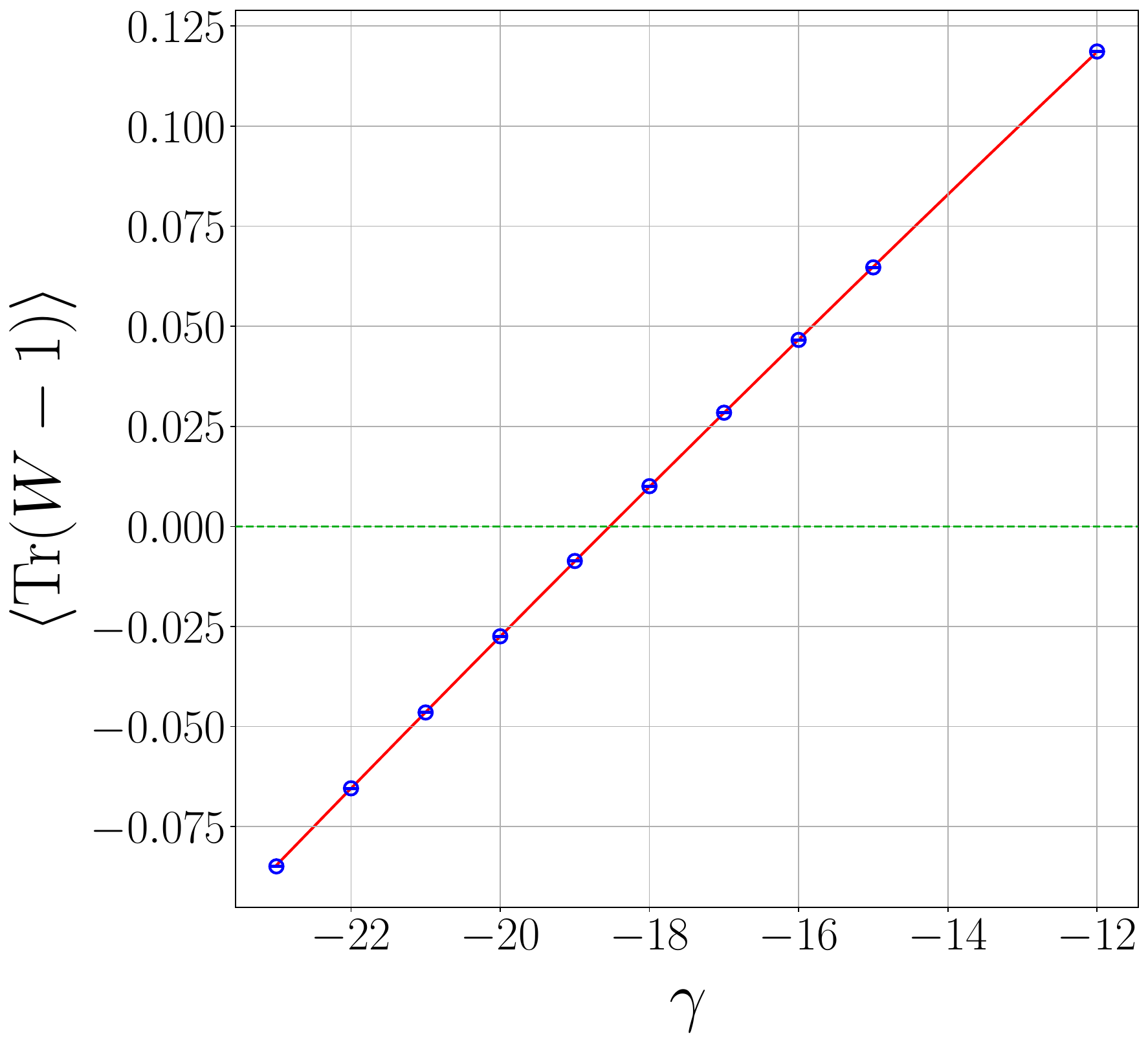}
    \end{subfigure}
    \hfill
    \begin{subfigure}{0.47\textwidth}
        \centering
    \includegraphics[width=\linewidth]{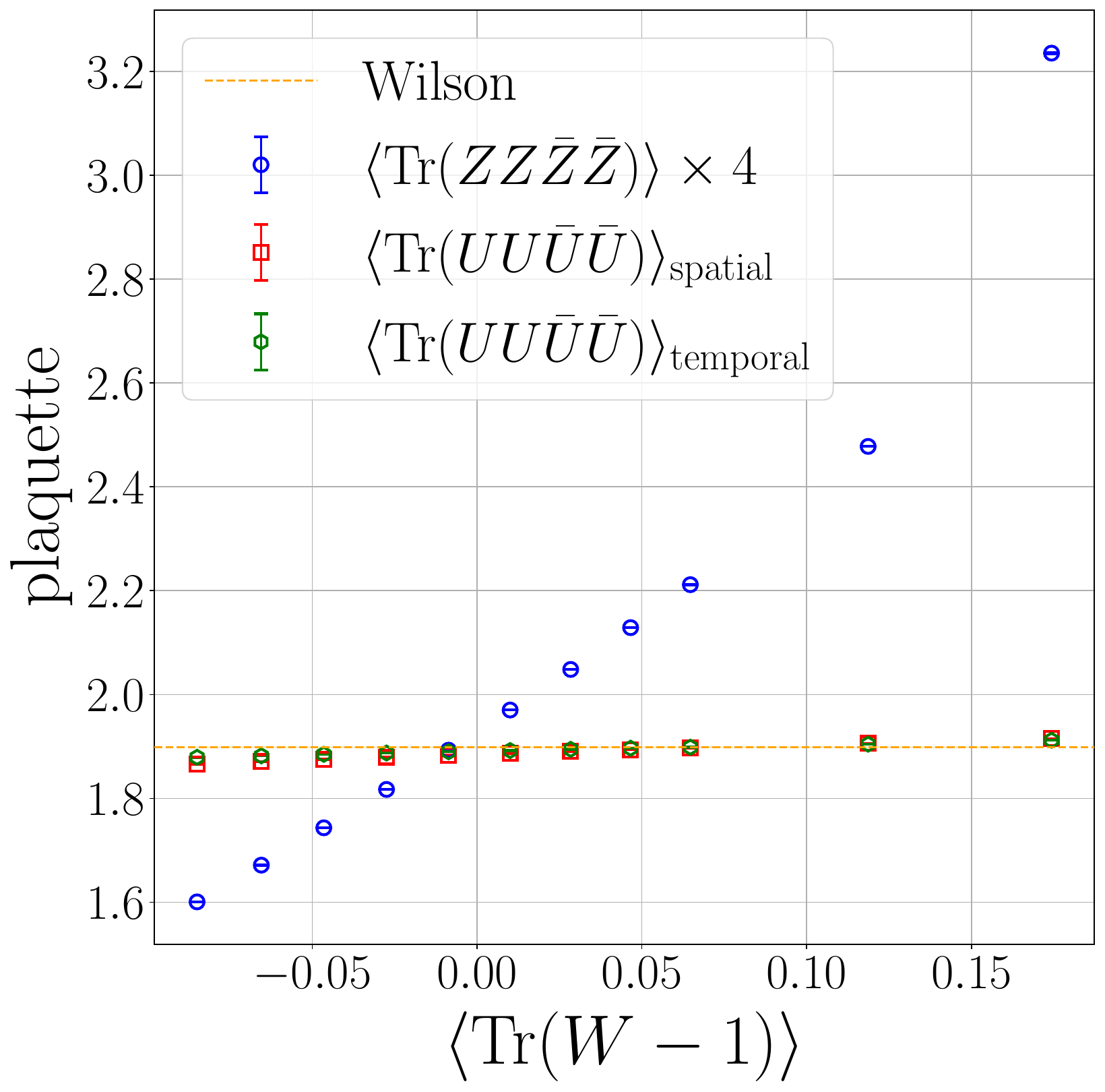}
    \end{subfigure}
    \caption{The $\hat{H}_2$ Hamiltonian embedded in $\mathbb{R}^8$, on $8^3$ and $a=a_t =0.1$, with $m^2=500$. [\textbf{Left}] $\langle \mathrm{Tr}(W-\mathbf{1}_N)\rangle$ versus $\gamma$. The red line is a quadratic fit.  The green dashed line indicates the target value of zero.  
[\textbf{Right}] Plaquette expectation value versus $\langle \mathrm{Tr}(W-\mathbf{1}_N)\rangle$. Symbols denote the three plaquette observables (see text for color conventions). } \label{fig:TrW_1_gamma_R8_H2}
\end{figure}

For each Hamiltonian embedded in $\mathbb{R}^8$, we fix all parameters and vary only the counter-term coefficient $\gamma$. Positive values of $\gamma$ are used for $\hat{H}$ and $\hat{H}_1$, while negative values are used for $\hat{H}_2$. The optimal $\gamma$ is determined by fitting $\langle\mathrm{Tr}(W-\mathbf{1}_N)\rangle$ to a quadratic function of $\gamma$ and extrapolating to the zero crossing (Figs.~\ref{fig:TrW_1_gamma_R8_H}--\ref{fig:TrW_1_gamma_R8_H2}, left panels).

With the tuned counter-term, plaquette values converge toward the Wilson action already at $m^2 = 50$ for $H$ and $H_1$, and $m^2 = 500$ for $H_2$ at lattice spacing $a_t = a = 0.1$ (right panels); results for $a_t = a = 0.3$ are shown in Appendix~\ref{app:CT_R8_larger_spacing}. These values are one to two orders of magnitude smaller than those required without the counter-term (Section~\ref{sec:Embedding_R8}).

\subsubsection{Adding counter-term: $\mathbb{R}^4$ embedding }\label{sec:results_ZZbar_counterterm_R4}

\begin{figure}[H]
    \centering
    \begin{subfigure}{0.50\textwidth}
        \centering
        \includegraphics[width=\linewidth]{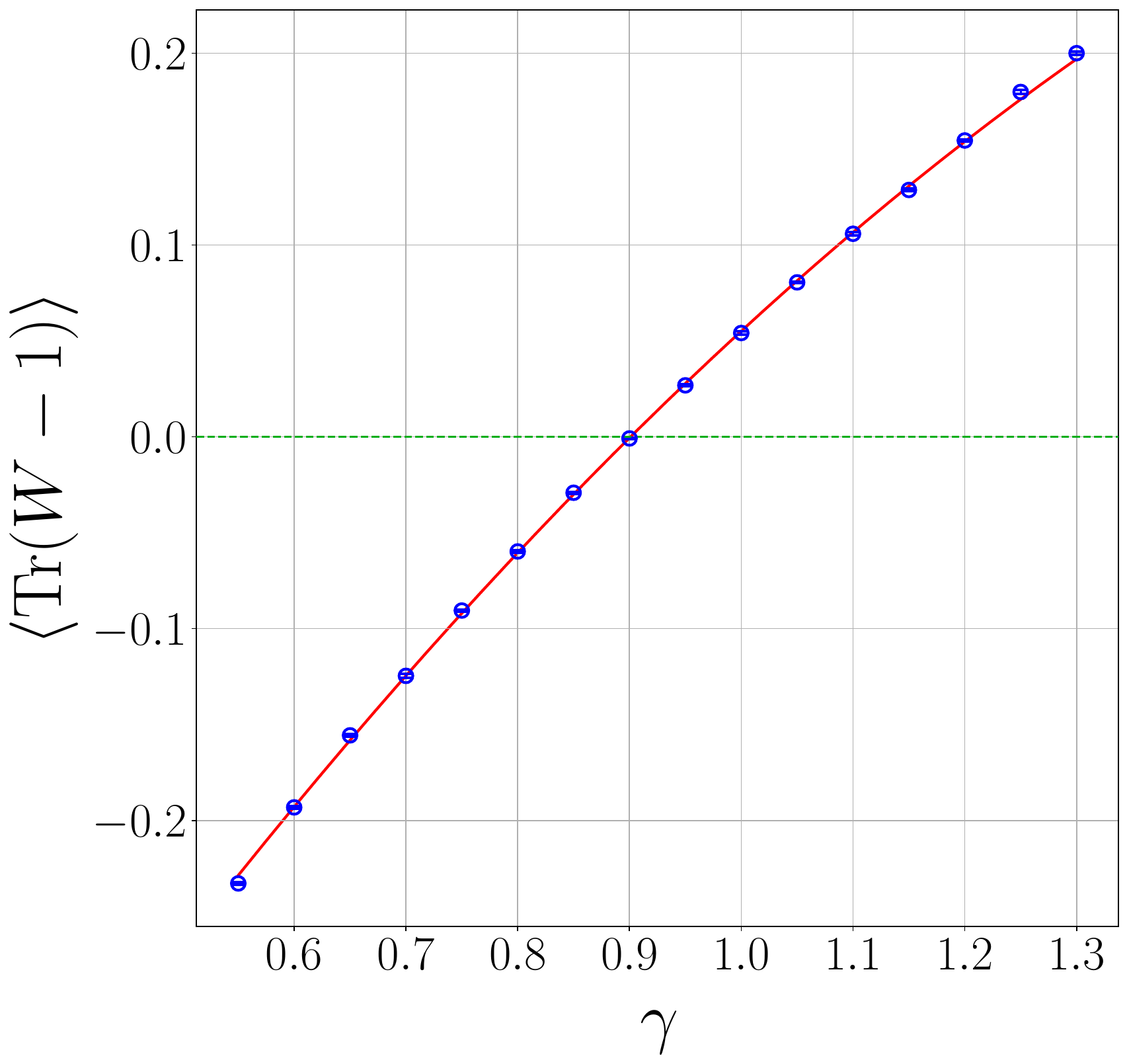}
    \end{subfigure}
    \hfill
    \begin{subfigure}{0.49\textwidth}
        \centering
    \includegraphics[width=\linewidth]{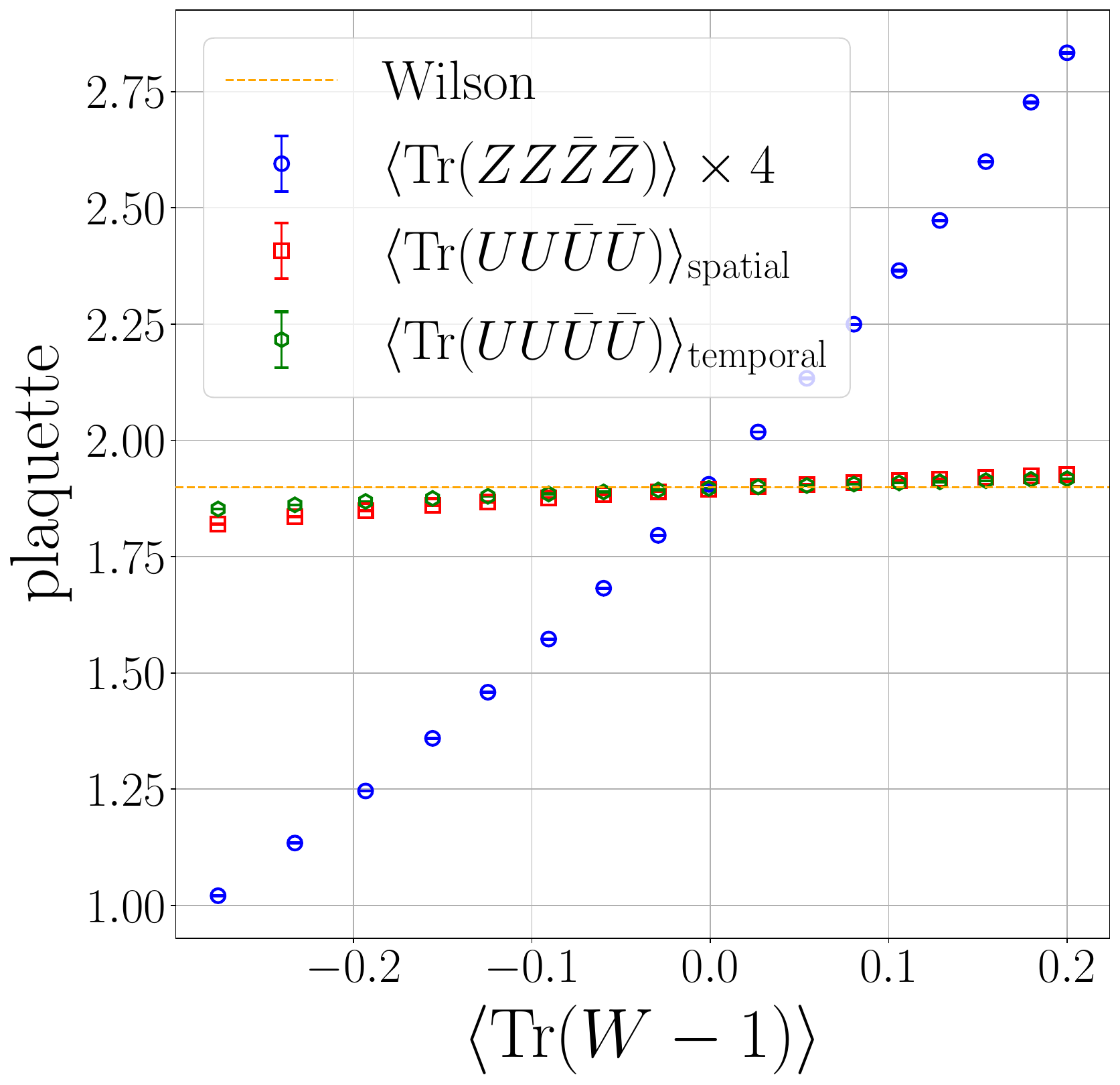}
    \end{subfigure}
    \caption{The $\hat{H}$ Hamiltonian embedded in $\mathbb{R}^4$, on $8^3$ lattice and lattice spacing $a=a_t =0.1$, with $m^2=50$. [\textbf{Left}] $\langle \mathrm{Tr}(W-\mathbf{1}_N)\rangle$ versus $\gamma$. The red line is a quadratic fit. The green dashed line indicates the target value of zero.  
[\textbf{Right}] Plaquette expectation value versus $\langle \mathrm{Tr}(W-\mathbf{1}_N)\rangle$. Symbols denote the three plaquette observables (see text for color conventions).
} \label{fig:TrW_1_gamma_R4_H}
\end{figure}

\begin{figure}[H]
    \centering
    \begin{subfigure}{0.50\textwidth}
        \centering
        \includegraphics[width=\linewidth]{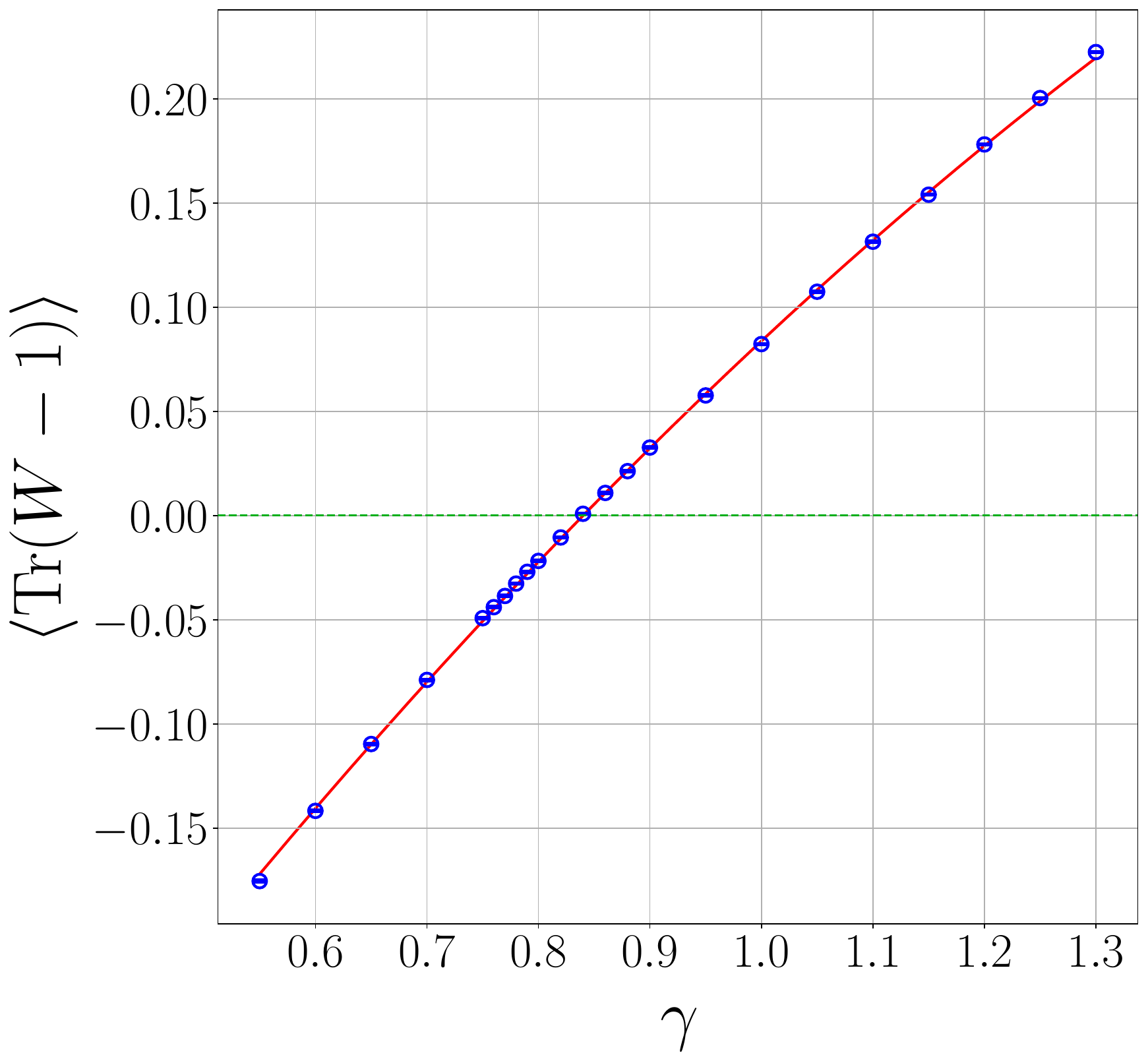}
    \end{subfigure}
    \hfill
    \begin{subfigure}{0.48\textwidth}
        \centering
    \includegraphics[width=\linewidth]{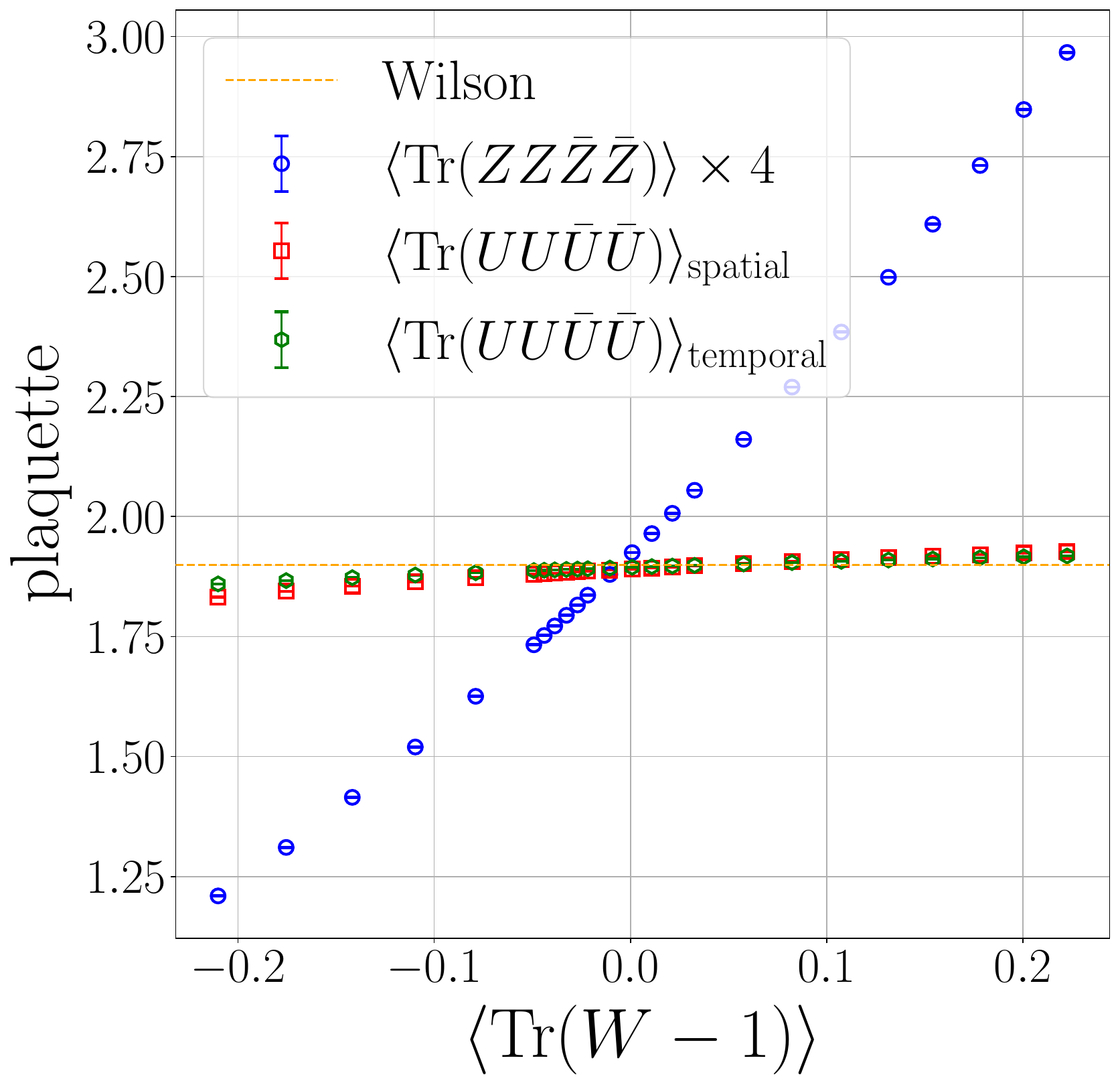}
    \end{subfigure}
    \caption{The $\hat{H}_1$ Hamiltonian embedded in $\mathbb{R}^4$, on $8^3$ and $a=a_t =0.1$, with $m^2=50$. [\textbf{Left}] $\langle \mathrm{Tr}(W-\mathbf{1}_N)\rangle$ versus $\gamma$. The red line is a quadratic fit.  The green dashed line indicates the target value of zero.  
[\textbf{Right}] Plaquette expectation value versus $\langle \mathrm{Tr}(W-\mathbf{1}_N)\rangle$. Symbols denote the three plaquette observables (see text for color conventions).
} \label{fig:TrW_1_gamma_R4_H1}
\end{figure}

\begin{figure}[H]
    \centering
    \begin{subfigure}{0.51\textwidth}
        \centering
        \includegraphics[width=\linewidth]{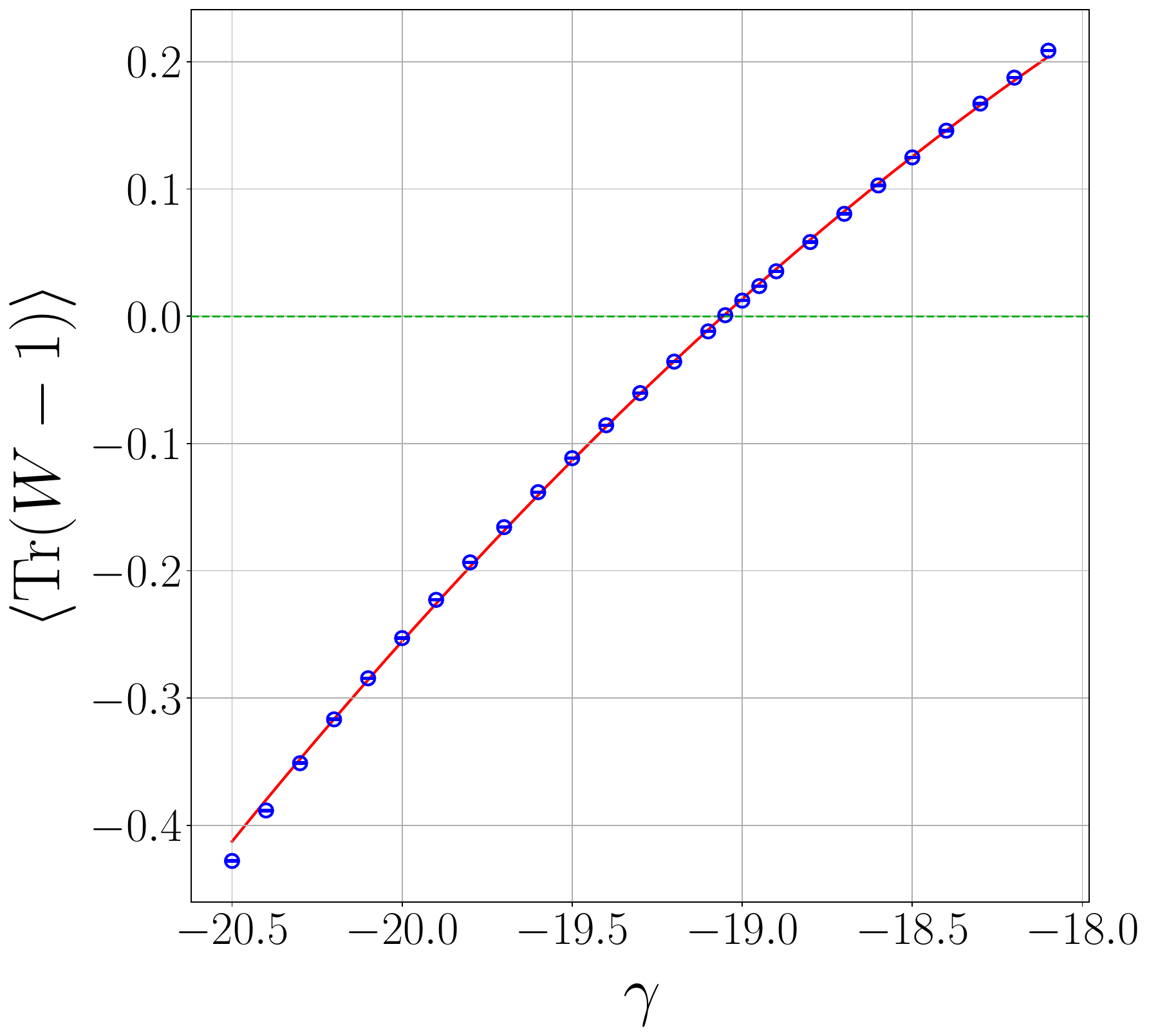}
    \end{subfigure}
    \hfill
    \begin{subfigure}{0.47\textwidth}
        \centering
    \includegraphics[width=\linewidth]{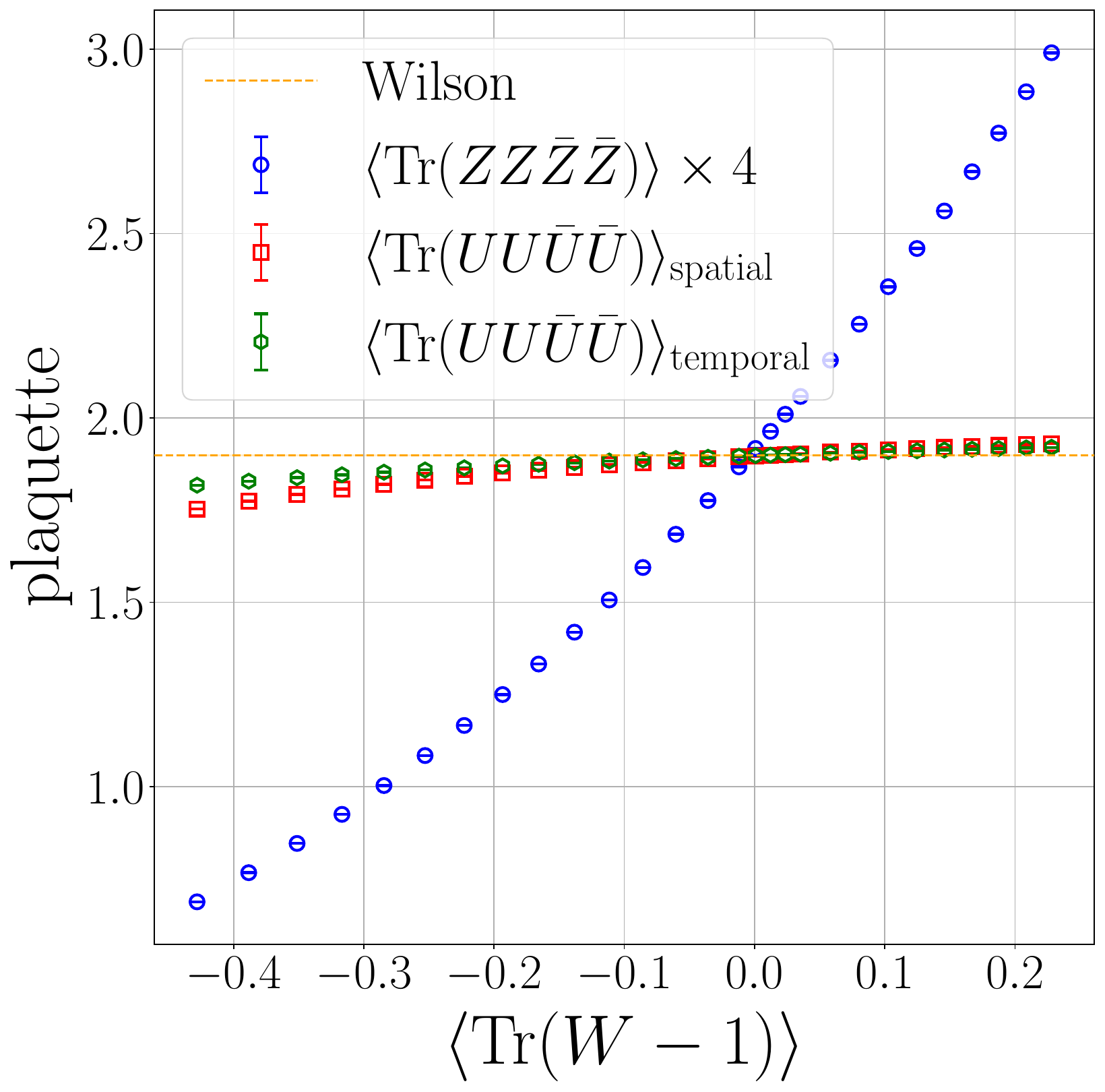}
    \end{subfigure}
    \caption{The $\hat{H}_2$ Hamiltonian embedded in $\mathbb{R}^4$, on $8^3$ and $a=a_t =0.1$, with $m^2=500$. [\textbf{Left}] $\langle \mathrm{Tr}(W-\mathbf{1}_N)\rangle$ versus $\gamma$. The red line is a quadratic fit. The green dashed line indicates the target value of zero.  
[\textbf{Right}] Plaquette expectation value versus $\langle \mathrm{Tr}(W-\mathbf{1}_N)\rangle$. Symbols denote the three plaquette observables (see text for color conventions).
} \label{fig:TrW_1_gamma_R4_H2}
\end{figure}

The same counter-term procedure applied to the $\mathbb{R}^4$ embedding yields analogous improvements (Figs.~\ref{fig:TrW_1_gamma_R4_H}--\ref{fig:TrW_1_gamma_R4_H2}). Plaquette values converge to the Wilson action at $m^2 = 50$ for $H$ and $H_1$, and $m^2 = 500$ for $H_2$, at $a_t = a = 0.1$; results for $a_t = a = 0.3$ are in Appendix~\ref{app:CT_R4_larger_spacing}. Again, the required $m^2$ is one to two orders of magnitude smaller than without the counter-term (Section~\ref{sec:Embedding_R4}).

\subsubsection{Tuning the effective lattice spacing}\label{sec:a_bare_tuning}
Numerical tuning of the bare lattice spacing  provides improved convergence to the Kogut-Susskind limit, as discussed in Section~\ref{sec:tune_bare_spacing}.
It is an alternative approach to the tuning of the linear counter-term $\gamma$.
The strategy exploits the fact that the scalar vev shifts the effective lattice spacing relative from the bare value: a Monte Carlo simulation at a chosen bare $a=a_t$ produces a measurable $c=\frac{1}{N}\langle\mathrm{Tr}\,W\rangle$, which fixes the effective spacings through $a_{\rm eff}=c^{-3}a$ and $a_{{\rm eff},t}=c^{-2}a_t$. The result at bare $(a,a_t)$ is then to be compared with the Wilson action evaluated at the matched effective spacings $(a_{\rm eff},a_{{\rm eff},t})$.

In practice, one performs a set of orbifold(-ish) runs at varying bare spacings, measures $c$ at each point, and interpolates to target a chosen Wilson lattice spacing of interest. We illustrate this procedure for the $\mathbb{R}^4$ embedding at fixed $m^2=80$ (and $m^2=150$ for $\hat{H}_2$), $\gamma=0$, varying $a=a_t$. The resulting effective spacings used to set up the matched Wilson simulations are collected in Tables~\ref{tab:a_eff_H}--\ref{tab:a_eff_H2}.

\begin{figure}[H]
    \centering
    %
    %
    \begin{subfigure}{0.32\textwidth}
        \centering
        \includegraphics[width=\linewidth]
        {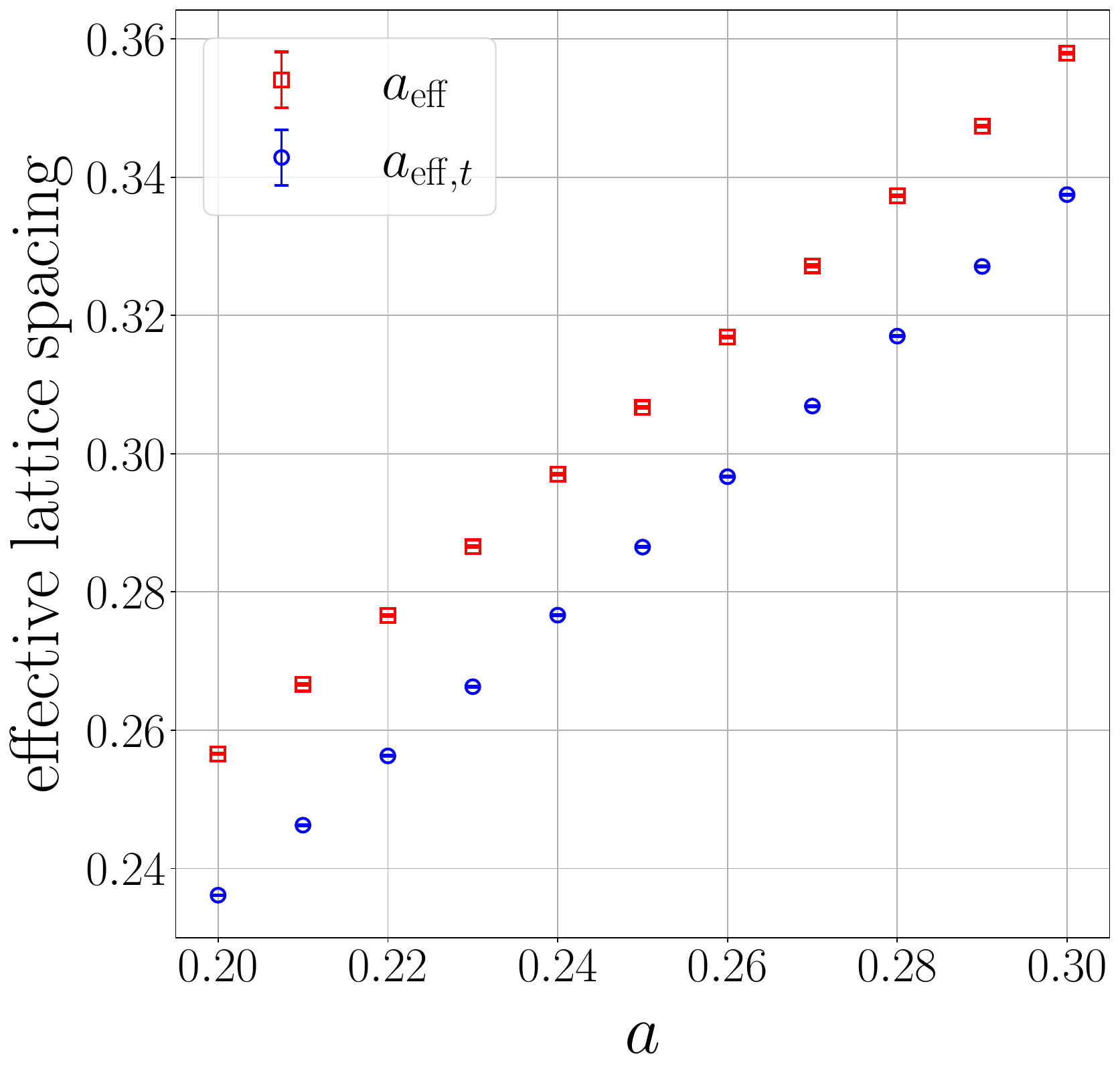}
    \end{subfigure}
    \hfill
        %
        %
    \begin{subfigure}{0.32\textwidth}
        \centering
\includegraphics[width=\linewidth]
{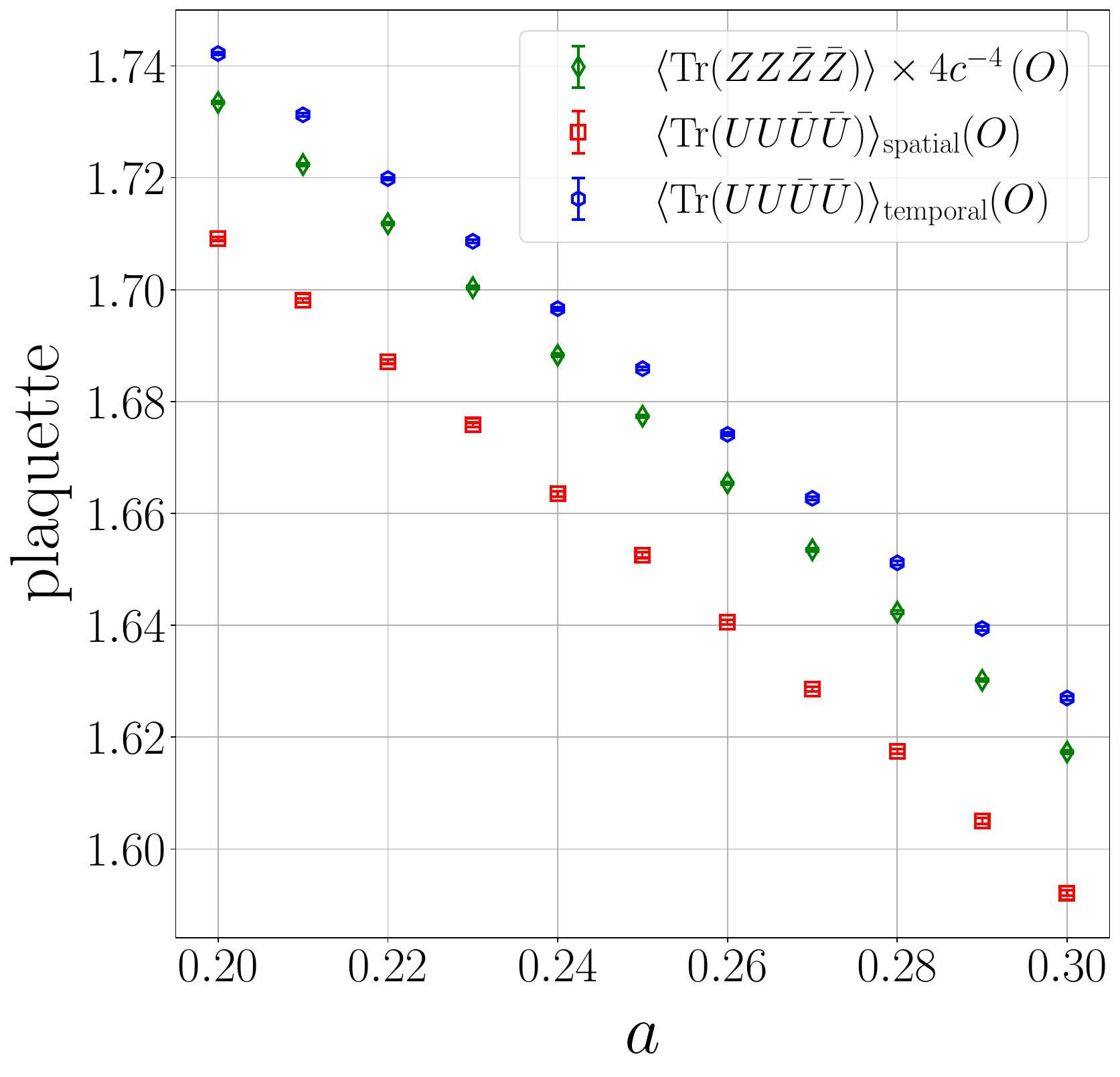}
    \end{subfigure}
    \hfill
        %
        %
    \begin{subfigure}{0.32\textwidth}
        \centering
\includegraphics[width=\linewidth]
{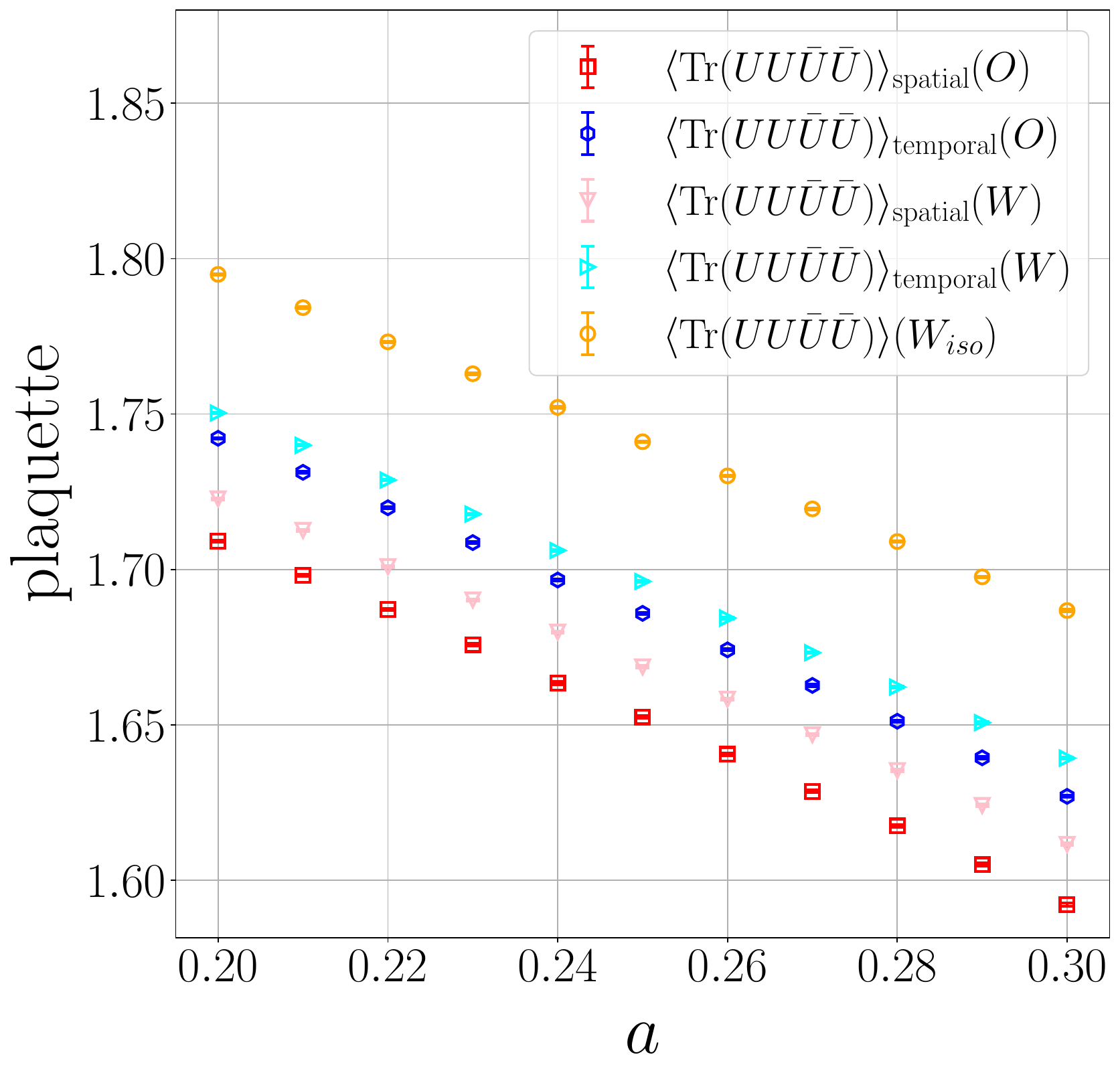}
    \end{subfigure} 
    \caption{ The original Hamiltonian $\hat{H}$, $\mathbb{R}^4$ embedding, $m^2 = 80$, $8^3$ lattice, $\gamma = 0$, with varying $a = a_t$.
    [\textbf{Left}] Effective spacing $a_{\rm eff}$ versus the lattice spacing $a$.
    [\textbf{Center}] Plaquette expectation value versus $a=a_t$. Three plaquette observables from $\hat{H}$ are shown. 
[\textbf{Right}] Plaquette expectation value versus $a=a_t$. Two plaquette observables from $\hat{H}$,    
    their counterparts from the anisotropic Wilson action with the corresponding effective lattice spacings, and the values from the isotropic Wilson action with the bare lattice spacing (for which temporal and spatial plaquettes take the same expectation value) are plotted together. (O) and (W) stands for Orbifold and Wilson, respectively.
} \label{fig:tuning_a_bare_H}
\end{figure}

\begin{figure}[H]
    \centering
    \begin{subfigure}{0.32\textwidth}
        \centering
        \includegraphics[width=1.01\linewidth]
        {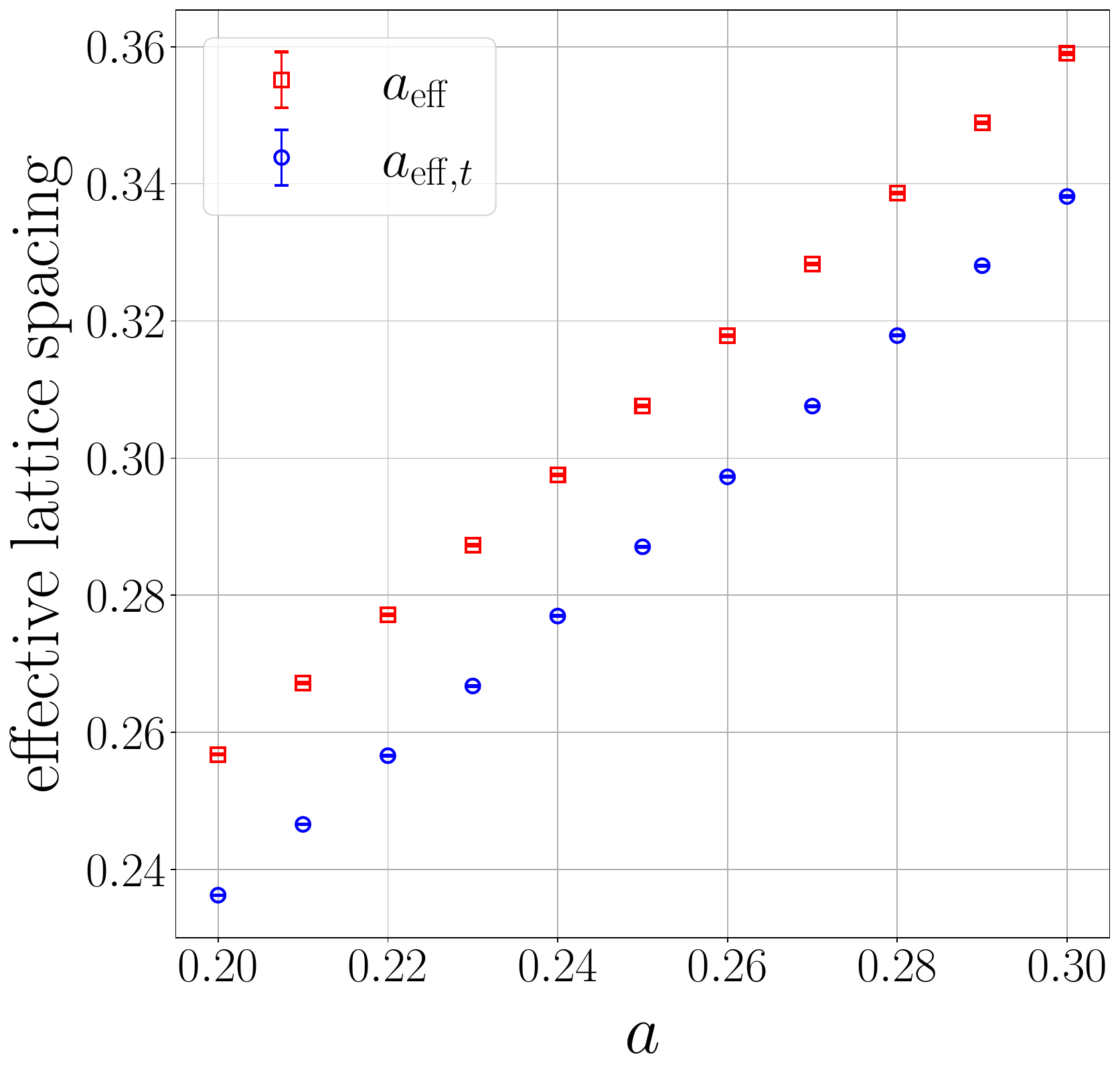}
    \end{subfigure}
    \hfill
    \begin{subfigure}{0.32\textwidth}
        \centering
\includegraphics[width=1.02\linewidth]
{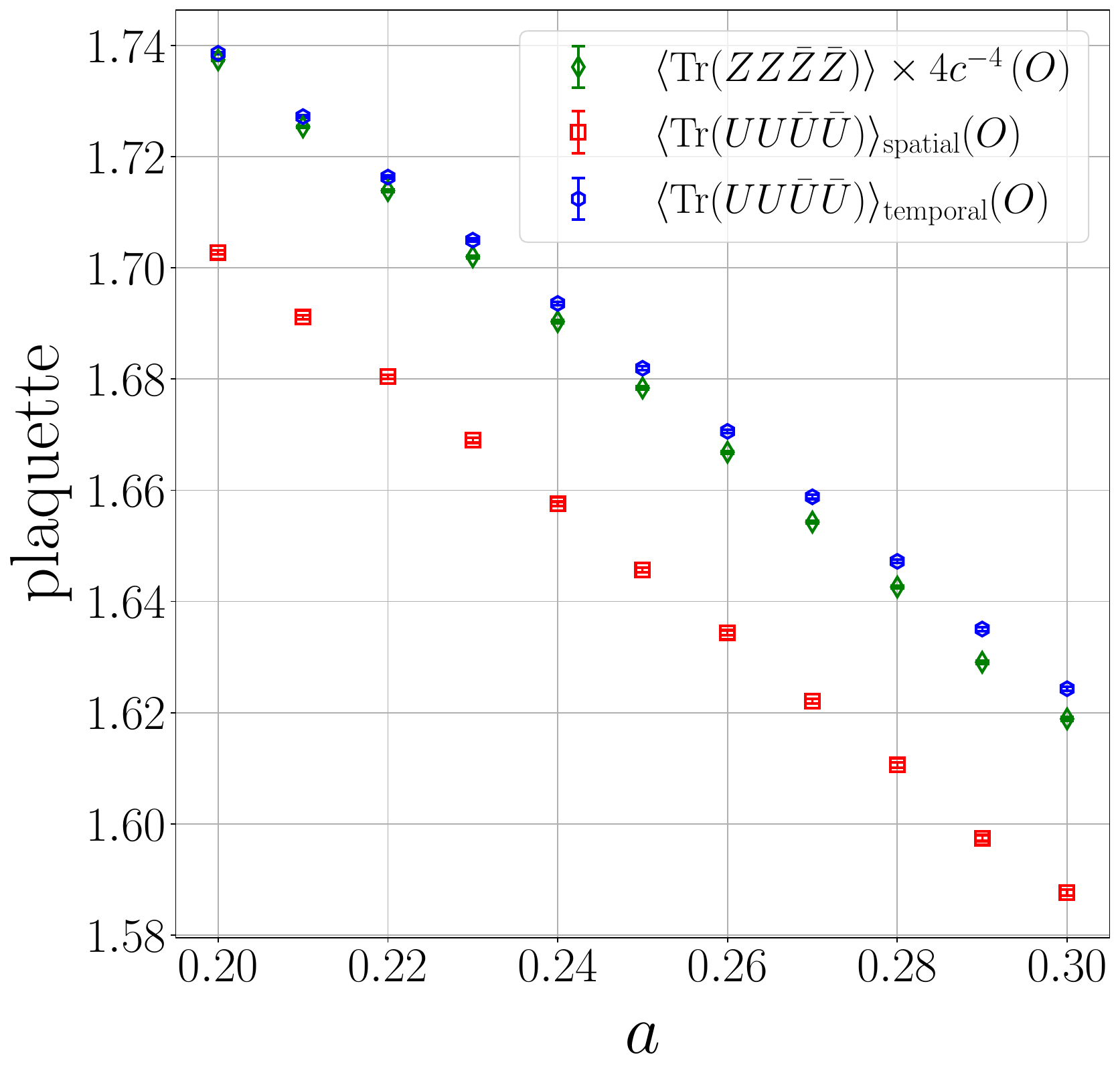}
    \end{subfigure}
    \hfill
    \begin{subfigure}{0.32\textwidth}
        \centering
\includegraphics[width=1.02\linewidth]
{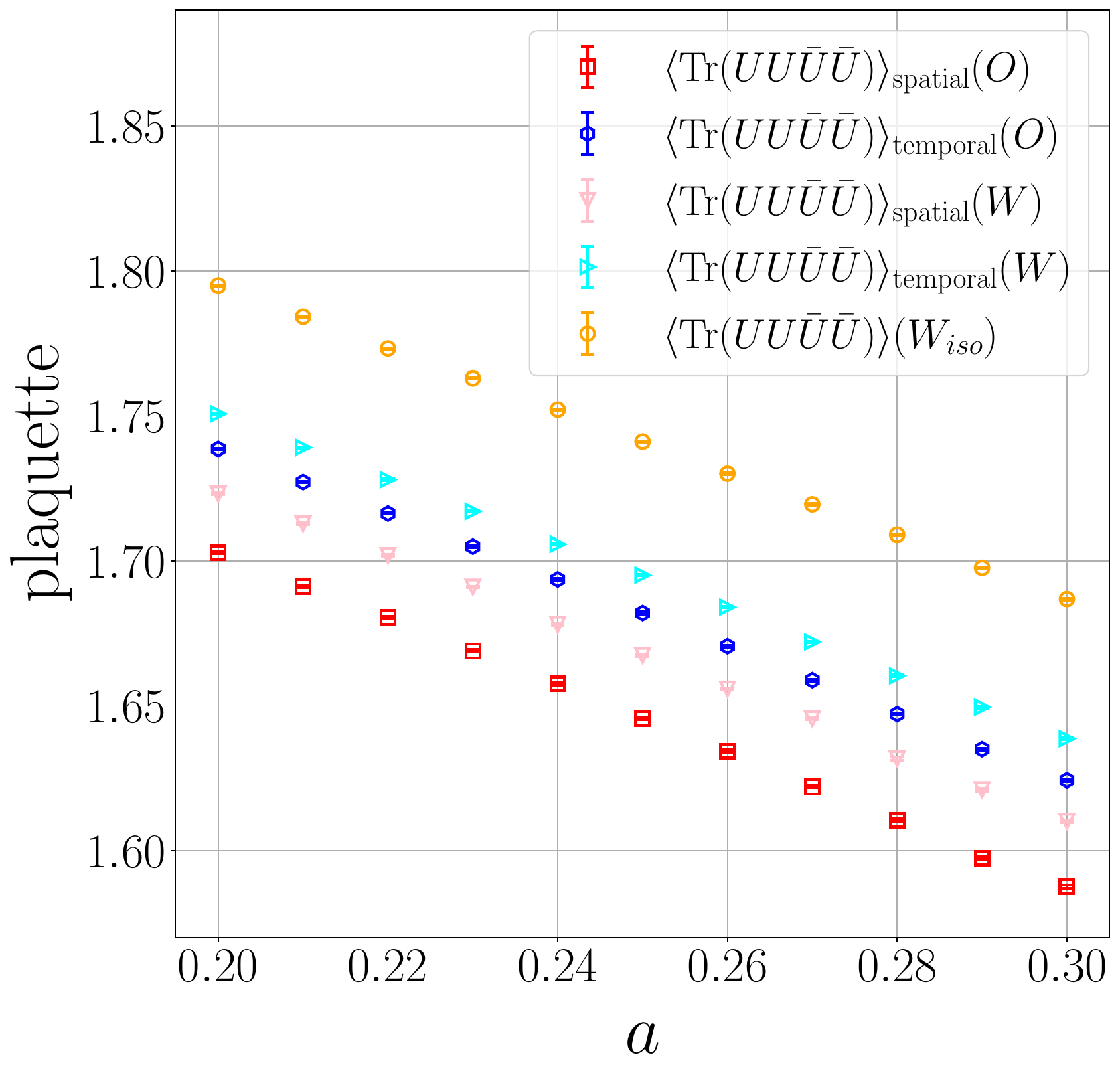}
    \end{subfigure}
    \caption{The $\hat{H}_1$ Hamiltonian, $\mathbb{R}^4$ embedding, $m^2 = 80$, $8^3$ lattice, $\gamma = 0$, with varying $a = a_t$.
    [\textbf{Center}] Plaquette expectation value versus $a=a_t$. Three plaquette observables from $\hat{H}_1$ are shown. 
[\textbf{Right}] Plaquette expectation value versus $a=a_t$. Two plaquette observables from $\hat{H}_1$,    
    their counterparts from the anisotropic Wilson action with the corresponding effective lattice spacings, and the values from the isotropic Wilson action with the bare lattice spacing (for which temporal and spatial plaquettes take the same expectation value) are plotted together. (O) and (W) stands for Orbifold-ish and Wilson, respectively.
}\label{fig:tuning_a_bare_H1}
\end{figure}

\begin{figure}[H]
    \centering
    \begin{subfigure}{0.32\textwidth}
        \centering
        \includegraphics[width=1.02\linewidth]{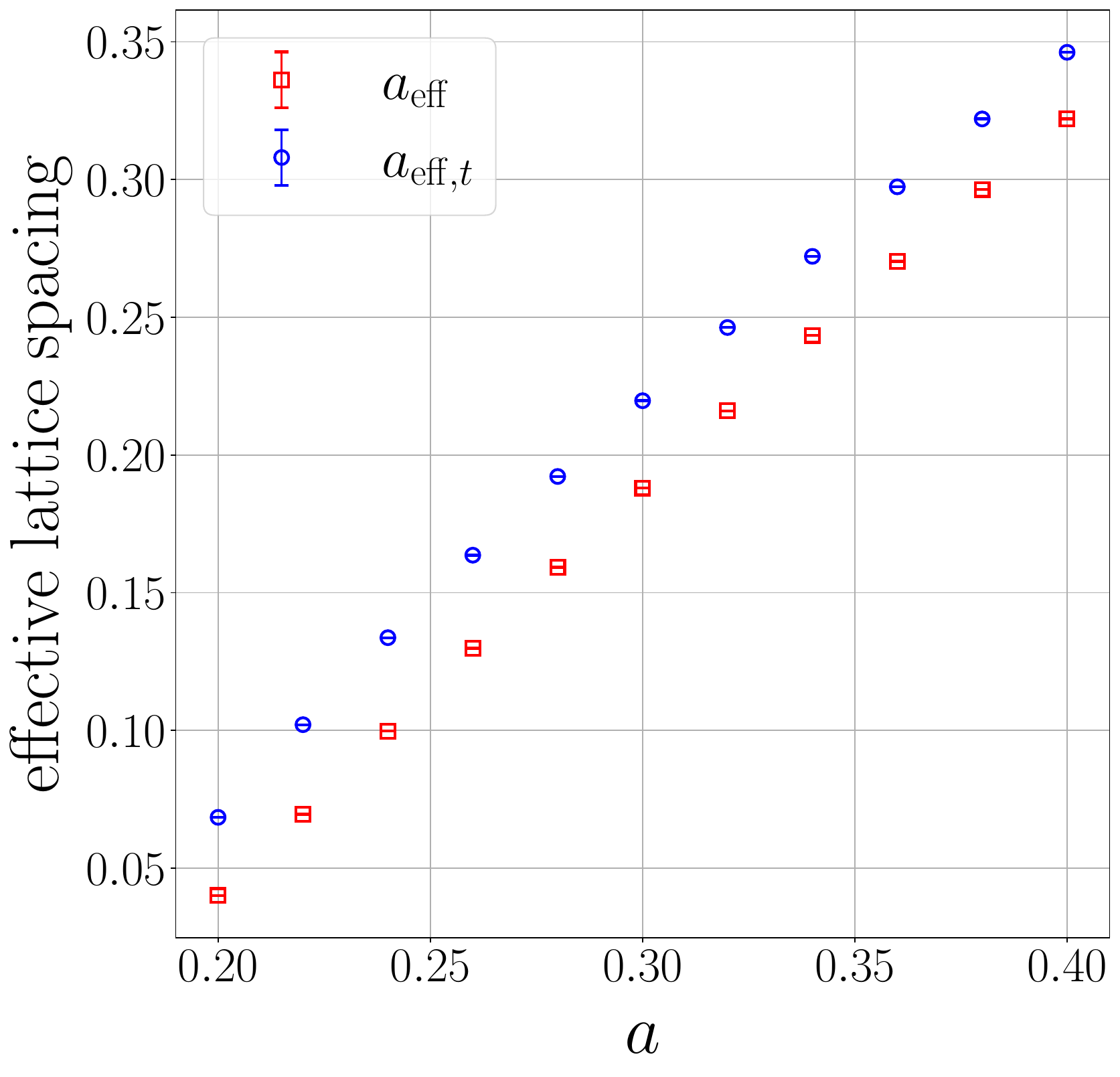}
    \end{subfigure}
    \hfill
    \begin{subfigure}{0.32\textwidth}
        \centering
\includegraphics[width=1.02\linewidth]{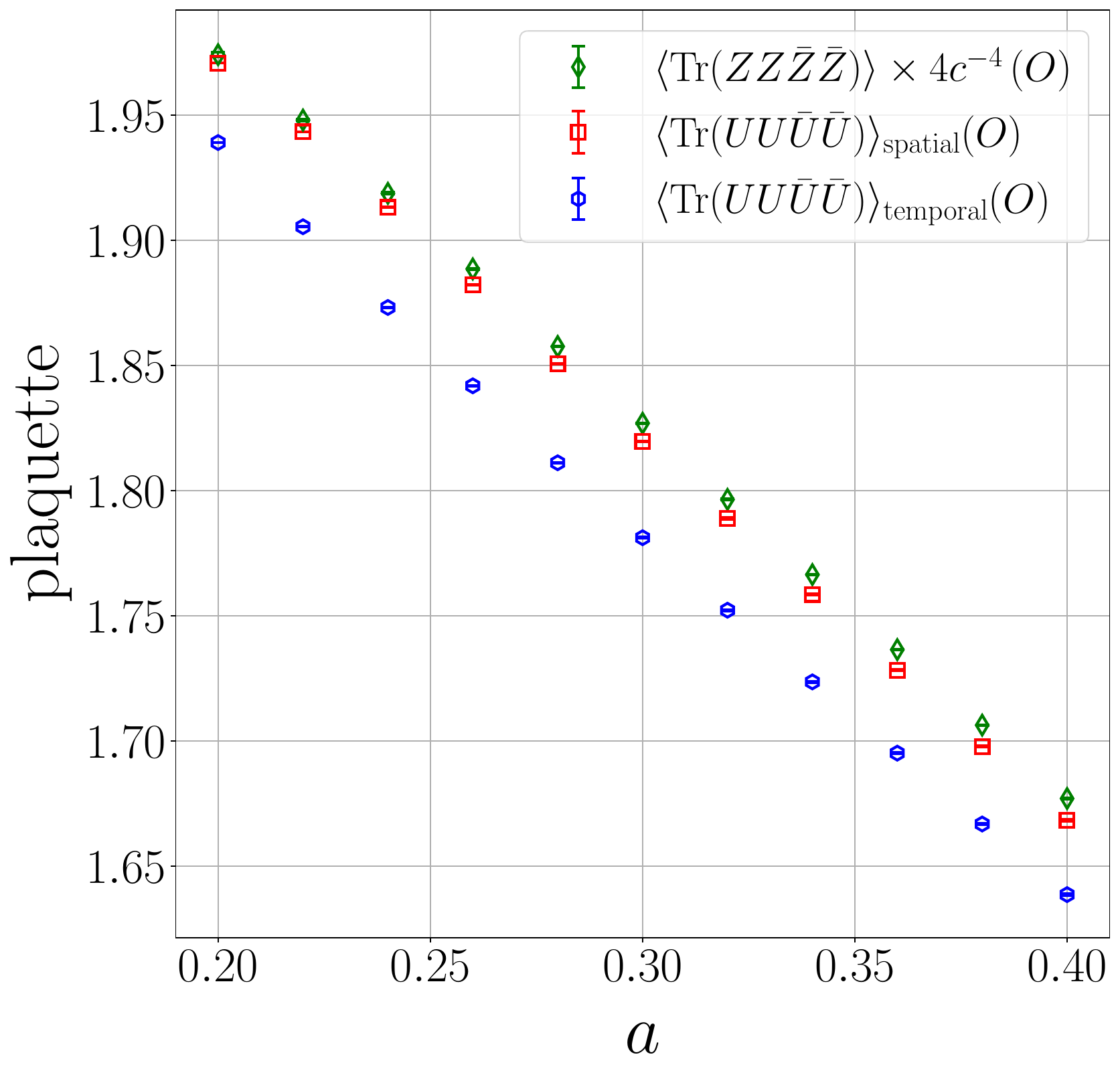}
    \end{subfigure}
    \hfill
        \begin{subfigure}{0.32\textwidth}
        \centering
\includegraphics[width=\linewidth]{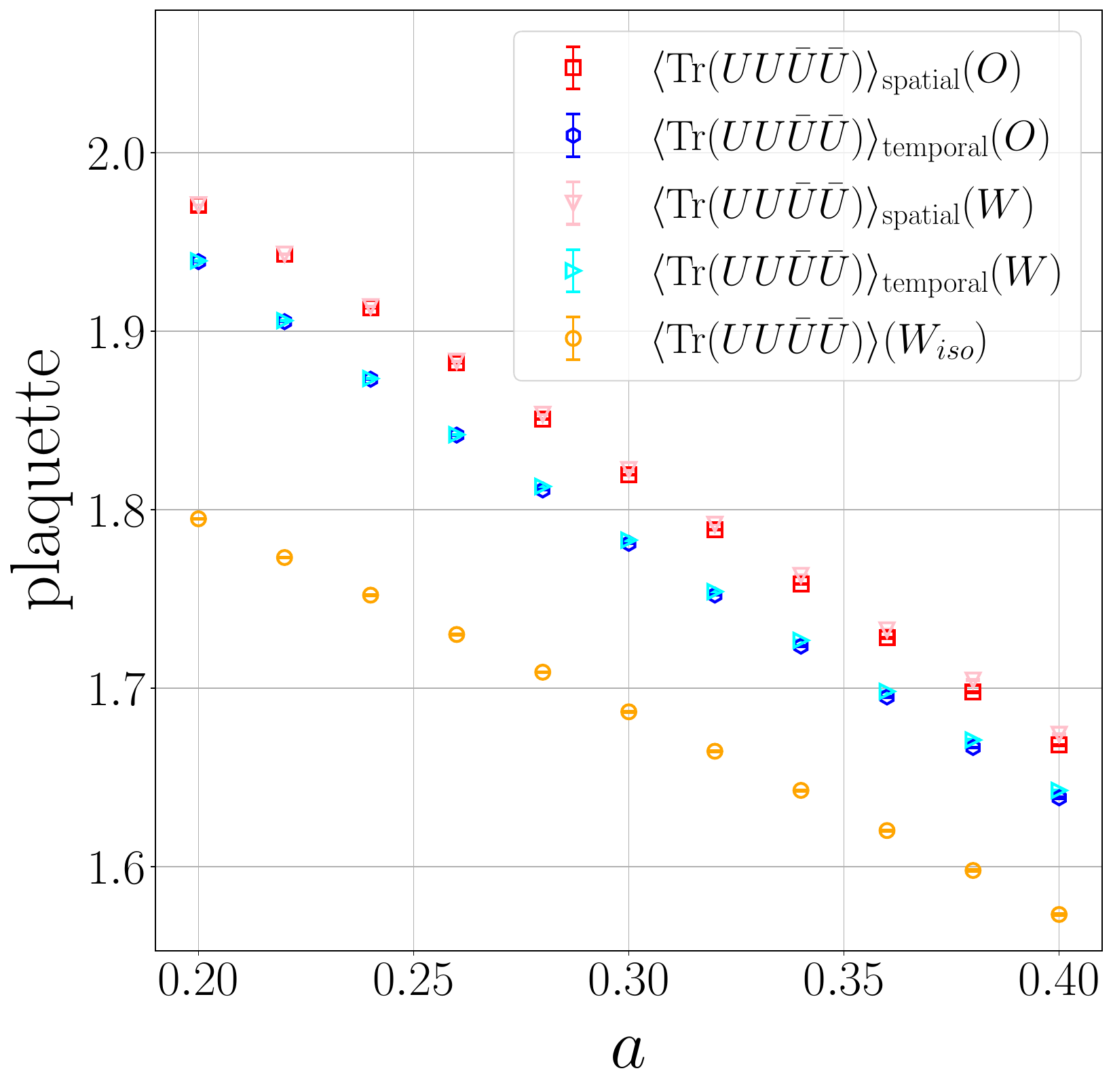}
    \end{subfigure}
    \caption{The $\hat{H}_2$ Hamiltonian, $\mathbb{R}^4$ embedding, $m^2 = 150$, $8^3$ lattice, $\gamma = 0$, with varying $a = a_t$.  
[\textbf{Left}] Effective spacing $a_{\rm eff}$ versus the lattice spacing $a$.
[\textbf{Center}] Plaquette expectation value versus $a=a_t$. Three plaquette observables from $\hat{H}_2$ are shown. 
[\textbf{Right}] Plaquette expectation value versus $a=a_t$. Two plaquette observables from $\hat{H}_2$,    
    their counterparts from the anisotropic Wilson action with the corresponding effective lattice spacings, and the values from the isotropic Wilson action with the bare lattice spacing (for which temporal and spatial plaquettes take the same expectation value) are plotted together. (O) and (W) stands for Orbifold-ish and Wilson, respectively.
}\label{fig:tuning_a_bare_H2}
\end{figure}

On the left panels of Figs.~\ref{fig:tuning_a_bare_H}--\ref{fig:tuning_a_bare_H2}, $a_{\rm eff}$ and $a_{{\rm eff},t}$ are shown against the bare values, $a=a_t$.

On the center panels, spatial and temporal plaquettes from the orbifold lattice are plotted. For $\langle\mathrm{Tr}(ZZ\bar{Z}\bar{Z})\rangle$, we multiplied by $4c^{-4}$ to adjust the normalization, taking into account the shift of the lattice spacing. If the scalar fluctuations were completely suppressed (including zero-point fluctuations) by a large mass, $\langle\mathrm{Tr}(ZZ\bar{Z}\bar{Z})\rangle\times 4c^{-4}$ would agree with $\langle\mathrm{Tr}(UU\bar{U}\bar{U})\rangle_{\rm spatial}$ exactly. For our purpose this is not required: it is enough that the scalar field does not affect the dynamics of the gauge field. The fact that $\langle\mathrm{Tr}(ZZ\bar{Z}\bar{Z})\rangle\times 4c^{-4}$ comes close to $\langle\mathrm{Tr}(UU\bar{U}\bar{U})\rangle_{\rm spatial}$ in Figs.~\ref{fig:tuning_a_bare_H}--\ref{fig:tuning_a_bare_H2} is already a good sign. A more meaningful test is whether $\langle\mathrm{Tr}(UU\bar{U}\bar{U})\rangle_{\rm spatial}$ and $\langle\mathrm{Tr}(UU\bar{U}\bar{U})\rangle_{\rm temporal}$ approach the counterparts of the Wilson action. Their convergence toward the Wilson limit can be further improved, if necessary, by slightly increasing the $m^2$ value, as shown in Appendix~\ref{app:CT_R4_larger_spacing}.

For $H$ and $H_1$, we have $a<a_{{\rm eff},t}<a_{\rm eff}$. In this setup, $\langle\mathrm{Tr}(UU\bar{U}\bar{U})\rangle_{\rm spatial}$ should be smaller than $\langle\mathrm{Tr}(UU\bar{U}\bar{U})\rangle_{\rm temporal}$, and we do see that pattern in Fig.~\ref{fig:tuning_a_bare_H} and Fig.~\ref{fig:tuning_a_bare_H1}. For $H_2$, on the other hand, we have $a>a_{{\rm eff},t}>a_{\rm eff}$. Then $\langle\mathrm{Tr}(UU\bar{U}\bar{U})\rangle_{\rm spatial}$ should be larger than $\langle\mathrm{Tr}(UU\bar{U}\bar{U})\rangle_{\rm temporal}$, and we do see that from Fig.~\ref{fig:tuning_a_bare_H2}.

On the right panels, $\langle\mathrm{Tr}UU\bar{U}\bar{U}\rangle$ computed from the orbifold lattice, anisotropic Wilson action with corresponding effective lattice spacings, and isotropic Wilson action with the bare lattice spacings are plotted. 
We can see that, by using the effective lattice spacings, the convergence to the Wilson action is significantly improved. Note that, for $H_2$, the effective lattice spacings can be far away from the bare values -- $a_{\rm eff}=0.040$ and $a_{{\rm eff},t}=0.068$ for $a=a_t=0.200$ -- and we still find the values close to those from the Wilson action. 

The convergence toward the Wilson limit can be further improved, if necessary, by slightly increasing the $m^2$ value, as shown in Appendix~\ref{app:CT_R4_larger_spacing}.

\begin{table}[hbtp]
\centering

\begin{subtable}{0.3\textwidth}
\centering
\begin{tabular}{|c||c|c|}
\hline
$a=a_t$ & $a_{{\rm eff}}$ & $a_{{\rm eff},t}$ \\ \hline\hline
0.200 & 0.257 & 0.236 \\
0.210 & 0.267 & 0.246 \\
0.220 & 0.277 & 0.256 \\
0.230 & 0.287 & 0.266 \\
0.240 & 0.297 & 0.277 \\
0.250 & 0.307 & 0.286 \\
0.260 & 0.317 & 0.297 \\
0.270 & 0.327 & 0.307 \\
0.280 & 0.337 & 0.317 \\
0.290 & 0.347 & 0.327 \\
0.300 & 0.358 & 0.337 \\
\hline
\end{tabular}
\caption{}
\label{tab:a_eff_H}
\end{subtable}
\hfill
\begin{subtable}{0.3\textwidth}
\centering
\begin{tabular}{|c||c|c|}
\hline
$a=a_t$ & $a_{\rm eff}$ & $a_{{\rm eff},t}$ \\ \hline\hline
0.200 & 0.257 & 0.236 \\
0.210 & 0.267 & 0.247 \\
0.220 & 0.277 & 0.257 \\
0.230 & 0.287 & 0.267 \\
0.240 & 0.298 & 0.277 \\
0.250 & 0.308 & 0.287 \\
0.260 & 0.318 & 0.297 \\
0.270 & 0.328 & 0.308 \\
0.280 & 0.339 & 0.318 \\
0.290 & 0.349 & 0.328 \\
0.300 & 0.359 & 0.338 \\ \hline  
\end{tabular}
\caption{}
\label{tab:a_eff_H1}
\end{subtable}
\hfill
\begin{subtable}{0.3\textwidth}
\centering
\begin{tabular}{|c||c|c|}
\hline
$a=a_t$ & $a_{\rm eff}$ & $a_{{\rm eff},t}$ \\ \hline\hline
0.200 & 0.040 & 0.068 \\
0.220 & 0.070 & 0.102 \\
0.240 & 0.100 & 0.134 \\
0.260 & 0.130 & 0.164 \\
0.280 & 0.159 & 0.192 \\
0.300 & 0.188 & 0.220 \\
0.320 & 0.216 & 0.246 \\
0.340 & 0.243 & 0.272 \\
0.360 & 0.270 & 0.297 \\
0.380 & 0.296 & 0.322 \\
0.400 & 0.322 & 0.346 \\
 \hline
\end{tabular}
\caption{}
\label{tab:a_eff_H2}
\end{subtable}

\caption{ Parameter choice for Wilson simulations for Fig.~\ref{fig:tuning_a_bare_H} (left), Fig.~\ref{fig:tuning_a_bare_H1} (center), and Fig.~\ref{fig:tuning_a_bare_H2} (right).
}
\end{table}\label{tab:a_eff_H_H1_m80_H2_m150}

\section{Conclusions and future directions}
\label{sec:conclusions_f_directions}
Quantum computers hold significant potential for solving computationally intensive problems, but realizing this promise depends on the development of scalable frameworks for quantum simulations in addition to hardware development. These frameworks are essential for encoding theories into the available quantum hardware and enabling large-scale computations challenging for classical computers. To date, the focus within the quantum computing community has largely been on exploring and learning from the capabilities of NISQ hardware, ignoring the scalability of fault-tolerant devices. However, to achieve quantum advantage in physically meaningful parameter regimes, scalable frameworks are mandatory.

In this work, we introduced several key improvements over previous formulations of the orbifold lattice Hamiltonian to further simplify the quantum simulation framework. First, we proposed two simplified Hamiltonians, $H_1$ and $H_2$. These Hamiltonians are obtained from the original orbifold Hamiltonian $H$ by systematically eliminating terms that become negligible in the Kogut-Susskind (KS) limit. As far as the KS limit is concerned, this leads to a significant reduction of gate counting. The KS limit of both $H_1$ and $H_2$ was studied using Monte Carlo simulations, and smooth convergence toward Wilson-action values for all considered observables was confirmed. This convergence was observed as the scalar mass $m^2$ was increased, even at the regularized level, and extrapolation to the infinite-mass limit confirmed quantitative agreement with the Wilson formulation. 

In addition, we introduced a novel encoding of SU(2) link variable into $\mathbb{R}^4$ that is analogous to the orbifold lattice. This halves the number of scalar degrees of freedom per link compared to the original encoding, $\mathrm{SU}(2)\subset\mathbb{R}^8$. This reduction directly translates into a halving of the required qubit count and a substantial decrease in circuit depth. Monte Carlo simulations of the Hamiltonians $H$, $H_1$, and $H_2$ in the $\mathbb{R}^4$ encoding demonstrated smooth convergence of action-related observables toward their Wilson counterparts in the KS limit, with no pathological behavior observed. These results identify the $\mathbb{R}^4$ formulations as particularly promising targets for near-term and future quantum simulations.\\

To address the practical challenge of requiring very large scalar masses $m^2$ in quantum simulations, we introduced two complementary strategies. The key idea behind these strategies is that the significant mass dependence is due to the shift of the effective lattice spacing, and hence, the mass dependence should be much milder if the lattice spacing is under control. The first strategy incorporates a linear counter-term, effectively lowering the required bare mass $m^2$. The second strategy involves tuning the bare lattice spacing in such a way that the effective lattice spacing improves agreement between the orbifold and Wilson formulations without requiring large values of $m^2$. Both approaches were benchmarked via Monte Carlo simulations and shown to achieve agreement with the Wilson action at the regularized level, employing scalar masses up to two orders of magnitude smaller than those required in simulations without the additional terms.

We have discussed also implementation details for the different Hamiltonians derived in this work in Section~\ref{sec:minimal_Hamiltonians}. The advantage of the reduced number of terms in $H_1$ and $H_2$ leads to a lower gate count and circuit depth. They are hence very interesting if these are the limiting factors of the implementation. The complete Hamiltonian $H$, on the other hand, contains the standard scalar kinetic and interaction terms of a Yang-Mills theory coupled to a scalar field.

A natural extension of this work is the formulation and study of the orbifold lattice in $3+1$ dimensions, including a detailed analysis of convergence to the KS limit and the explicit construction of quantum circuits. Another important direction is the extension of the optimized Hamiltonians $H_1$ and $H_2$ to the SU(3) gauge group. While smooth convergence to the KS limit is expected, it will be valuable to investigate how the approach to the KS limit scales with $1/m^2$ in this case.\\

As a preliminary step toward large-scale simulations, it will be important to develop and test quantum simulation protocols for toy models, ranging from simple two-link systems to plaquettes arranged in square and cubic geometries, before advancing to full $(2+1)$- and $(3+1)$-dimensional lattices. The near-term goals would be related to studying dynamical properties of Yang-Mills theories by implementing real-time evolution using orbifold lattice quantum circuits on quantum hardware, with possible mid-size system validation provided by tensor network methods. Minor tweaks on the simulation protocols may reduce the cost of quantum simulations further. For example, we can use the gauge degrees of freedom to take the wave function localized about $W=\textbf{1}$ and $U=\textbf{1}$~\cite{Hanada:2025goy}. Toward the continuum limit, the wave function becomes more and more peaked, so that we can save the number of qubits by covering only the vicinity of this point in the position basis.  

\vspace{8 pt}
\begin{center}
\section*{Acknowledgment}
\end{center}
We thank Johann Ostmeyer for helpful discussions and Enrico Rinaldi for collaboration in the early stages of this work.

M.~H. thanks the STFC for the support through the consolidated grant ST/Z001072/1.  E.~M. was supported by UK Research and Innovation Future Leader Fellowship {MR/X015157/1}. The numerical simulations were undertaken on Barkla High Performance Computing facilities at the University of Liverpool.\\

\vspace{1 cm}
\appendix

\section{Additional numerical results at different lattice sizes and spacings}\label{app:R8_larger_lattice}

In this appendix, we collect additional numerical results obtained at different lattice sizes and lattice spacings. These simulations provide further support for the results and conclusions presented in the main text.

\subsection{Simulations of the embedding of SU(2) into $\mathbb{R}^8$ on a larger lattice size}\label{app:R8A_larger_lattice}
In this appendix, we present all the numerical simulations conducted for the embedding of $SU(2)$ into $\mathbb{R}^8$ on a larger lattice size. While it is remarkable that numerical results in Section~\ref{sec:Embedding_R8} using $8^3$ lattice size demonstrate a smooth convergence of the Hamiltonians $\hat{H}$, $\hat{H}_1$ and $\hat{H}_2$ to the KS value already at the regularized level, and for large value of $m^2$, here we provide additional numerical results for larger lattice size to address any potential concern arising from the use of smaller lattice sizes.\\
The following plots show the results of Monte Carlo simulations for $16^3$ lattice with two lattices spacing $a_t = a = 0.1$ and $0.3$. The observables show a similar convergence to what was shown in Section~\ref{sec:Embedding_R8} for $8^3$, confirming the absence of lattice artifacts from smaller lattice sizes when extracting the KS limit.

\begin{figure}[H]
        \centering    {$a=a_t=0.1$}
        \vspace{0.2 cm}
        
    \centering
    \begin{subfigure}{0.22\textwidth}
        \centering
        \includegraphics[width=\linewidth]{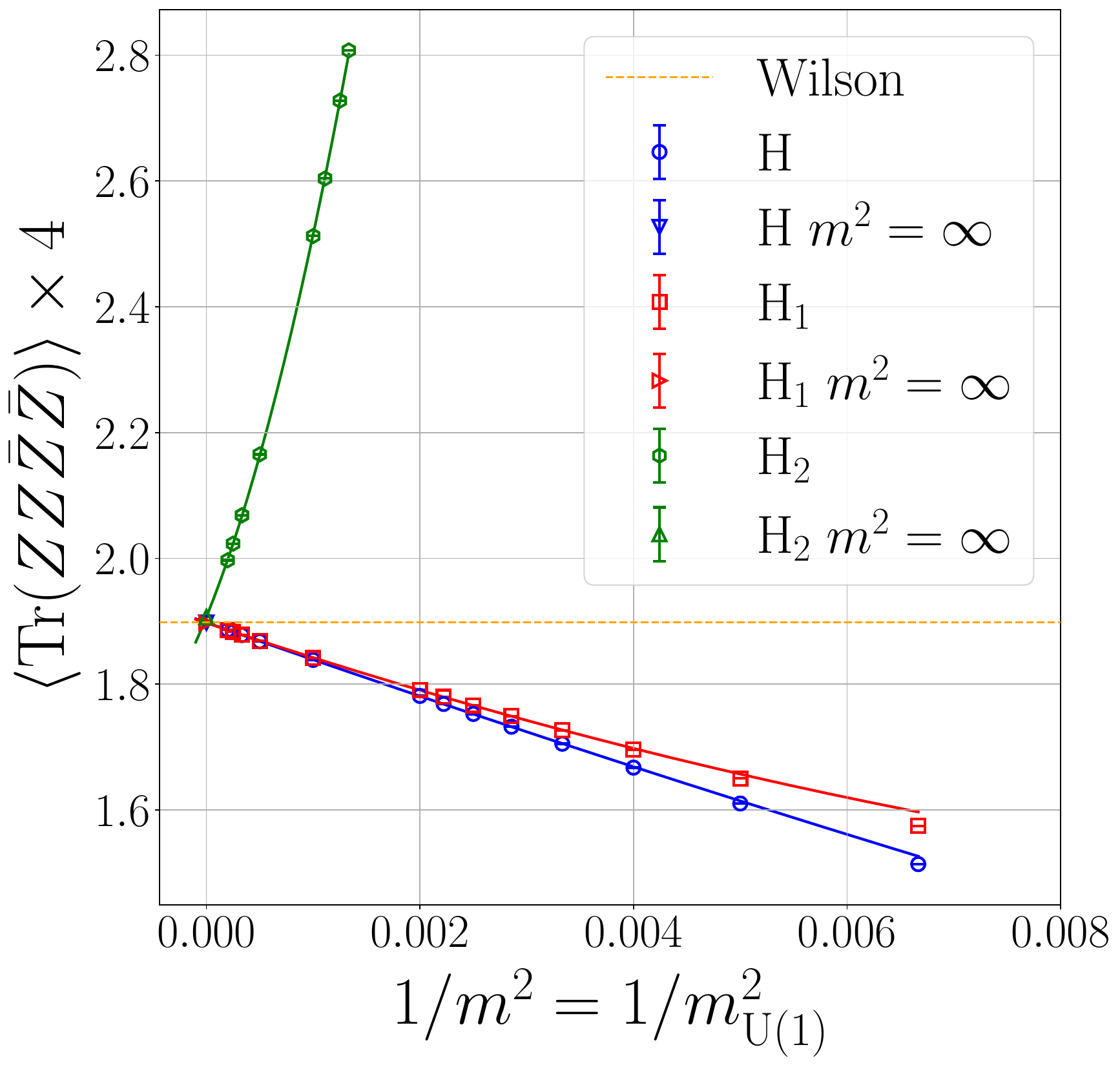}
    \end{subfigure}
    \hfill
        \begin{subfigure}{0.22\textwidth}
        \centering
        \includegraphics[width=1.02\linewidth]{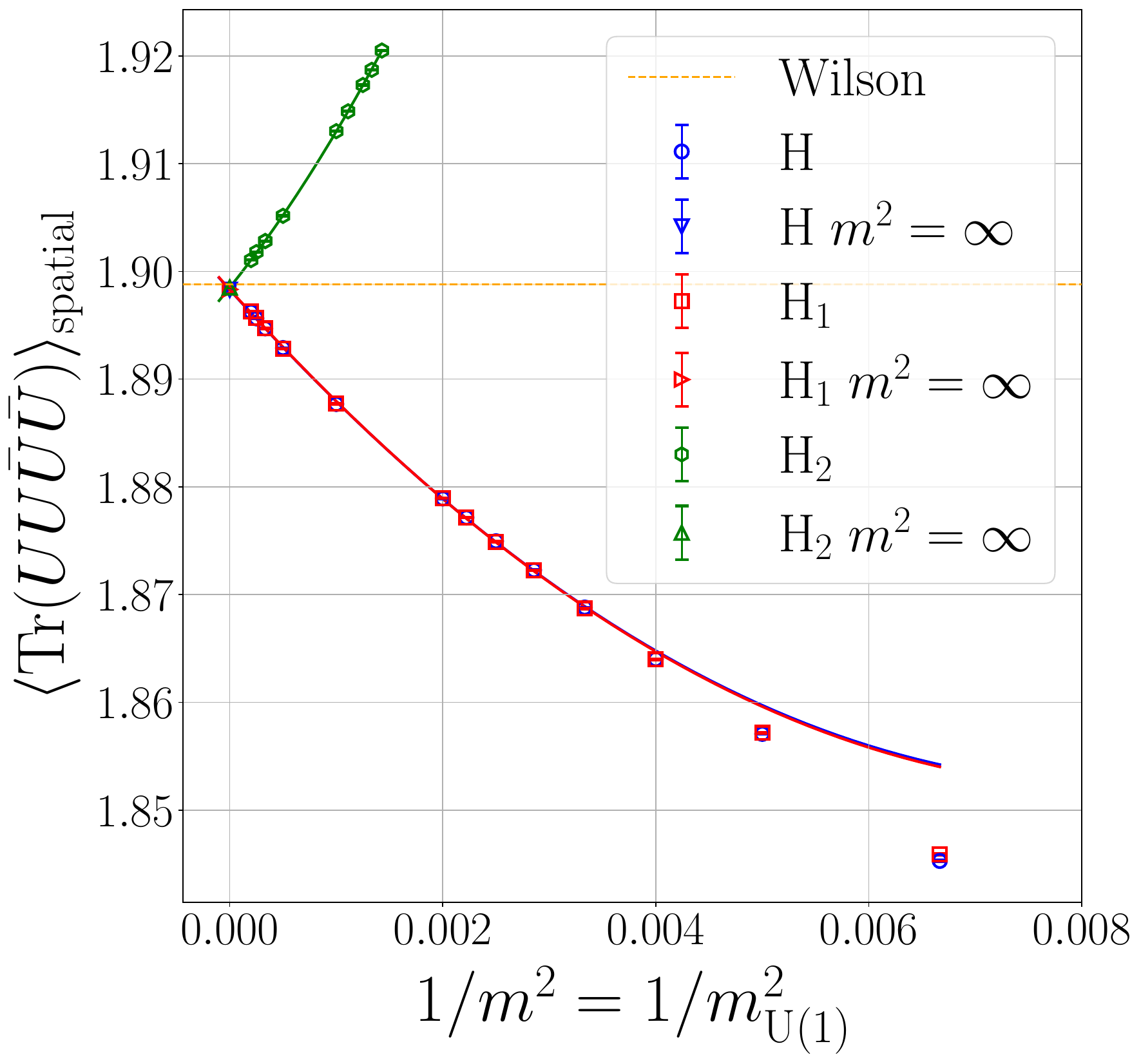}
    \end{subfigure}
    \hfill
        \begin{subfigure}{0.22\textwidth}
        \centering
        \includegraphics[width=1.02\linewidth]{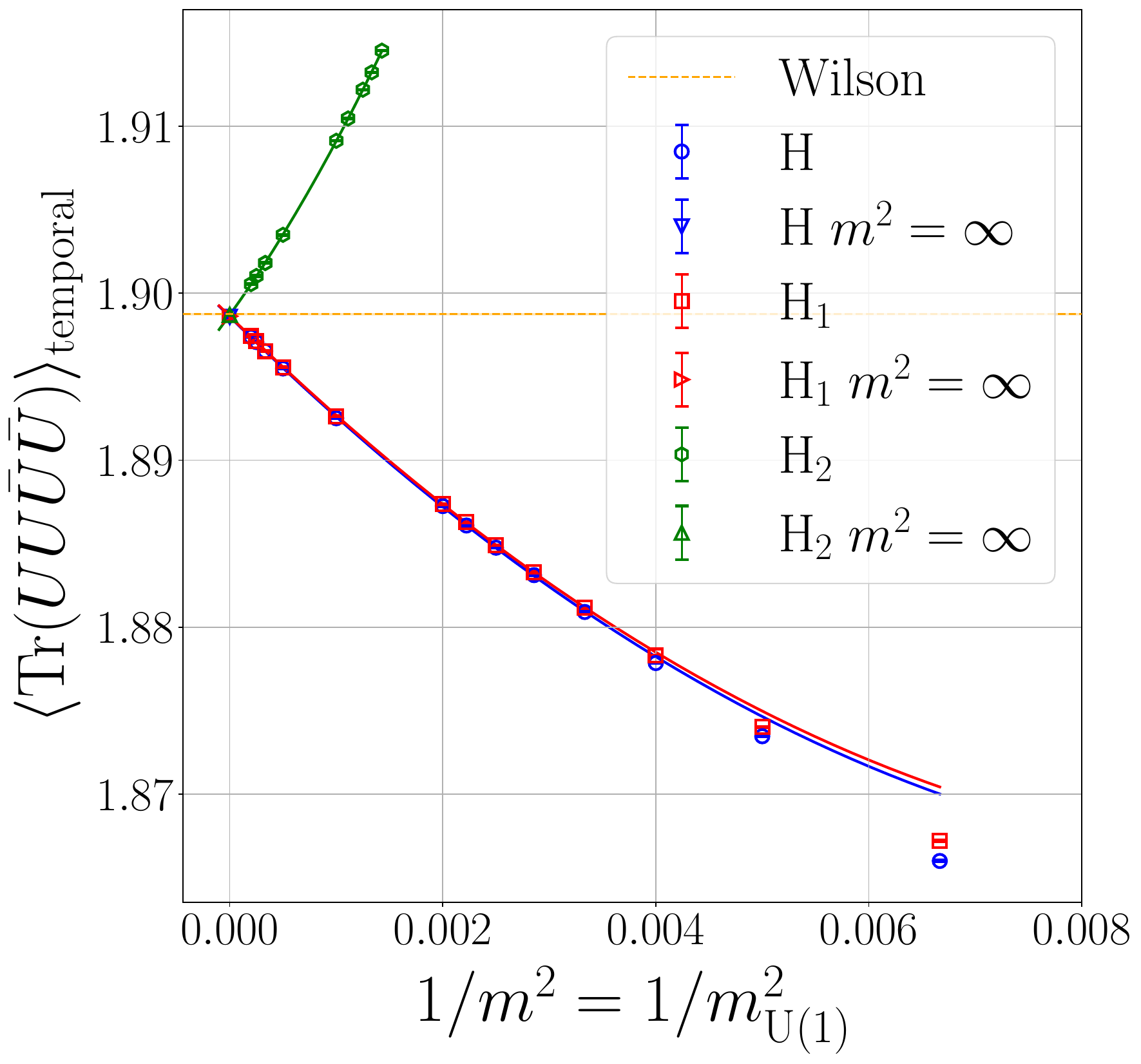}
    \end{subfigure}
    \hfill
        \begin{subfigure}{0.22\textwidth}
        \centering
        \includegraphics[width=1.03\linewidth]{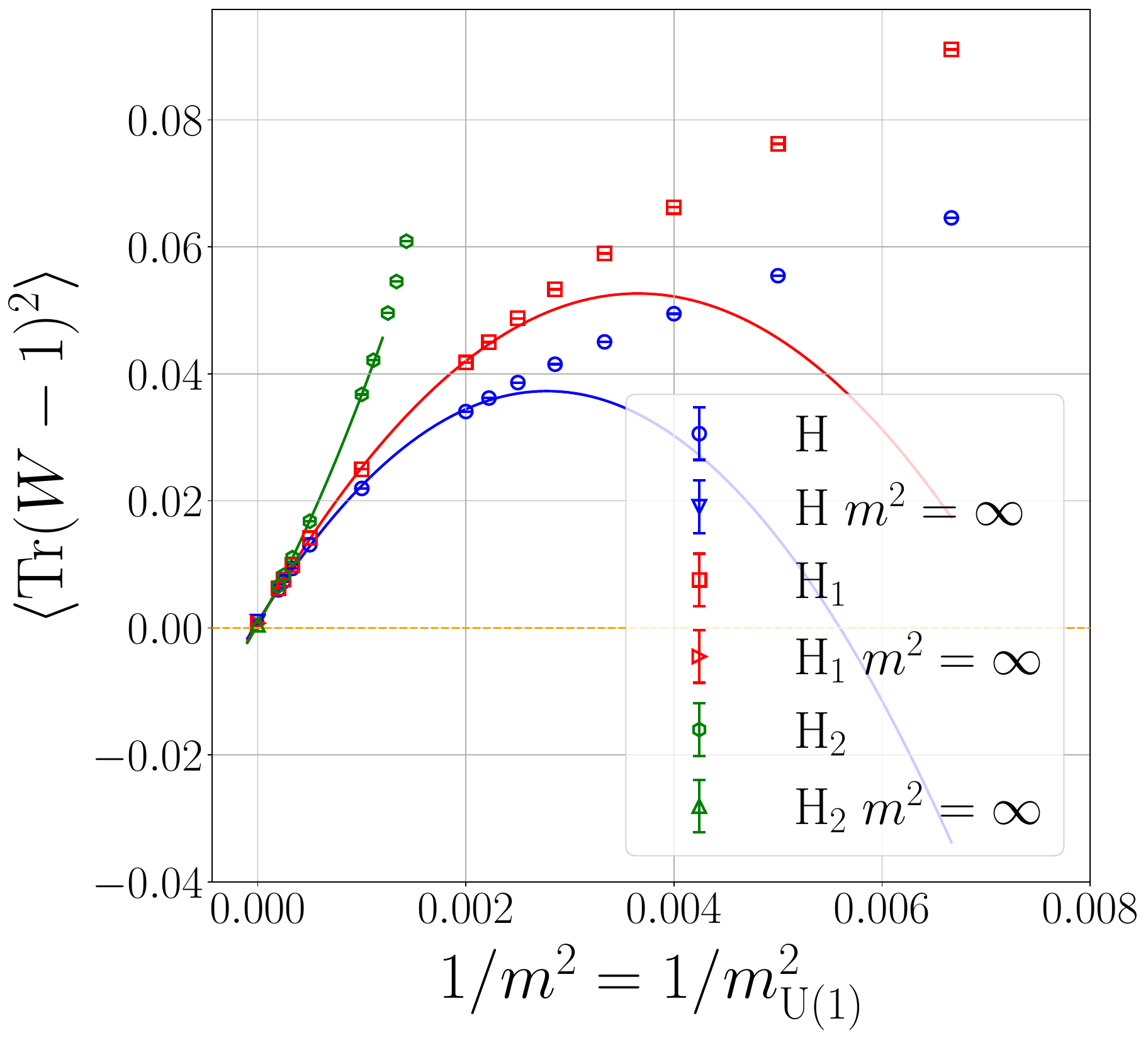}
    \end{subfigure}
    
    \vspace{0.2 cm}
        \centering    {$a=a_t=0.3$}
        \vspace{0.2 cm}
       
    \begin{subfigure}{0.22\textwidth}
        \centering
    \includegraphics[width=\linewidth]{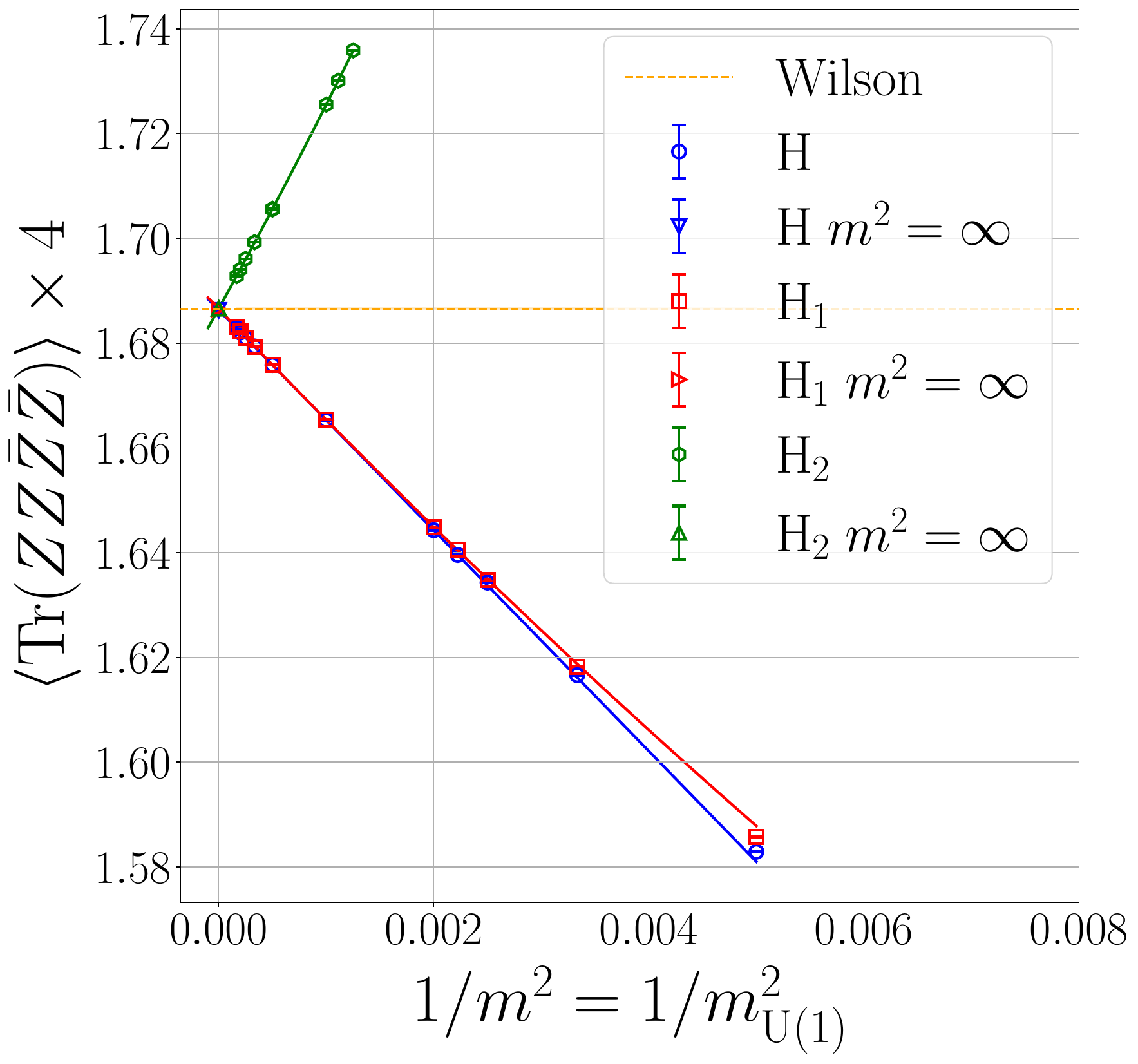}
    \end{subfigure}
        \hfill
     \begin{subfigure}{0.22\textwidth}
        \centering
    \includegraphics[width=\linewidth]{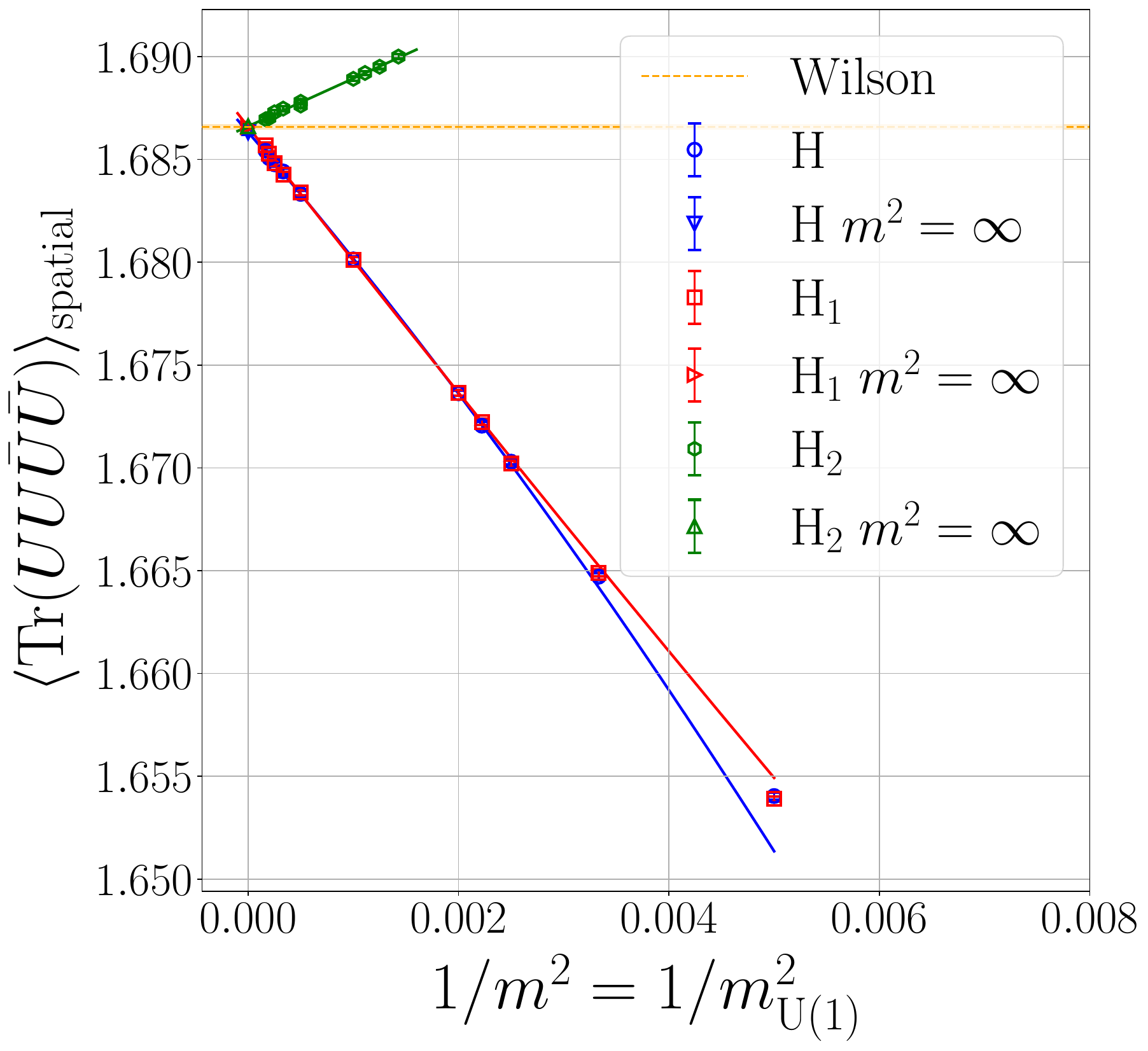}
    \end{subfigure}   
        \hfill
        \begin{subfigure}{0.22\textwidth}
        \centering
    \includegraphics[width=\linewidth]{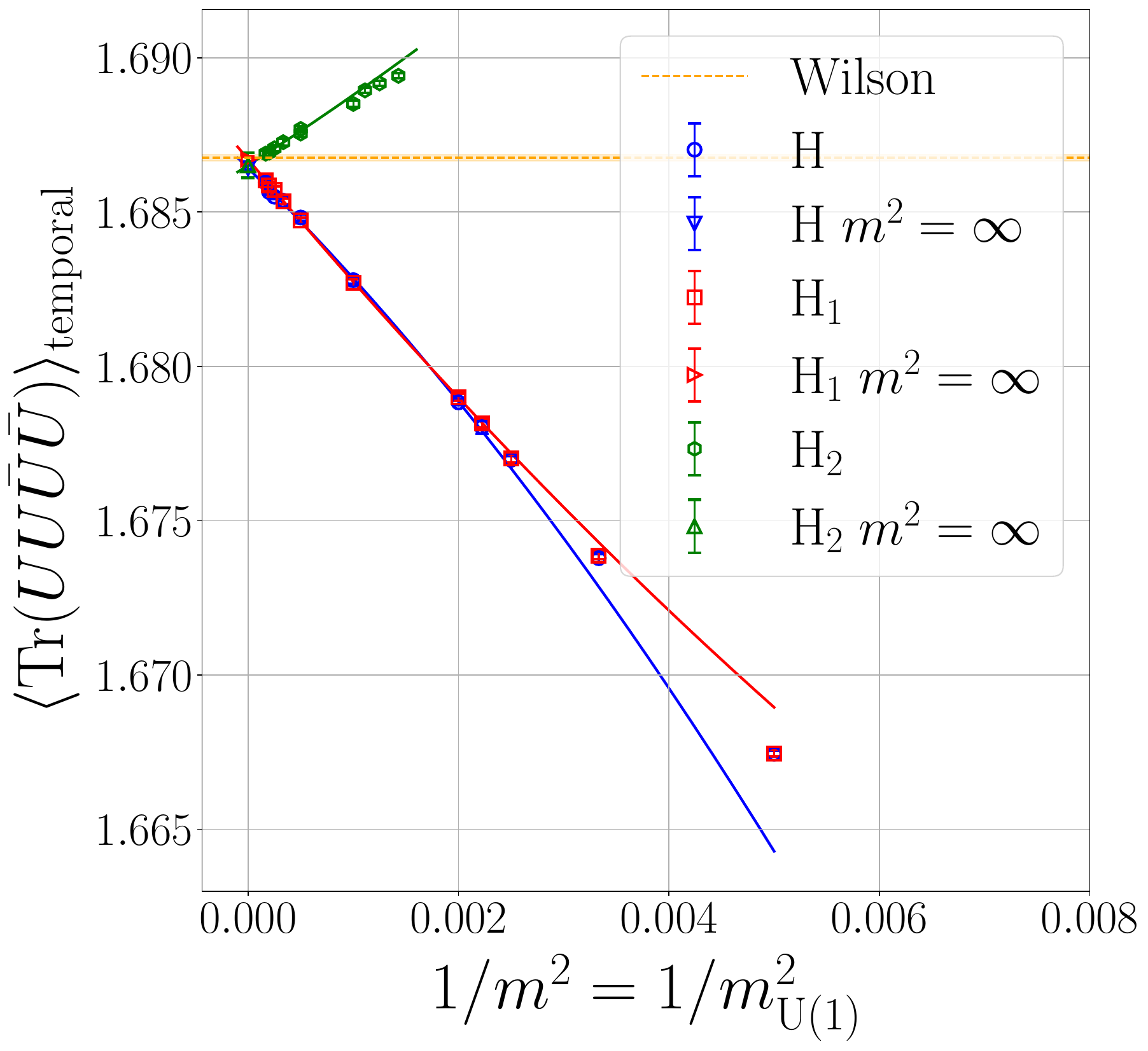}
    \end{subfigure}
        \hfill
        \begin{subfigure}{0.22\textwidth}
        \centering
    \includegraphics[width=\linewidth]{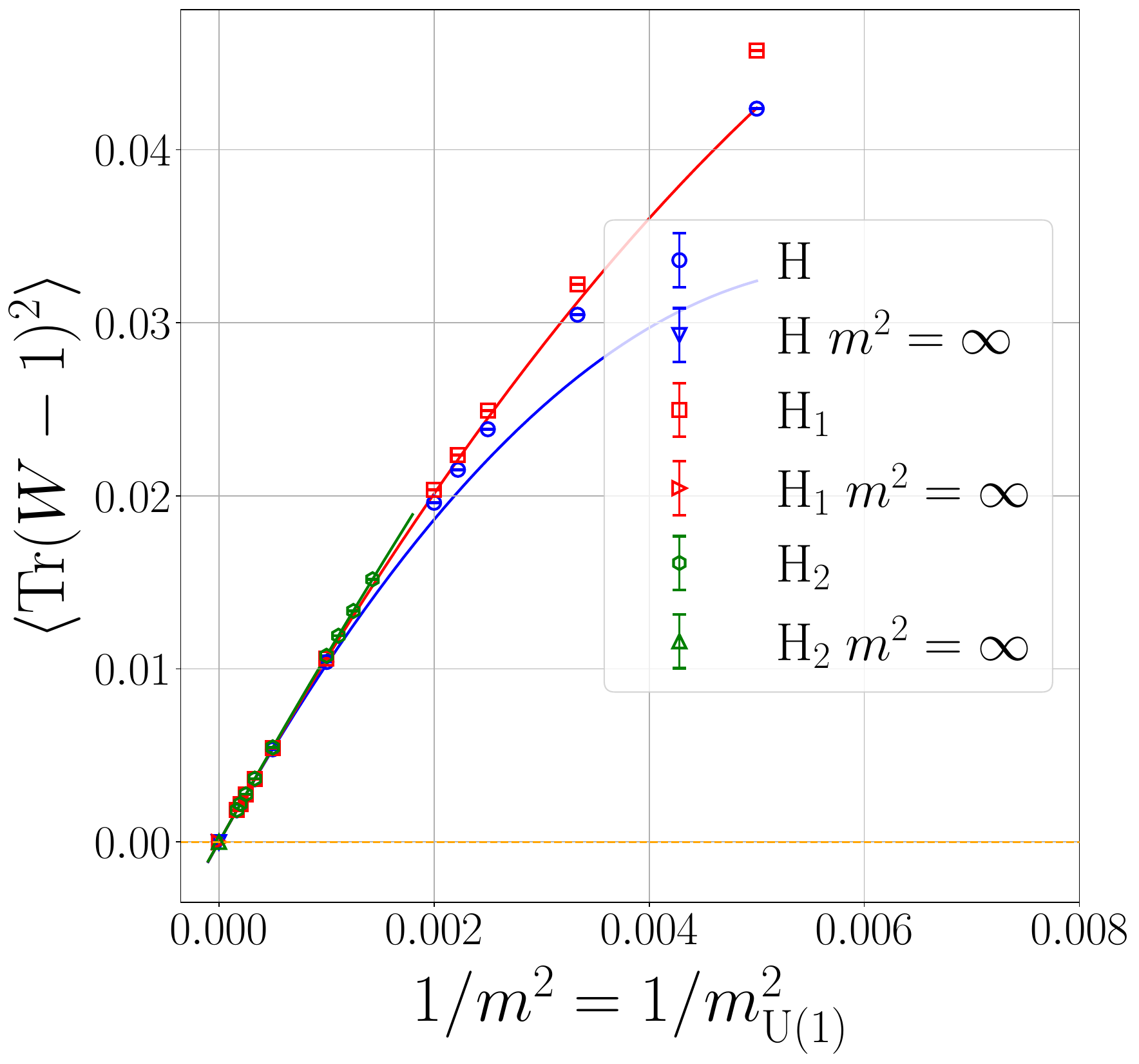}
    \end{subfigure}
    %
    \caption{
Plot of various quantities as function of $1/m^2$, for $\hat{H}$, $\hat{H}_1$, and $\hat{H}_2$ embedded in $\mathbb{R}^8$. Similar to Figures~\ref{fig:Pla_Z_8q_atas_0.1_0.3_R8}, 
\ref{fig:spatial_Pla_U_8q_atas_0.1_0.3_R8}, 
\ref{fig:temporal_Pla_U_8q_atas_0.1_0.3_R8}, 
and~\ref{fig:TrWm1sq_8q_atas_0.1_0.3_R8} but for lattice size $16^3$.
[\textbf{Top}] $a=a_t=0.1$. 
[\textbf{Bottom}] $a=a_t=0.3$. } 
\end{figure}

\subsection{Simulations of the embedding of SU(2) into $\mathbb{R}^4$ on a larger lattice size}\label{app:R4_larger_lattice}
In this appendix, we present all the numerical simulations conducted for the embedding of $SU(2)$ into $\mathbb{R}^4$ on larger lattice size.\\
The following plots show the results of Monte Carlo simulations for $16^3$ lattice with two lattices spacing $a_t = a = 0.1$ and $0.3$. The observables show a similar convergence to what was shown in Section~\ref{sec:Embedding_R4} for $8^3$, confirming the absence of lattice artifacts from smaller lattice sizes.

\begin{figure}[H]
        \centering    {$a=a_t=0.1$}
        \vspace{0.2 cm}
        
    \centering
    \begin{subfigure}{0.22\textwidth}
        \centering
        \includegraphics[width=\linewidth]{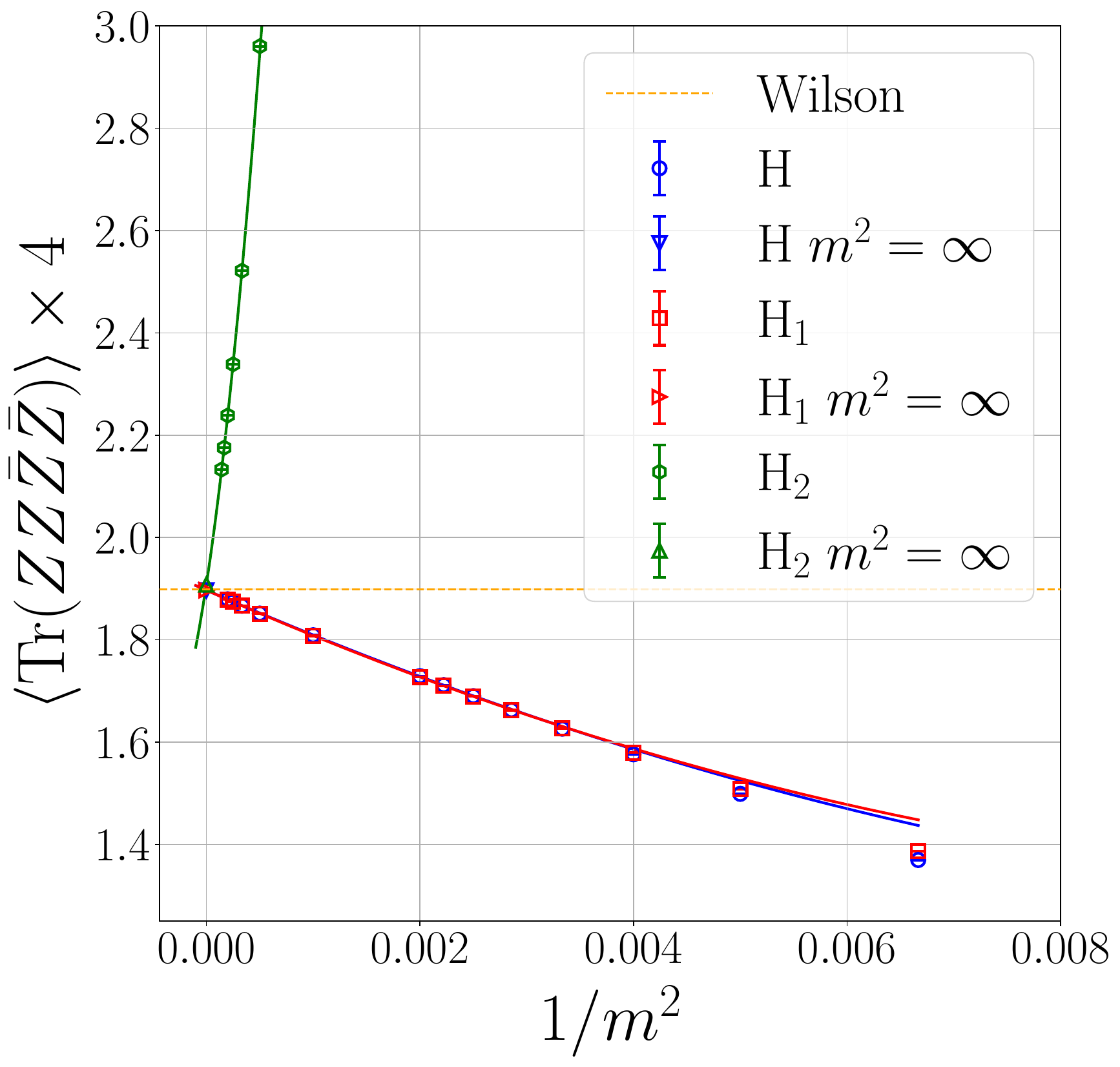}
    \end{subfigure}
    \hfill
        \begin{subfigure}{0.22\textwidth}
        \centering
        \includegraphics[width=1.02\linewidth]{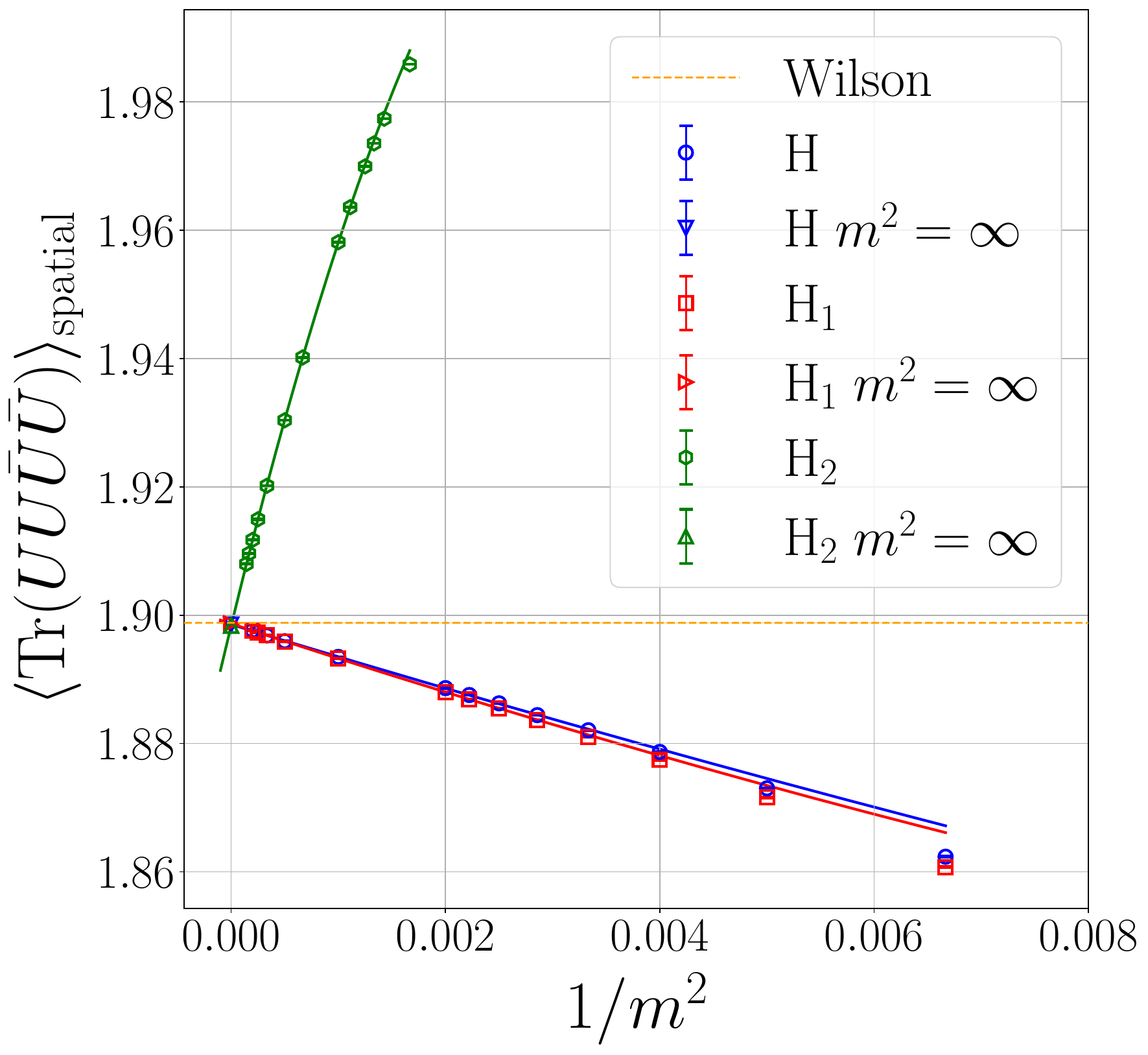}
    \end{subfigure}
    \hfill
        \begin{subfigure}{0.22\textwidth}
        \centering
        \includegraphics[width=1.02\linewidth]{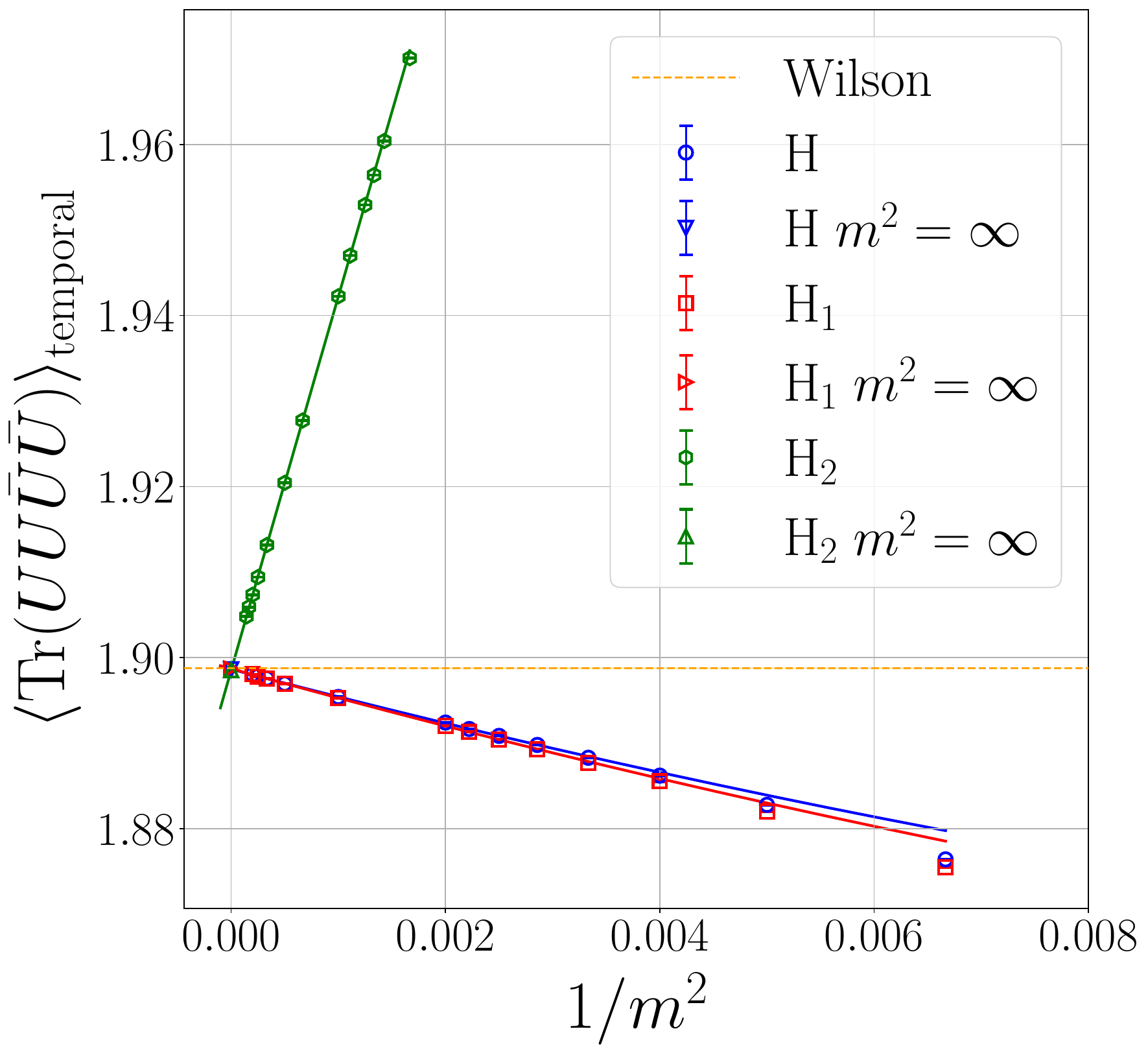}
    \end{subfigure}
        \hfill
        \begin{subfigure}{0.22\textwidth}
        \centering
        \includegraphics[width=1.02\linewidth]{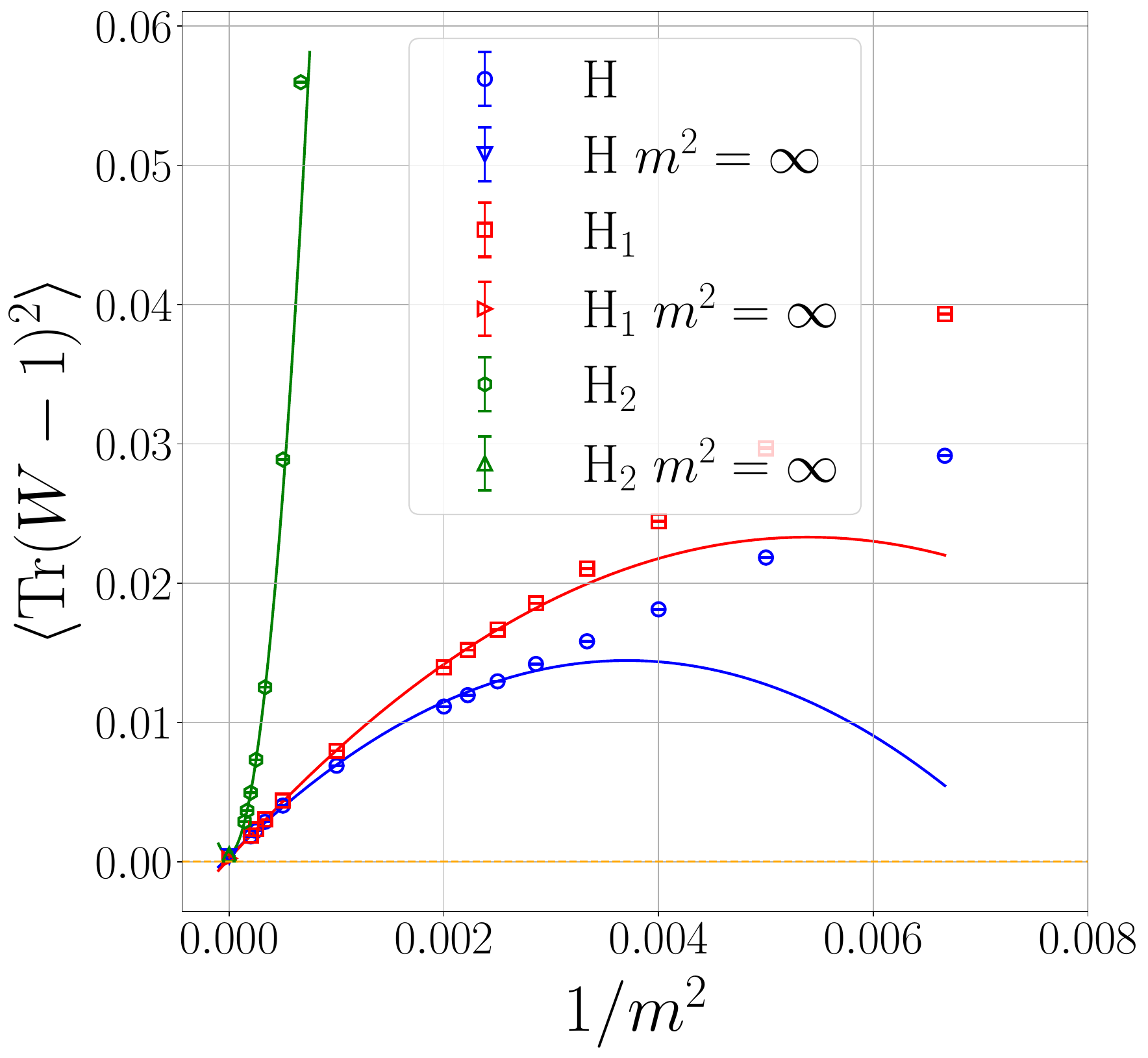}
    \end{subfigure}
    
        \centering    {$a=a_t=0.3$}
        \vspace{0.2 cm}
    
    \begin{subfigure}{0.22\textwidth}
        \centering
    \includegraphics[width=\linewidth]{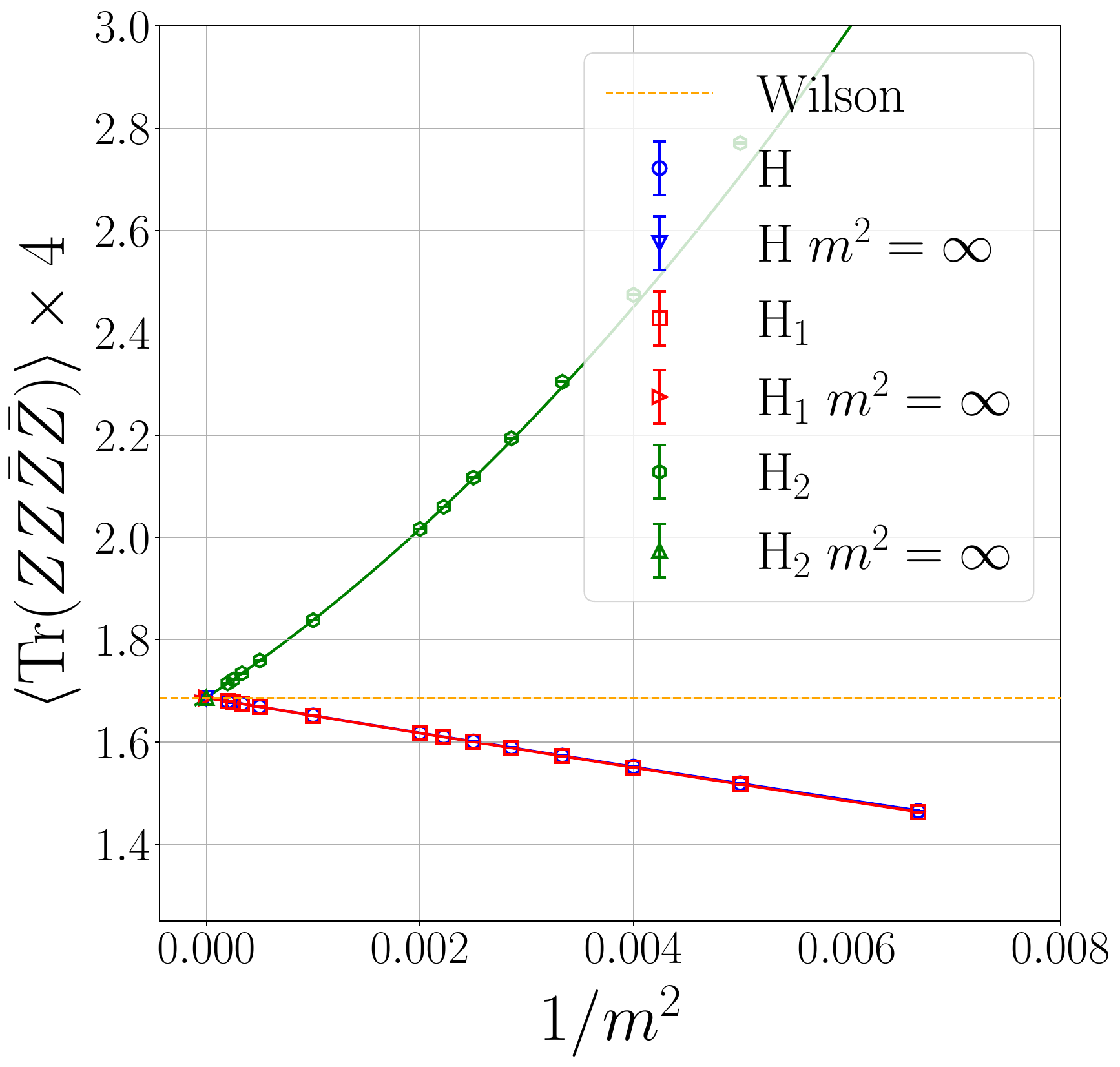}
    \end{subfigure}
        \hfill
        \begin{subfigure}{0.22\textwidth}
        \centering
    \includegraphics[width=1.02\linewidth]{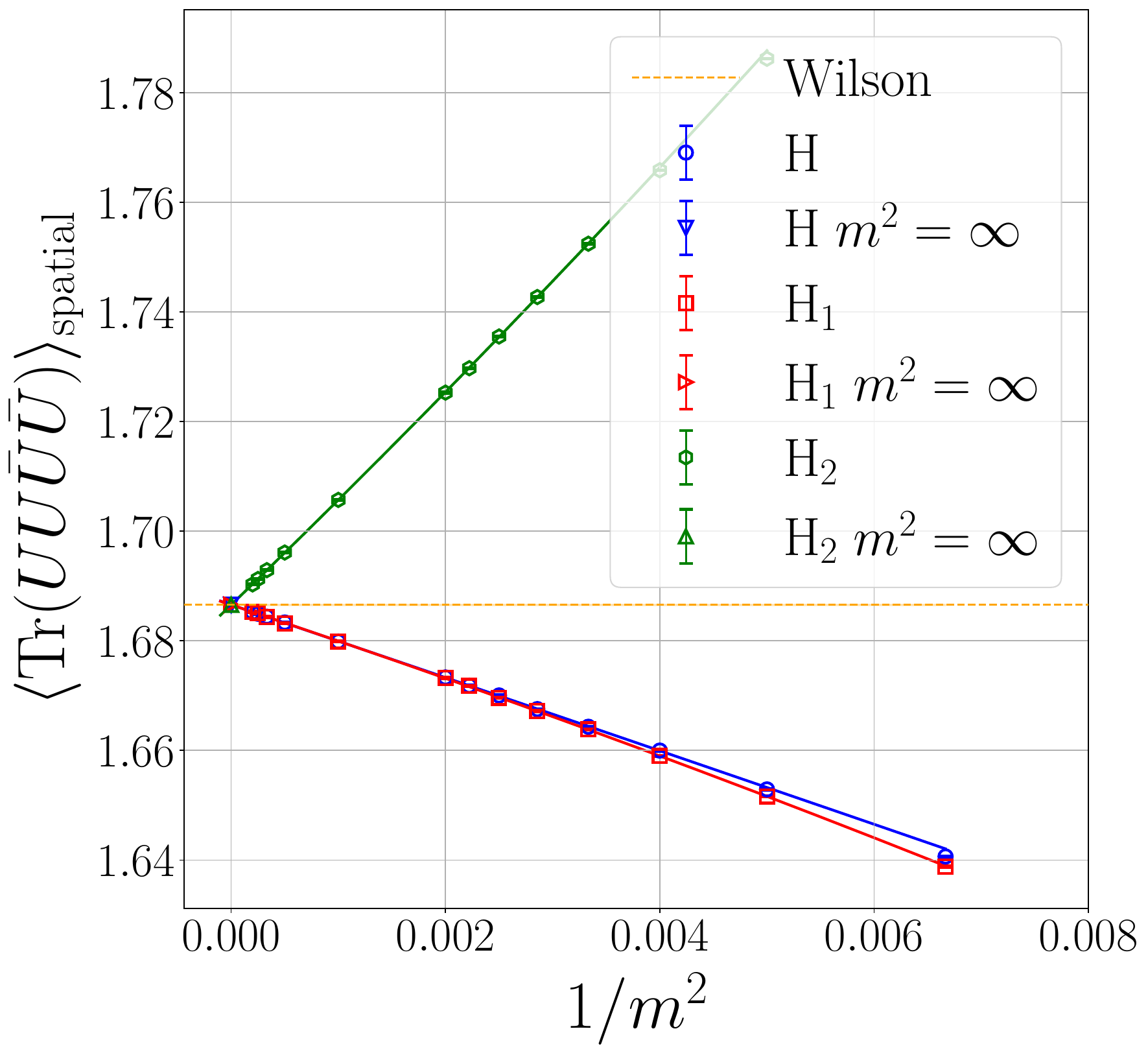}
    \end{subfigure}
        \hfill
        \begin{subfigure}{0.22\textwidth}
        \centering
    \includegraphics[width=1.02\linewidth]{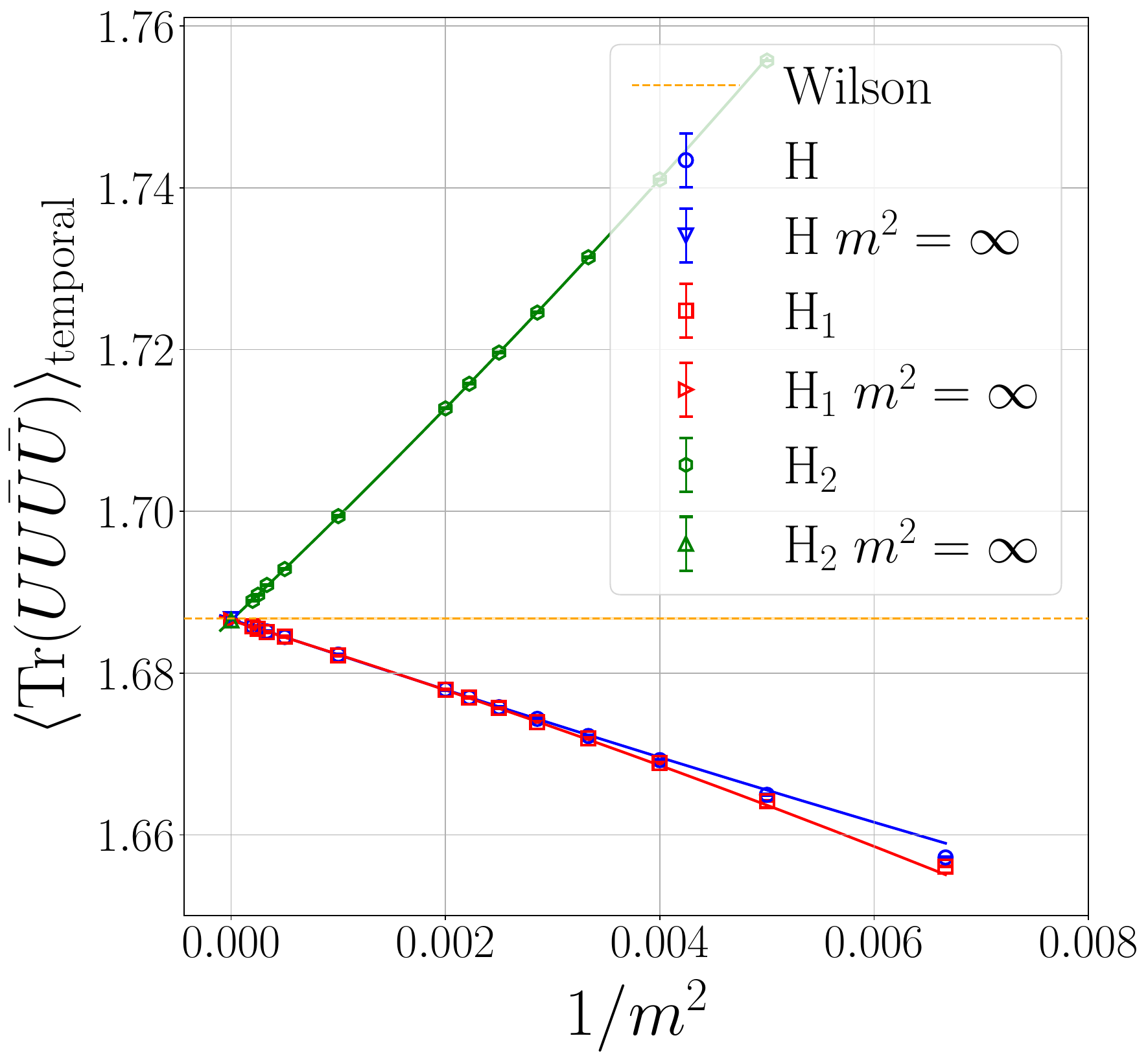}
    \end{subfigure}
        \hfill
        \begin{subfigure}{0.22\textwidth}
        \centering
    \includegraphics[width=1.025\linewidth]{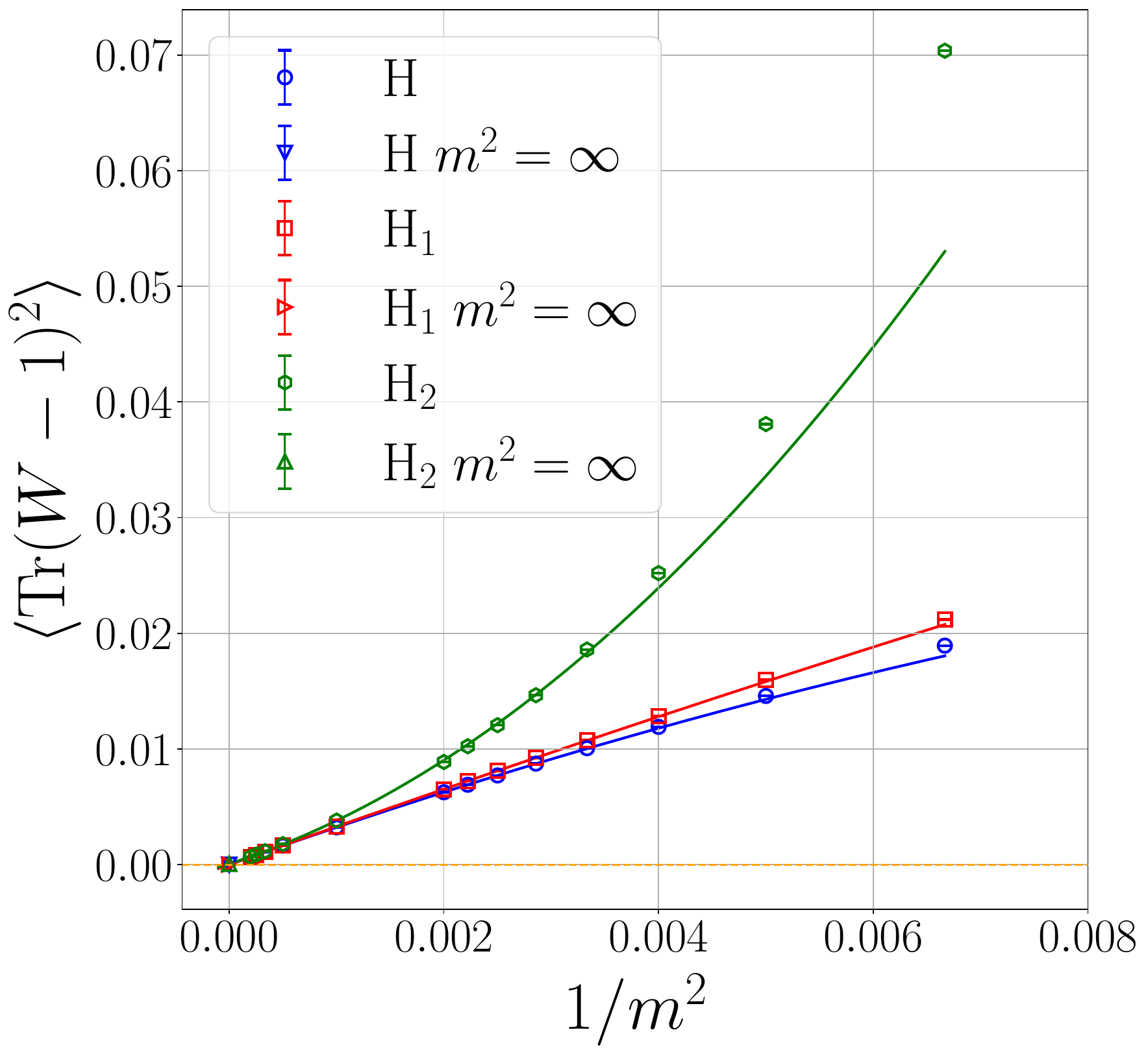}
    \end{subfigure}
            \hfill

    \caption{
Plot of various quantities as function of $1/m^2$, for $\hat{H}$, $\hat{H}_1$, and $\hat{H}_2$ embedded in $\mathbb{R}^4$. Similar to Figures~\ref{fig:Pla_Z_8q_atas_0.1_0.3}, 
\ref{fig:spatial_Pla_U_8q_atas_0.1_0.3}, 
\ref{fig:temporal_Pla_U_8q_atas_0.1_0.3}, 
and~\ref{fig:TrWm1sq_8q_atas_0.1_0.3}  but for lattice size $16^3$.
[\textbf{Top}] $a=a_t=0.1$. 
[\textbf{Bottom}] $a=a_t=0.3$. } 
\label{fig:Pla_Z_16q_atas_0.1_0.3}
\end{figure}

\subsection{Counter-term for $\mathbb{R}^8$  on larger lattice spacing }\label{app:CT_R8_larger_spacing}
In this appendix we present supplementary numerical results from simulations performed at a larger lattice spacing for the SU$(2)$ theory embedded in $\mathbb{R}^8$ with the counter-term.
The following plots show the results of Monte Carlo simulations on an $8^3$ lattice with lattice spacings $a_t = a = 0.3$, which are larger than those used in Section~\ref{sec:results_ZZbar_counterterm_R8}, where $a_t = a = 0.1$. The plaquette values exhibit a clear tendency to converge towards the Wilson action already at $m^2 = 50$ for $\hat{H}$ and $\hat{H}_1$, and at $m^2 = 500$ for $\hat{H}_2$. This convergence is slightly weaker compared to the results obtained in Section~\ref{sec:results_ZZbar_counterterm_R8} for $a_t = a = 0.1$, indicating that an increase in the lattice spacing requires a corresponding increase in the parameter $m^2$.

Overall, the required values of $m^2$ are reduced by approximately one order of magnitude compared to simulations employing the orbifold action without the counter-term, as discussed in Section~\ref{sec:Embedding_R8}.

\begin{figure}[H]
    \centering
    \begin{subfigure}{0.31\textwidth}
        \includegraphics[width=\linewidth]{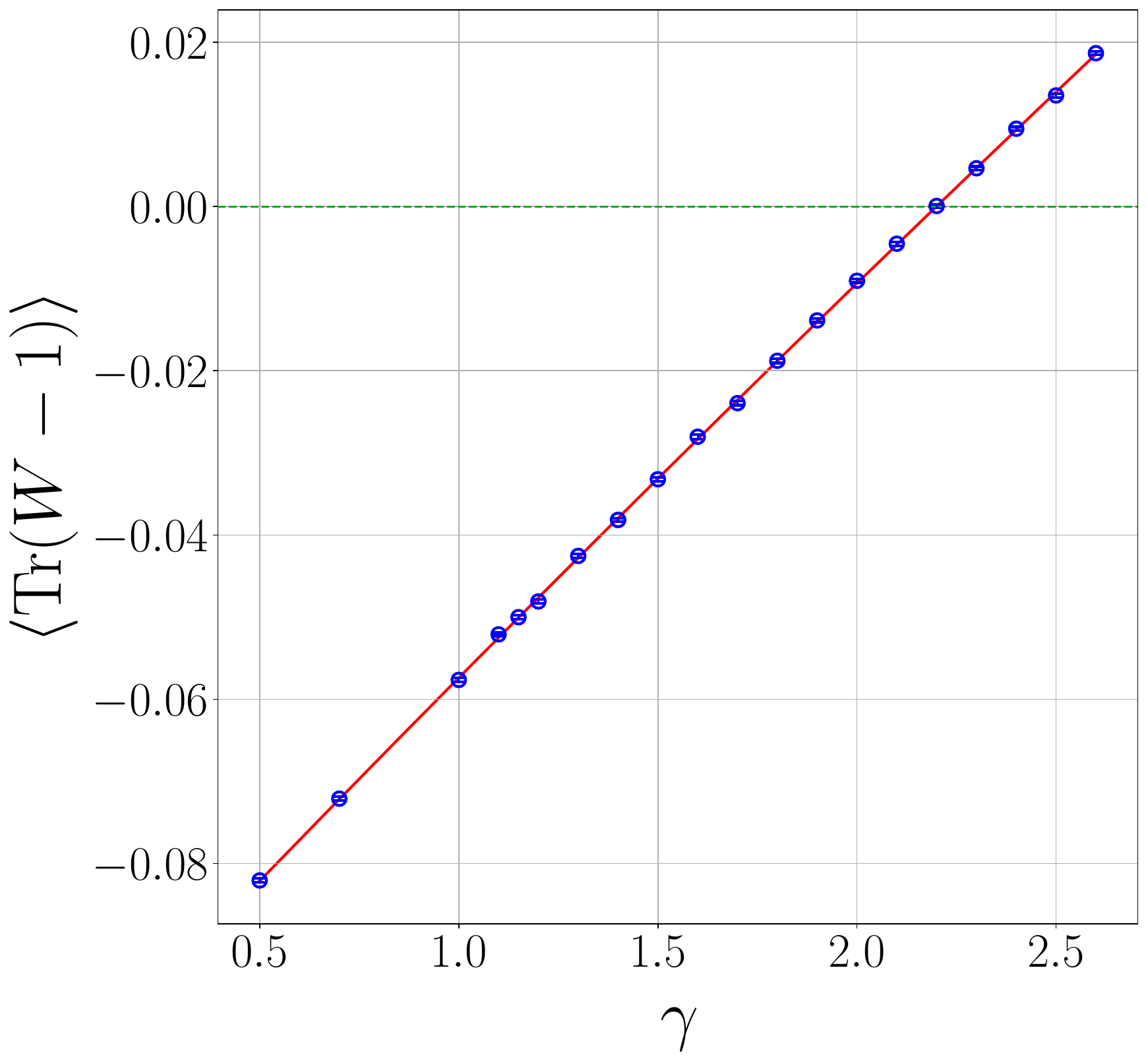}
    \end{subfigure}
    \hfill
    \begin{subfigure}{0.31\textwidth}
        \includegraphics[width=\linewidth]{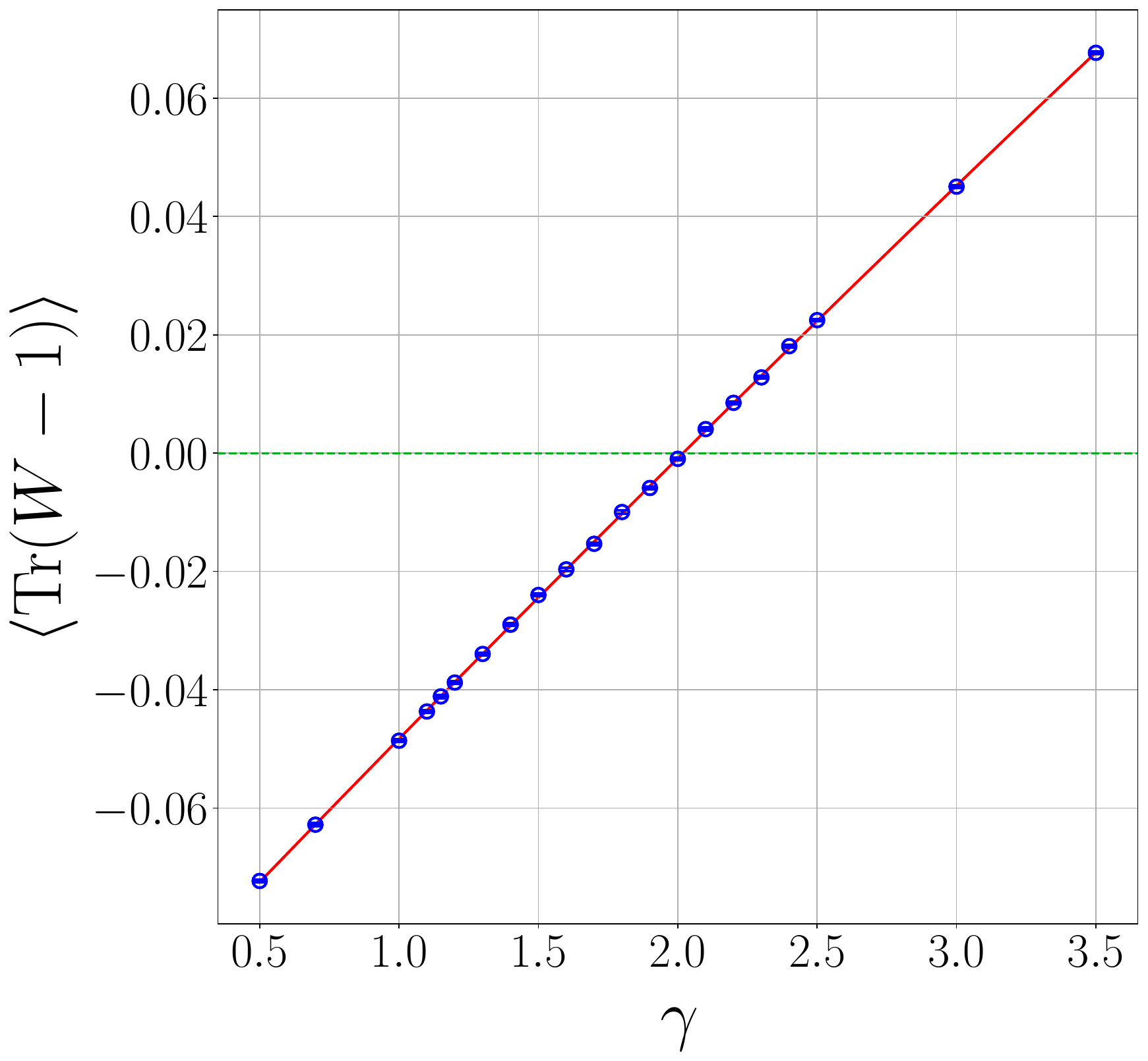}
    \end{subfigure}
    \hfill
    \begin{subfigure}{0.31\textwidth}
        \includegraphics[width=\linewidth]{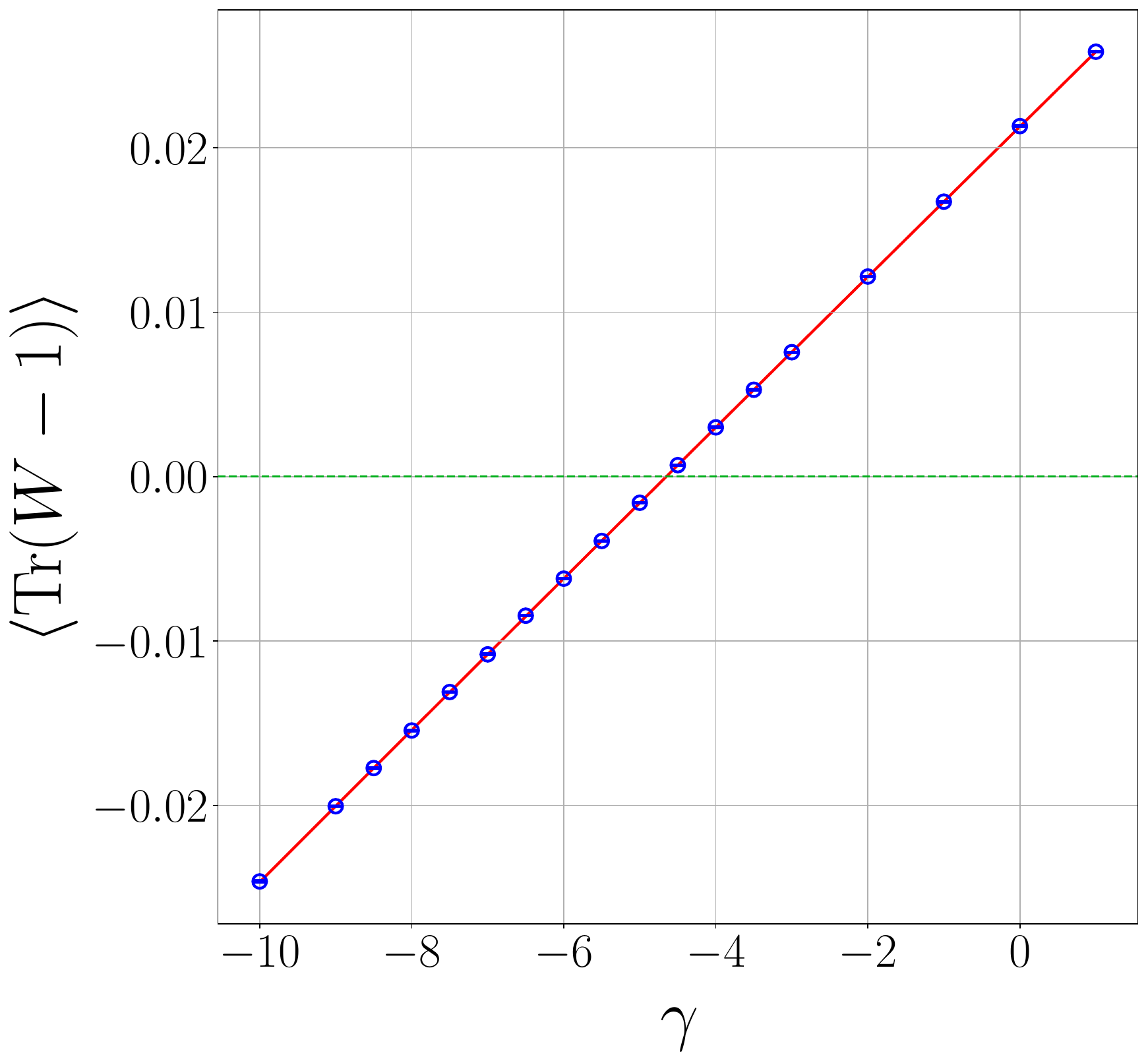}
    \end{subfigure}

    \vspace{1em}

    \begin{subfigure}{0.3\textwidth}
        \includegraphics[width=\linewidth]{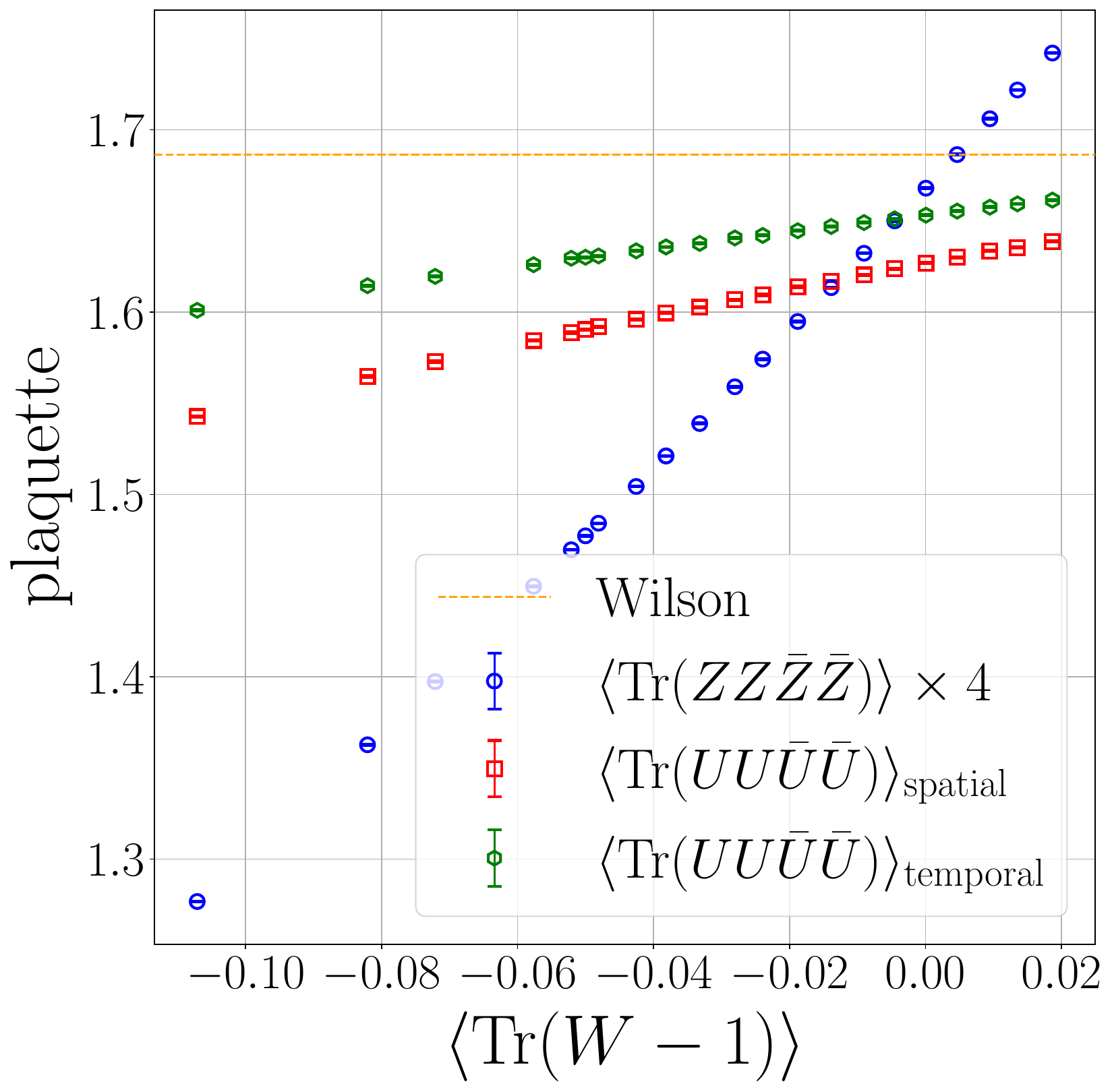}
       \caption*{$\hat{H}$, $8^3$, $m^2 = 50$}
    \end{subfigure}
    \hfill
    \begin{subfigure}{0.3\textwidth}
        \includegraphics[width=1.03\linewidth]{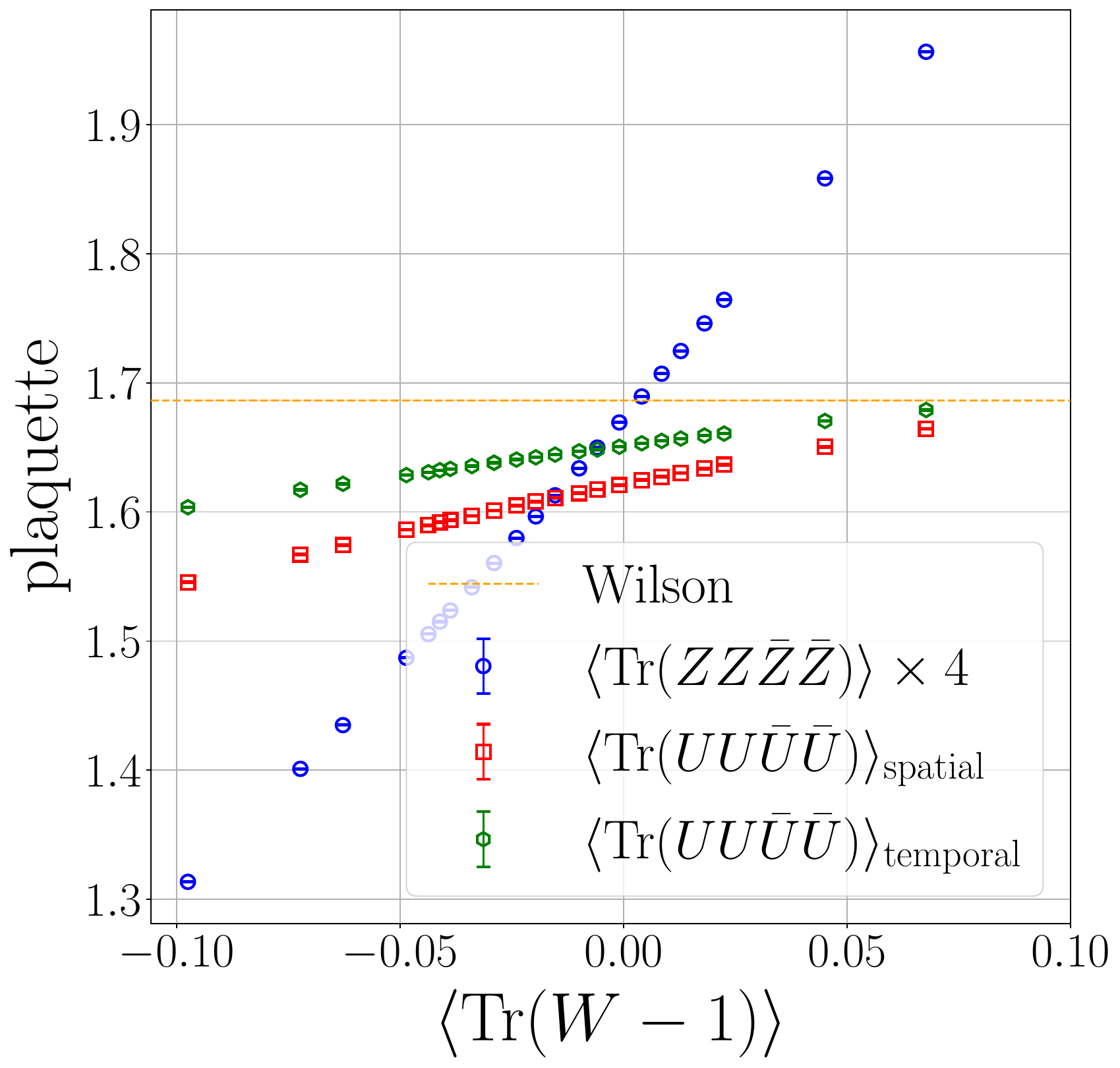}
        \caption*{$\hat{H}_1$, $8^3$, $m^2 = 50$}
    \end{subfigure}
    \hfill
    \begin{subfigure}{0.3\textwidth}
        \includegraphics[width=1.02\linewidth]{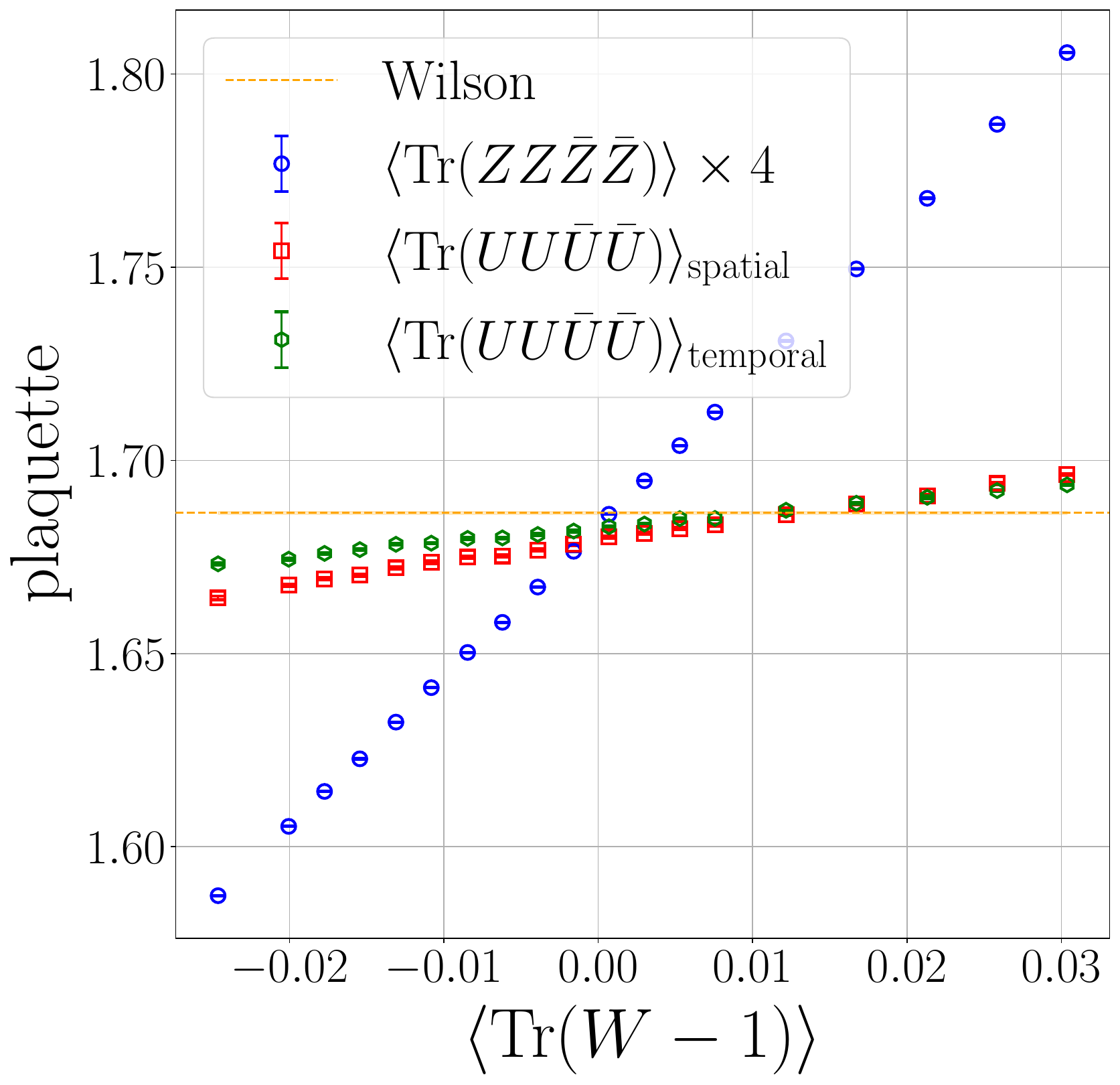}
       \caption*{$\hat{H}_2$, $8^3$, $m^2 = 500$}
    \end{subfigure}

    \caption{
        [\textbf{Left}] Same as Fig.~\ref{fig:TrW_1_gamma_R8_H}, showing the original Orbifold Hamiltonian $\hat{H}$ embedded in $\mathbb{R}^8$ with $\gamma$ counter-term, $m^2 = 50$, and lattice spacing $a = a_t = 0.3$.
        [\textbf{Center}] Same as Fig.~\ref{fig:TrW_1_gamma_R8_H1}, showing $\hat{H}_1$ with the same parameters.
        [\textbf{Right}] Same as Fig.~\ref{fig:TrW_1_gamma_R8_H2}, showing $\hat{H}_2$ with $m^2 = 500$ and $a = a_t = 0.3$.
    }
\end{figure}

\subsection{Counter-term for $\mathbb{R}^4$ on larger lattice spacing }\label{app:CT_R4_larger_spacing}
In this appendix we present supplementary numerical results from simulations performed at a larger lattice spacing for the $SU(2)$ theory embedded in $\mathbb{R}^4$ with the counter-term.

The following plots show the results of Monte Carlo simulations on an $8^3$ lattice with lattice spacings $a_t = a = 0.3$, which are larger than those used in Section~\ref{sec:results_ZZbar_counterterm_R4}, where $a_t = a = 0.1$. The plaquette values exhibit a clear tendency to converge towards the Wilson action already at $m^2 = 50$ for $\hat{H}$ and $\hat{H}_1$, and at $m^2 = 500$ for $\hat{H}_2$. This convergence is slightly weaker compared to the results obtained in Section~\ref{sec:results_ZZbar_counterterm_R4} for $a_t = a = 0.1$, indicating that an increase in the lattice spacing requires a corresponding increase in the parameter $m^2$.

Overall, the required values of $m^2$ are reduced by approximately one order of magnitude compared to simulations employing the orbifold action without the counter-term, as discussed in Section~\ref{sec:Embedding_R4}.

\begin{figure}[H]
    \centering
    \begin{subfigure}{0.31\textwidth}
        \centering
        \includegraphics[width=\linewidth]{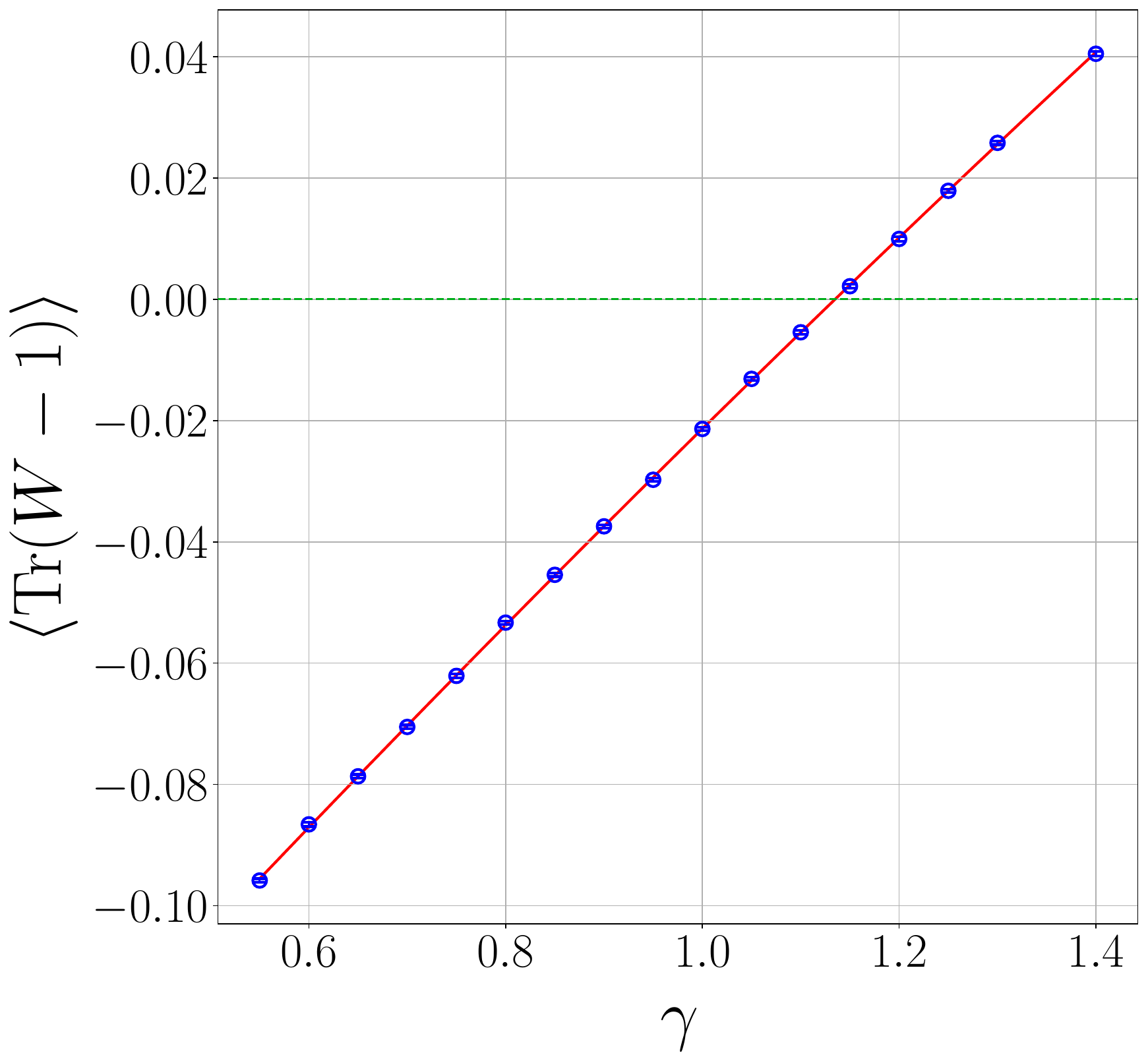}
    \end{subfigure}
    \hfill
        \begin{subfigure}{0.31\textwidth}
        \centering
        \includegraphics[width=\linewidth]{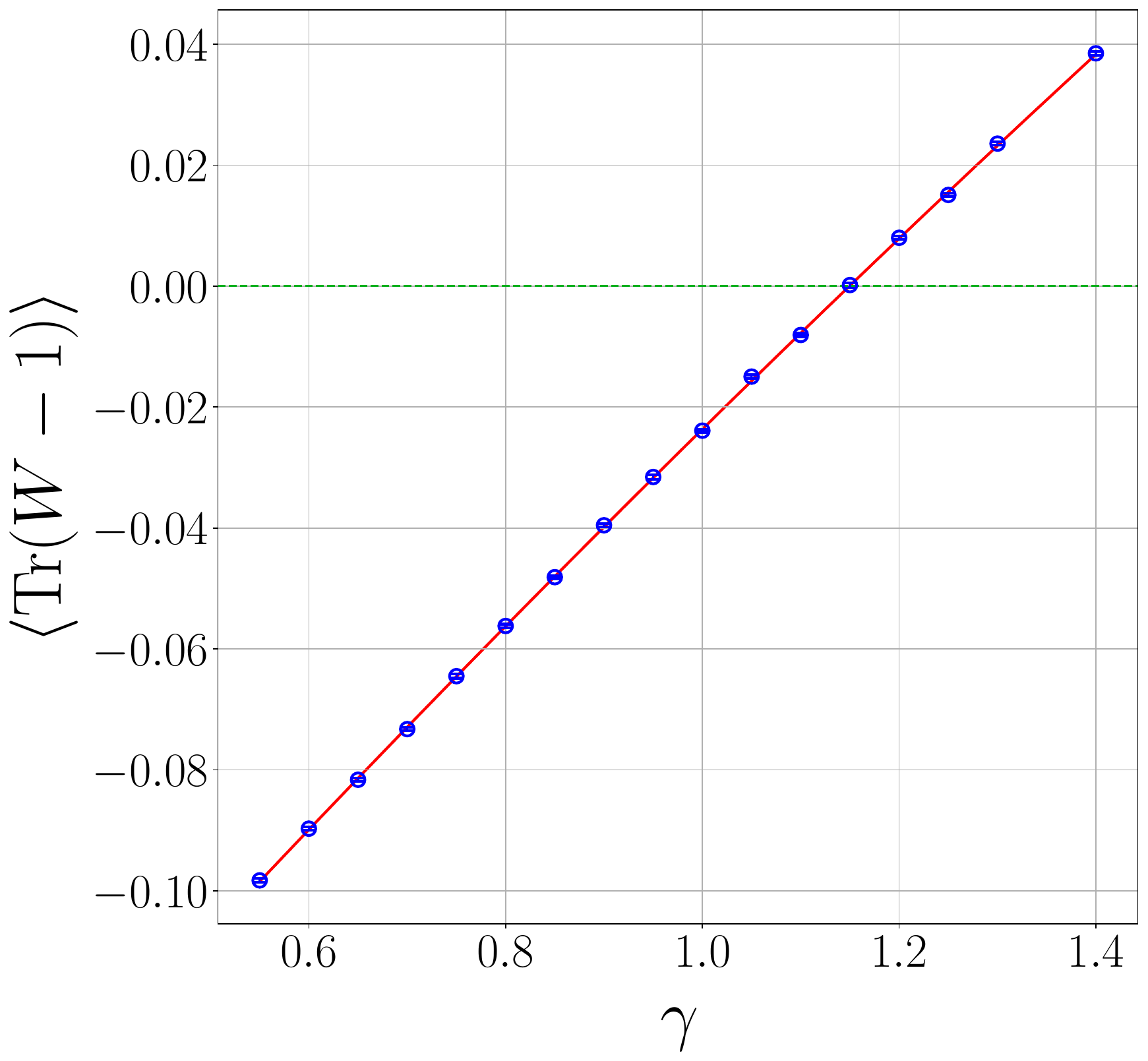}
    \end{subfigure}
    \hfill
        \begin{subfigure}{0.31\textwidth}
        \centering
        \includegraphics[width=\linewidth]{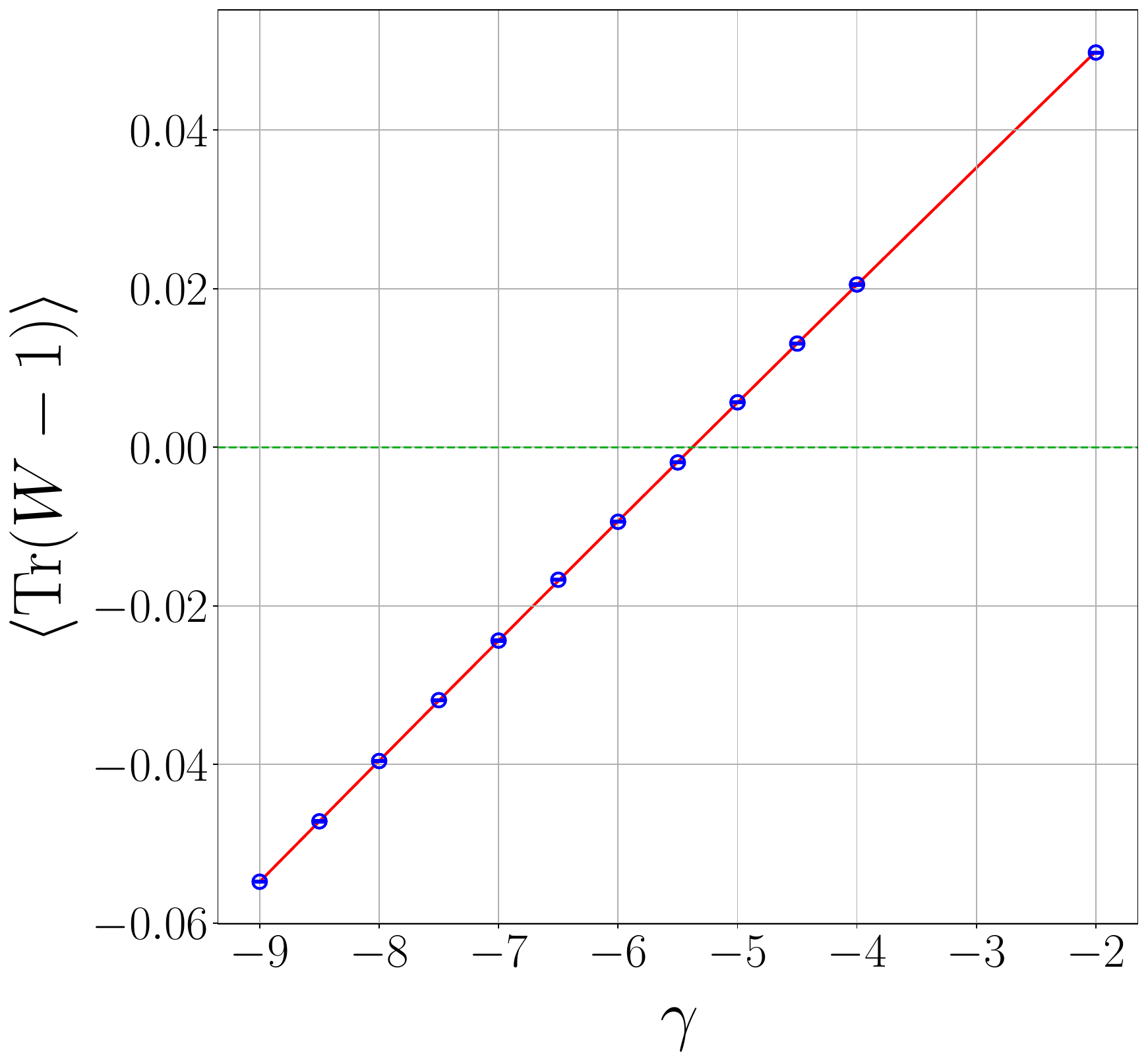}
    \end{subfigure}
    \hfill
    \begin{subfigure}{0.3\textwidth}
        \centering
    \includegraphics[width=\linewidth]{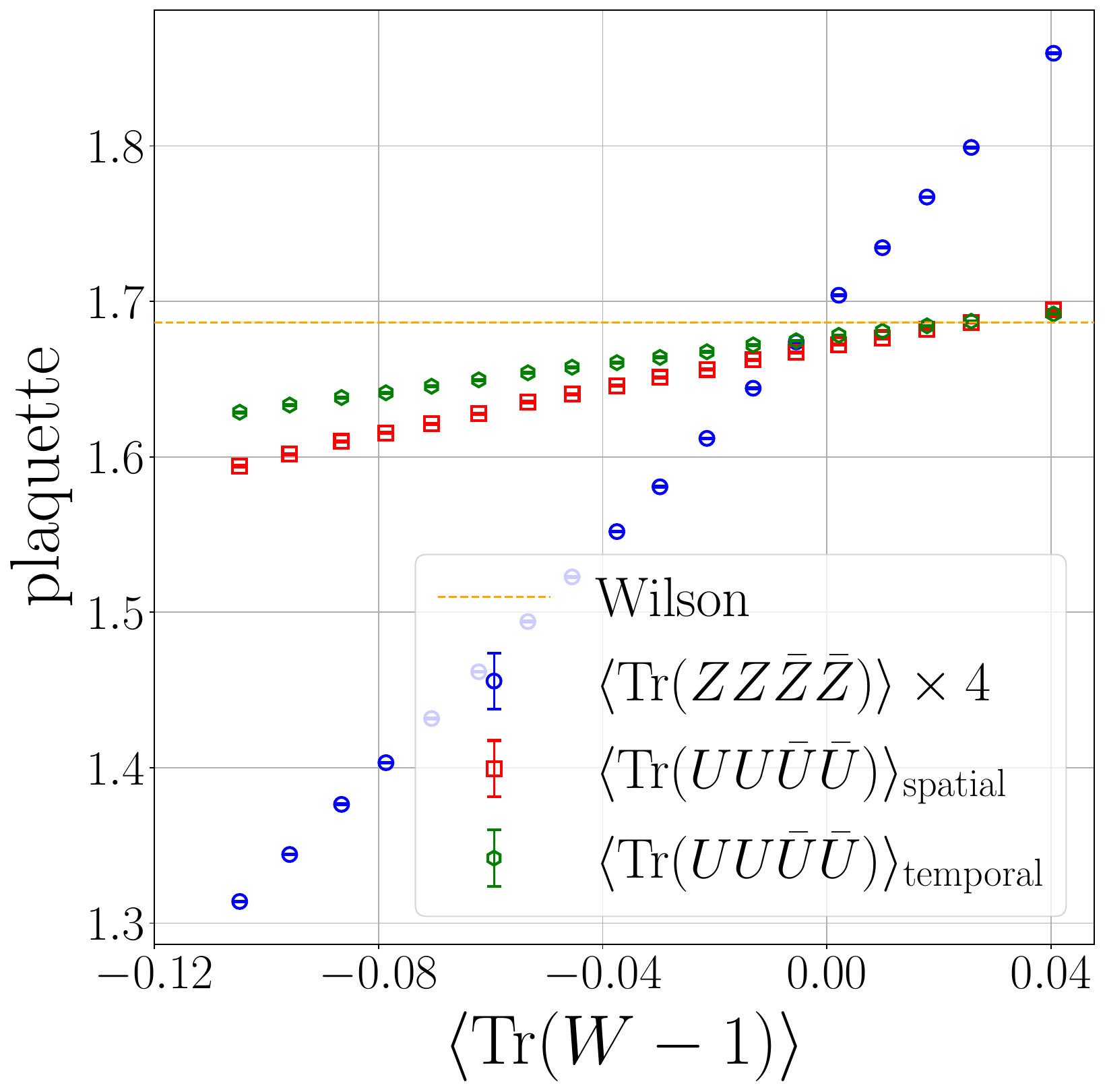}
     \caption*{$\hat{H}$, $a_t=a =0.3$, $m^2 = 50$}
    \end{subfigure}
     \hfill
         \begin{subfigure}{0.3\textwidth}
        \centering
    \includegraphics[width=1.01\linewidth]{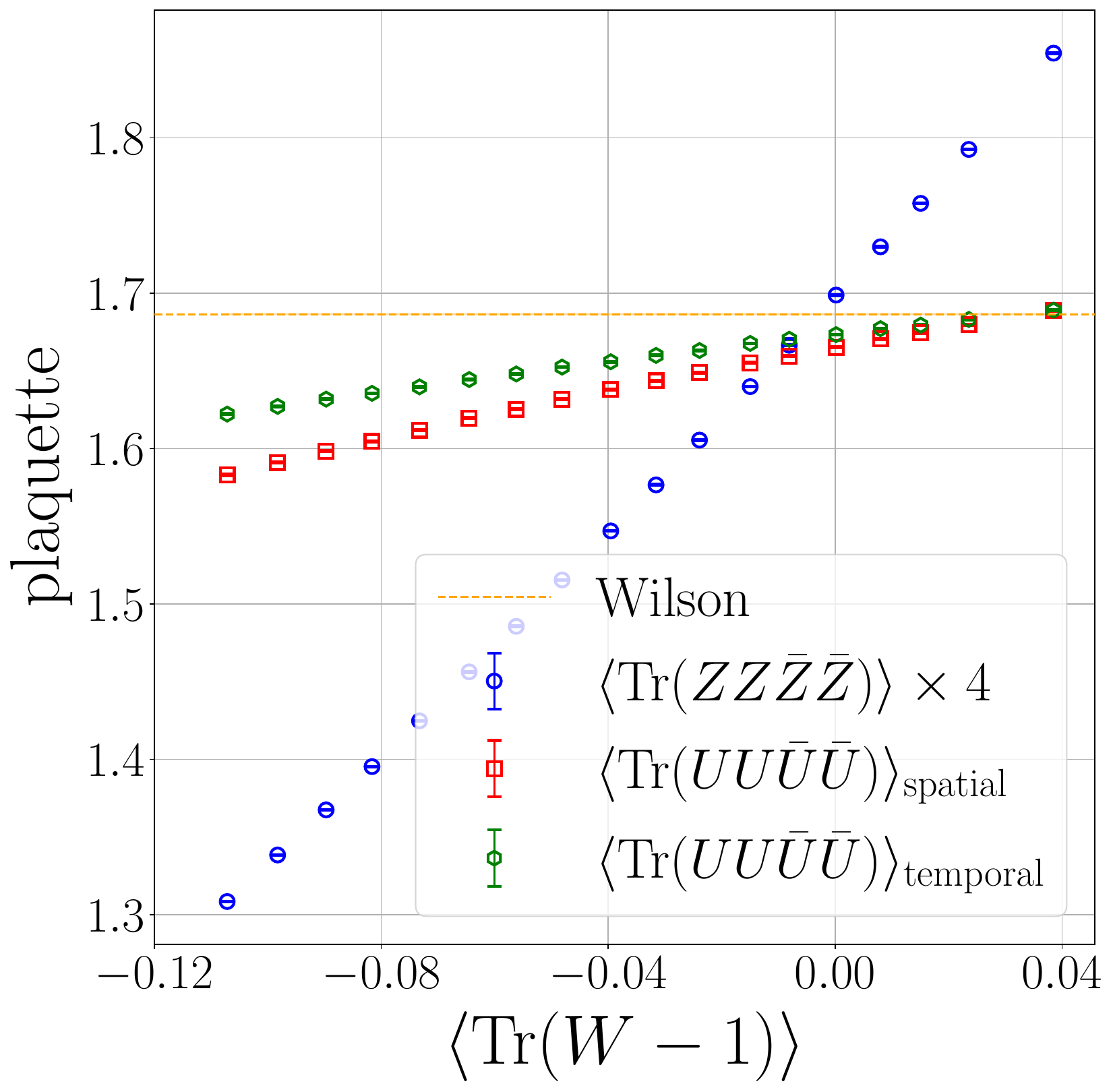}
      \caption*{$\hat{H}_1$, $a_t=a =0.3$, $m^2 = 50$}
    \end{subfigure}
    \hfill
    \begin{subfigure}{0.3\textwidth}
        \centering
    \includegraphics[width=\linewidth]{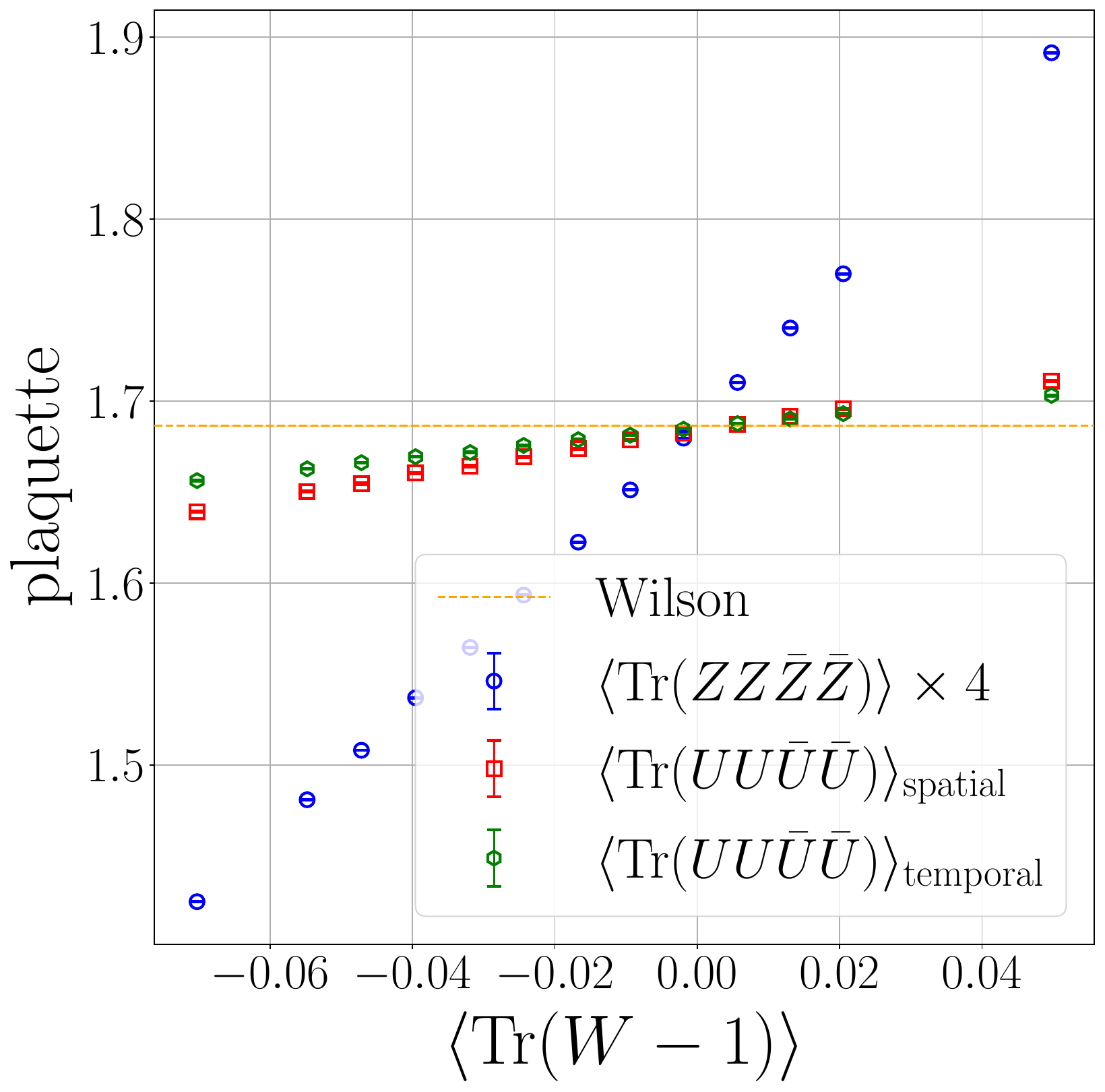}
       \caption*{$\hat{H}_2$, $a_t=a =0.3$, $m^2 = 500$}
    \end{subfigure}
    \caption{[\textbf{Left}] Same as Fig.~\ref{fig:TrW_1_gamma_R4_H}, showing $\hat{H}$ embedded in $\mathbb{R}^4$ with $\gamma$ counter-term, $m^2=50$ but for lattice spacing $a = a_t = 0.3$.
    [\textbf{Center}]
    Same as Fig.~\ref{fig:TrW_1_gamma_R4_H1}, showing $\hat{H}_1$ embedded in $\mathbb{R}^4$ with $\gamma$ counter-term, $m^2=50$ but for lattice spacing $a = a_t = 0.3$.  .
    [\textbf{Right}] Same as Fig.~\ref{fig:TrW_1_gamma_R4_H2}, showing $\hat{H}_2$ embedded in $\mathbb{R}^4$ with $\gamma$ counter-term, $m^2=500$ but for lattice spacing $a = a_t = 0.3$. 
   The lattice size is $8^3$ for all plots.  
    } \label{fig:TrW_1_gamma_R4_H1_gam}
\end{figure}

\subsection{Using effective lattice spacing for $\mathbb{R}^4$ with larger mass}\label{app:CT_R4_larger_spacing}
In this appendix we present additional numerical results on the tuning of the effective lattice spacing for simulations performed at a larger scalar mass parameter $m^2$ in the $SU(2)$ theory embedded in $\mathbb{R}^4$, extending the analysis discussed in Section~\ref{sec:a_bare_tuning}.

The results shown below are obtained from Monte Carlo simulations on an $8^3$ lattice at varying lattice spacings $a = a_t$, with scalar mass $m^2 = 150$, which is significantly almost double than the value $m^2 = 80$ used in Fig.~\ref{fig:tuning_a_bare_H} and Fig.~\ref{fig:tuning_a_bare_H1}.

\begin{figure}[H]
    \centering
    %
    %
    \begin{subfigure}{0.32\textwidth}
        \centering
        \includegraphics[width=\linewidth]
        {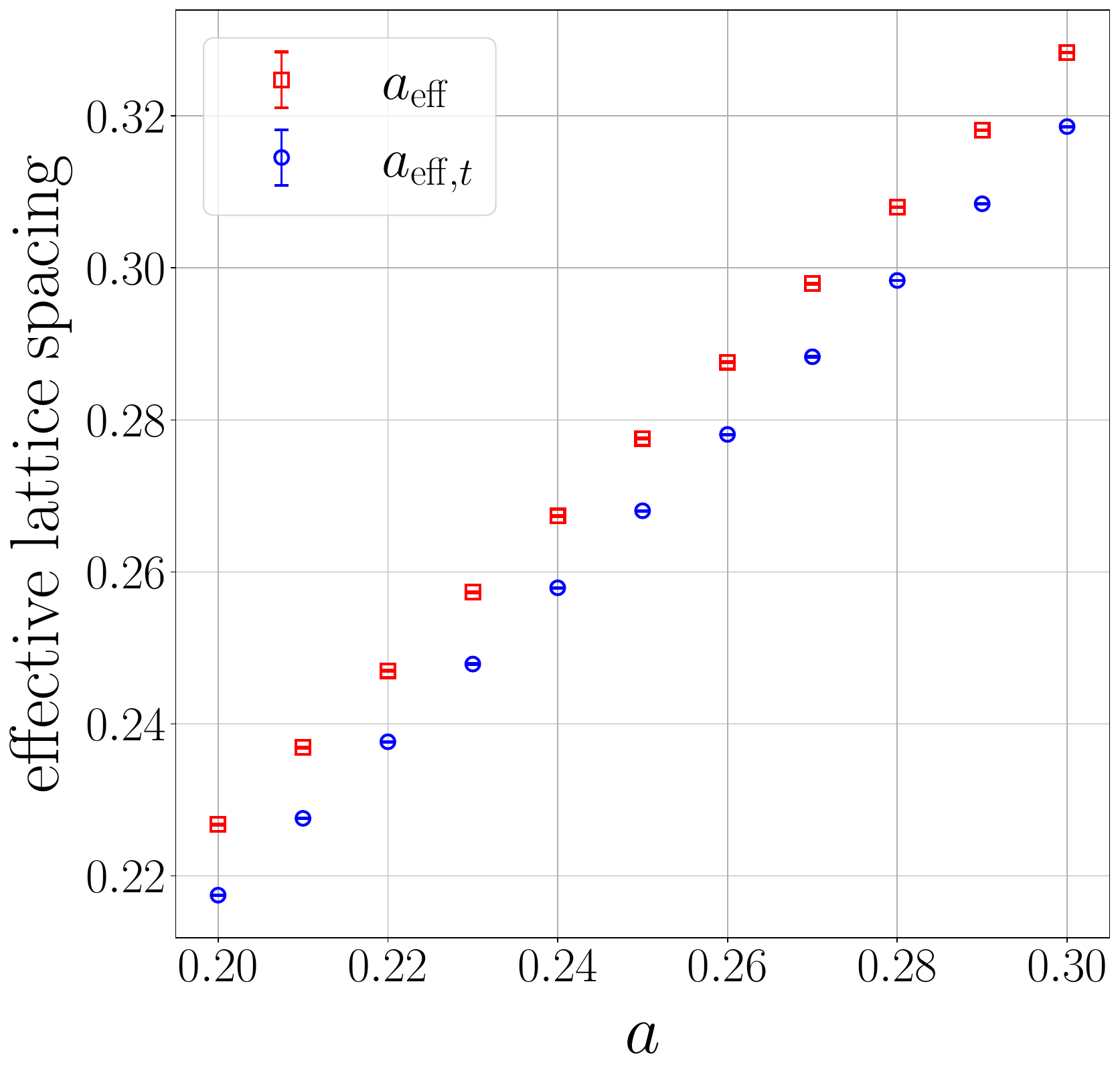}
    \end{subfigure}
    \hfill
        %
        %
    \begin{subfigure}{0.32\textwidth}
        \centering
\includegraphics[width=\linewidth]
{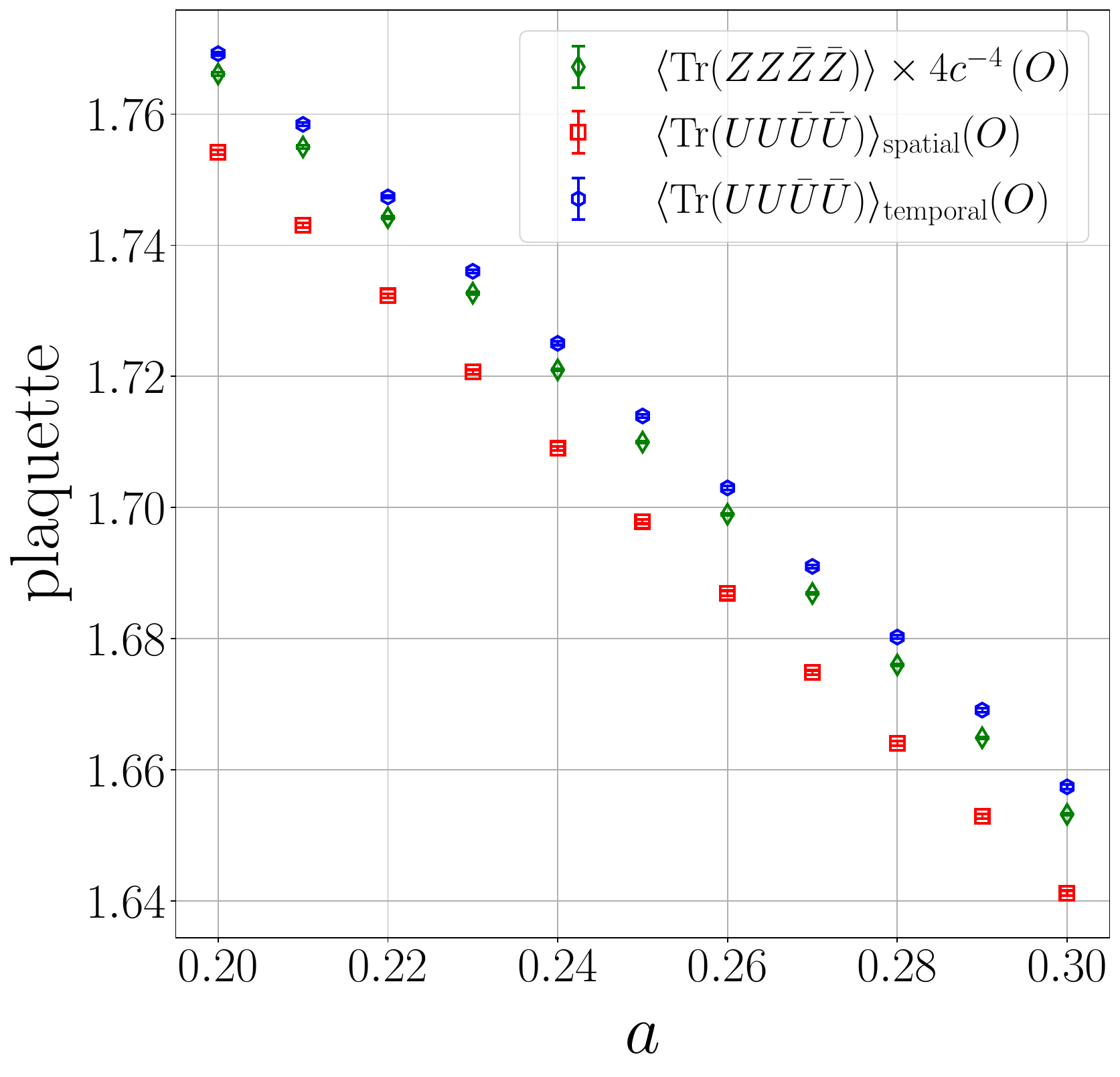}
    \end{subfigure}
    \hfill
        %
        %
    \begin{subfigure}{0.32\textwidth}
        \centering
\includegraphics[width=\linewidth]{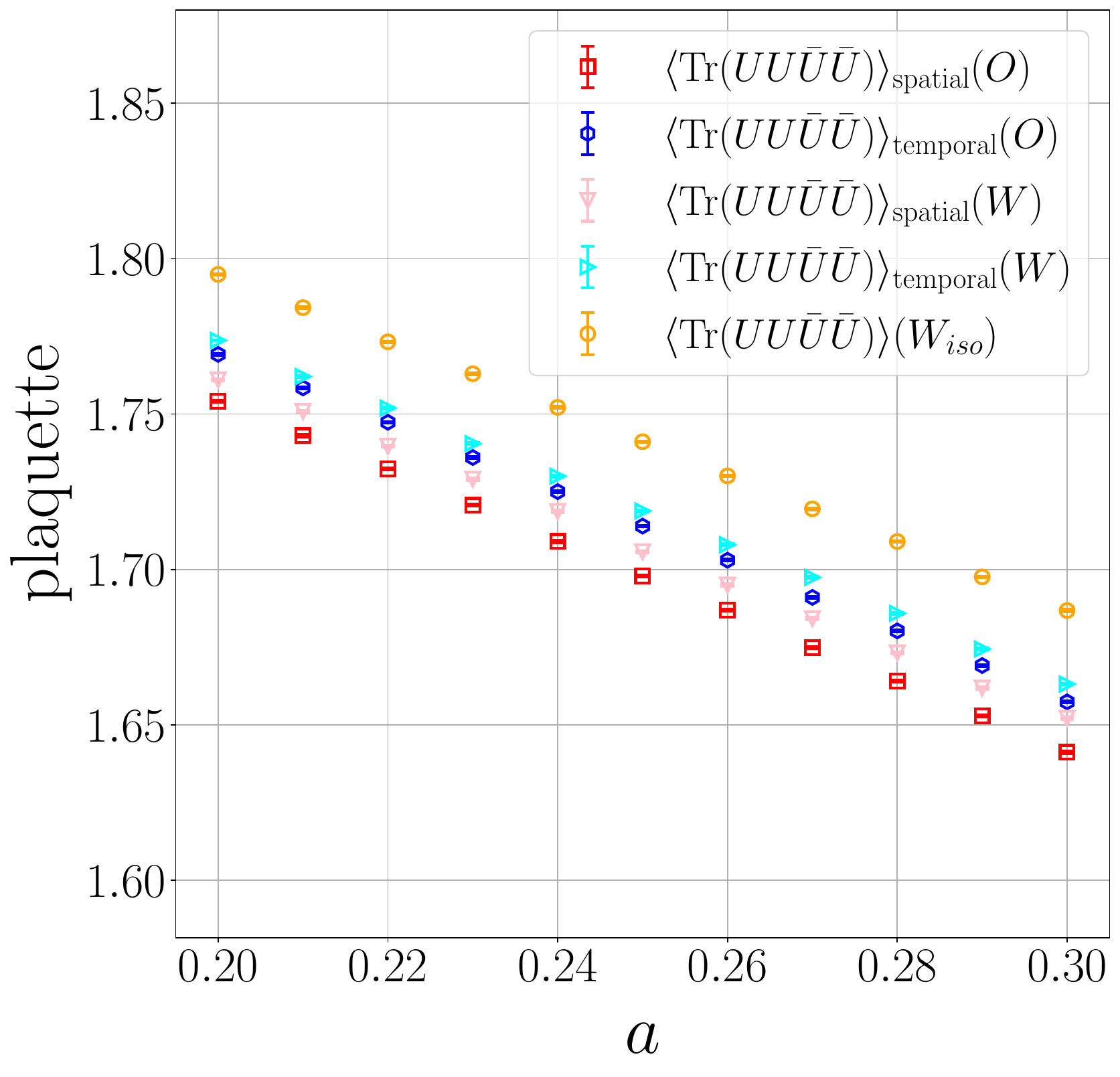}
    \end{subfigure} 
    \caption{ The original Hamiltonian $\hat{H}$, $\mathbb{R}^4$ embedding, $m^2 = 150$, $8^3$ lattice, $\gamma = 0$, with varying $a = a_t$.
    [\textbf{Left}] Effective spacing $a_{\rm eff}$ versus the lattice spacing $a$.
    [\textbf{Center}] Plaquette expectation value versus $a=a_t$. Three plaquette observables from $\hat{H}$ are shown. 
[\textbf{Right}] Plaquette expectation value versus $a=a_t$. Two plaquette observables from $\hat{H}$,    
    their counterparts from the anisotropic Wilson action with the corresponding effective lattice spacings, and the values from the isotropic Wilson action with the bare lattice spacing (for which temporal and spatial plaquettes take the same expectation value) are plotted together. (O) and (W) stands for Orbifold and Wilson, respectively.
} \label{fig:tuning_a_bare_H_m150}
\end{figure}

\begin{figure}[H]
    \centering
    %
    %
    \begin{subfigure}{0.32\textwidth}
        \centering
        \includegraphics[width=\linewidth]
        {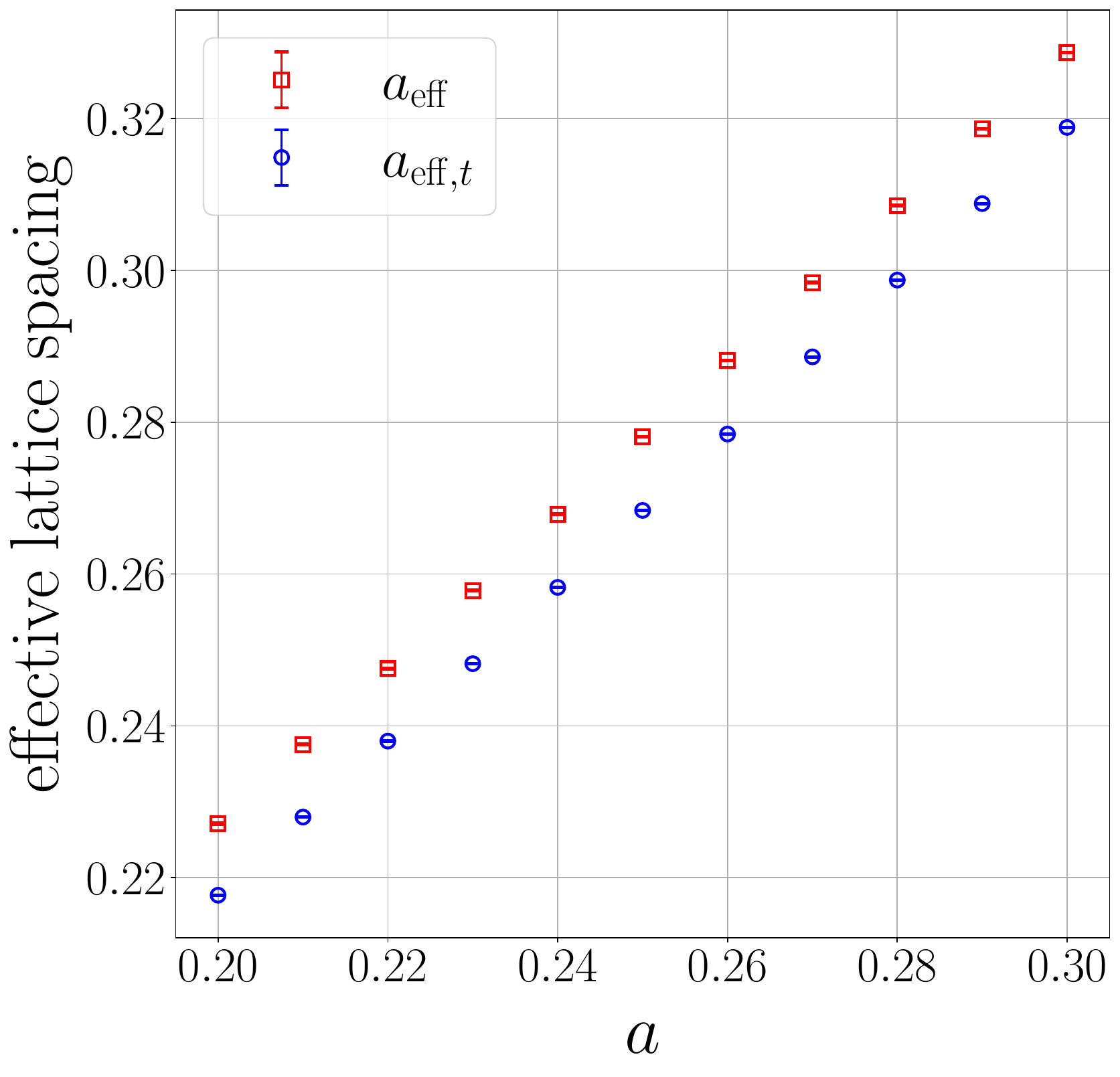}
    \end{subfigure}
    \hfill
        %
        %
    \begin{subfigure}{0.32\textwidth}
        \centering
\includegraphics[width=\linewidth]
{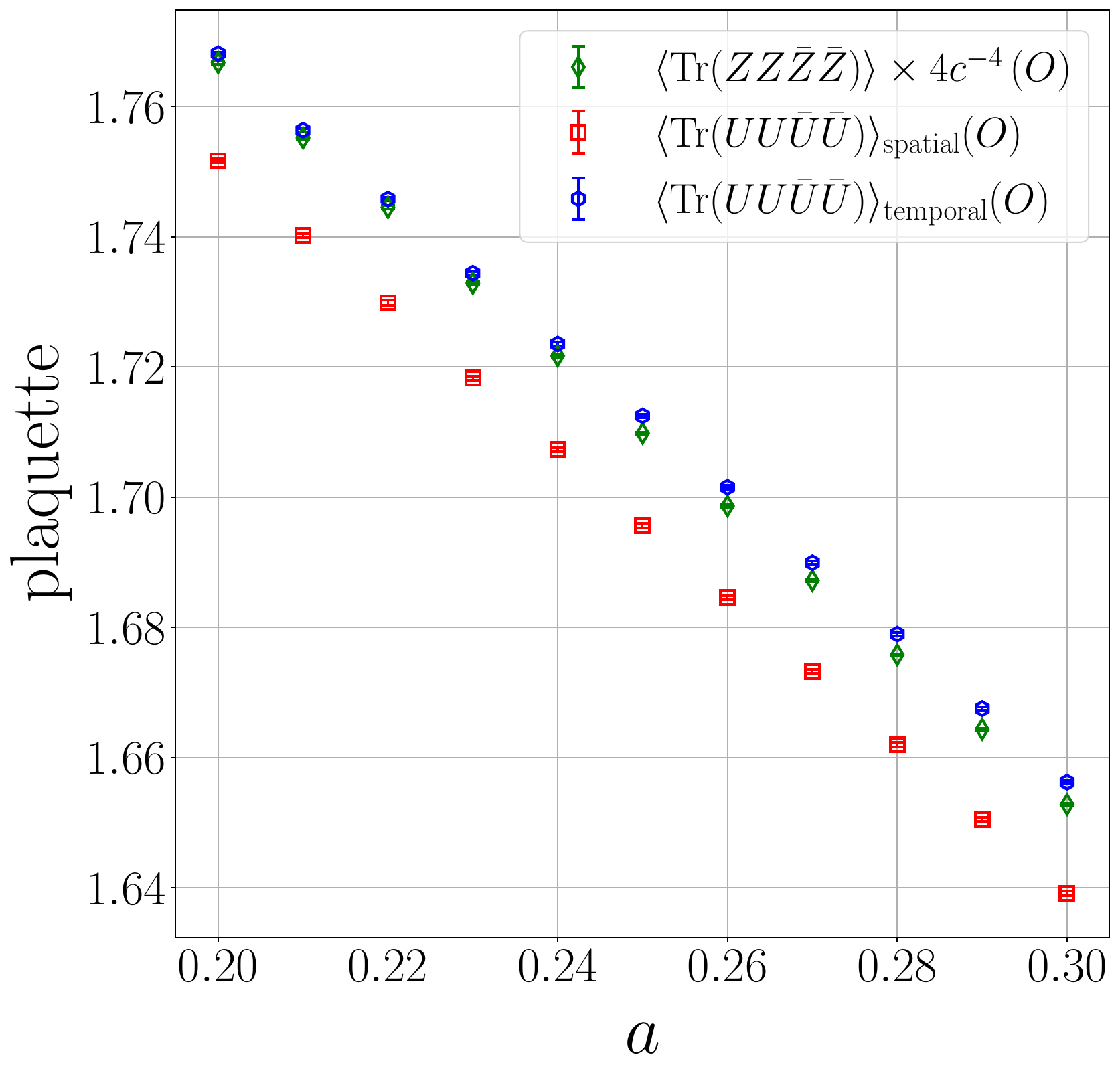}
    \end{subfigure}
    \hfill
        %
        %
    \begin{subfigure}{0.32\textwidth}
        \centering
\includegraphics[width=\linewidth]
{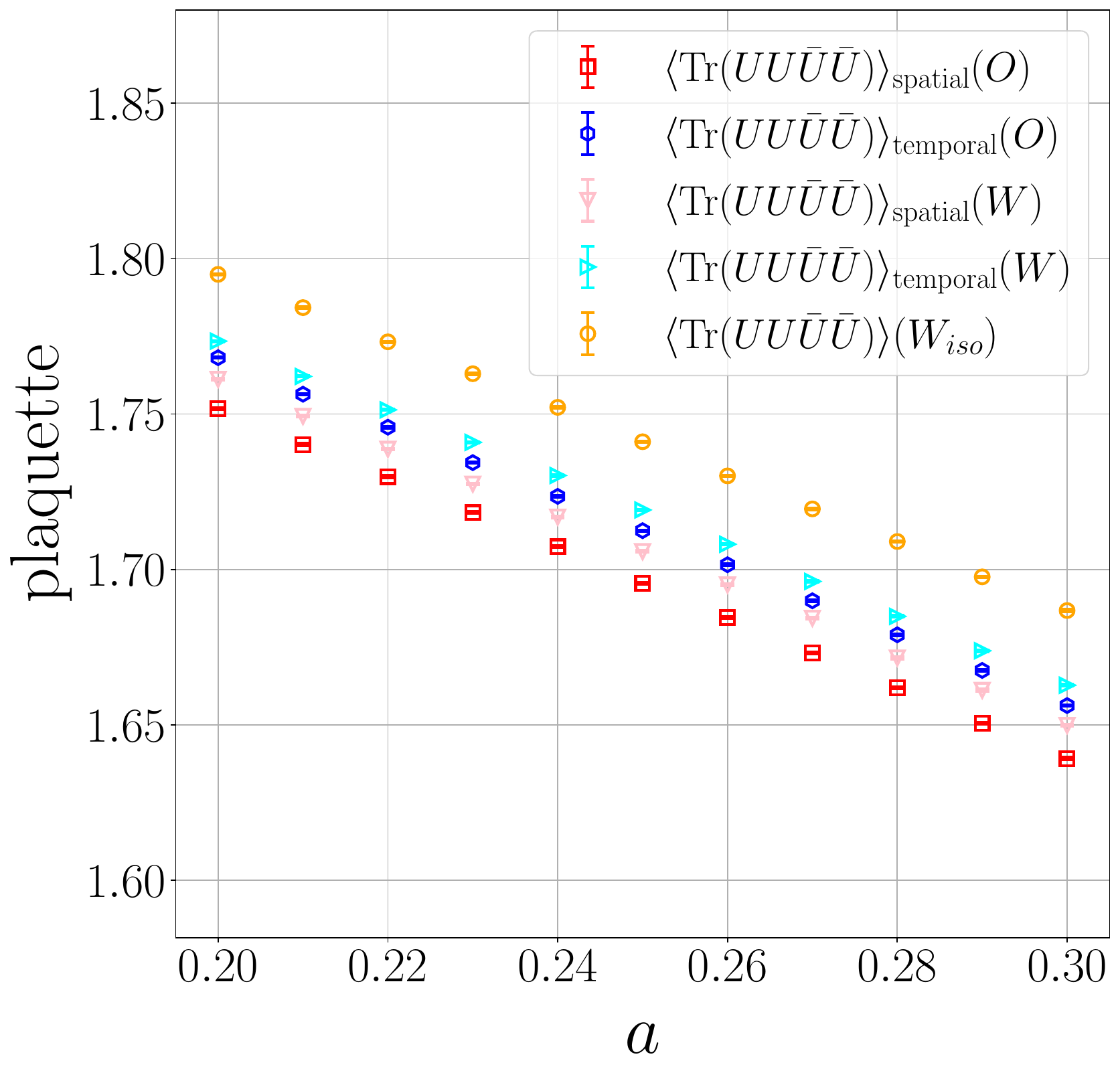}
    \end{subfigure} 
    \caption{ The original Hamiltonian $\hat{H_1}$, $\mathbb{R}^4$ embedding, $m^2 = 150$, $8^3$ lattice, $\gamma = 0$, with varying $a = a_t$.
    [\textbf{Left}] Effective spacing $a_{\rm eff}$ versus the lattice spacing $a$.
    [\textbf{Center}] Plaquette expectation value versus $a=a_t$. Three plaquette observables from $\hat{H}$ are shown. 
[\textbf{Right}] Plaquette expectation value versus $a=a_t$. Two plaquette observables from $\hat{H}$,    
    their counterparts from the anisotropic Wilson action with the corresponding effective lattice spacings, and the values from the isotropic Wilson action with the bare lattice spacing (for which temporal and spatial plaquettes take the same expectation value) are plotted together. (O) and (W) stands for Orbifold and Wilson, respectively.
} \label{fig:tuning_a_bare_H1_m150}
\end{figure}

Figures~\ref{fig:tuning_a_bare_H_m150} (for $\hat{H}$) and \ref{fig:tuning_a_bare_H1_m150} (for $\hat{H}_1$) show that the Orbifold-ish plaquette values exhibit a reduced deviation from the Wilson action plaquette values compared to the lower-mass case in Section~\ref{sec:a_bare_tuning}. This behavior indicates that a combination of effective lattice spacing tuning and increase in the scalar mass parameter $m^2$ improves the matching between the two formulations.

\begin{table}[hbtp]
\centering
\begin{subtable}{0.3\textwidth}
\centering
\begin{tabular}{|c||c|c|}
\hline
$a=a_t$ & $a_{{\rm eff}}$ & $a_{{\rm eff},t}$ \\ \hline\hline
0.20 & 0.227 & 0.217 \\
0.21 & 0.237 & 0.228 \\
0.22 & 0.247 & 0.238 \\
0.23 & 0.257 & 0.248 \\
0.24 & 0.267 & 0.258 \\
0.25 & 0.278 & 0.268 \\
0.26 & 0.288 & 0.278 \\
0.27 & 0.298 & 0.288 \\
0.28 & 0.308 & 0.298 \\
0.29 & 0.318 & 0.308 \\
0.30 & 0.328 & 0.319 \\
\hline
\end{tabular}
\caption{}
\label{tab:a_eff_H_m150}
\end{subtable}
\hfill
\begin{subtable}{0.3\textwidth}
\centering
\begin{tabular}{|c||c|c|}
\hline
$a=a_t$ & $a_{\rm eff}$ & $a_{{\rm eff},t}$ \\ \hline\hline
0.20 & 0.227 & 0.218 \\
0.21 & 0.238 & 0.228 \\
0.22 & 0.248 & 0.238 \\
0.23 & 0.258 & 0.248 \\
0.24 & 0.268 & 0.258 \\
0.25 & 0.278 & 0.268 \\
0.26 & 0.288 & 0.278 \\
0.27 & 0.298 & 0.289 \\
0.28 & 0.309 & 0.299 \\
0.29 & 0.319 & 0.309 \\
0.30 & 0.329 & 0.319 \\
\hline  
\end{tabular}
\caption{}
\label{tab:a_eff_H1_m150}
\end{subtable}
\hfill
\caption{ [\textbf{Left}] Parameter choice for Wilson simulations for Fig.~\ref{fig:tuning_a_bare_H_m150}.     
[\textbf{Right}] Parameter choice for Wilson simulations for Fig.~\ref{fig:tuning_a_bare_H1_m150}.
}
\end{table}

\bibliographystyle{utphys}
\bibliography{reference}

\end{document}